\documentclass[twocolumn,prl,showpacs,preprintnumbers,amsmath,amssymb,superscriptaddress,bibliography]{revtex4-1}
\usepackage{graphicx}
\usepackage{setspace}

\usepackage{color}
\usepackage{ulem} 
\usepackage{enumerate}  
\graphicspath{{./figures/}}

\usepackage{amsmath,amsthm,amssymb}
\usepackage{mathrsfs}

\usepackage[%dvipdfmx,
colorlinks=true, citecolor=blue, urlcolor=blue, linkcolor=blue,
setpagesize=false, bookmarks=false, breaklinks=true
]{hyperref}
\usepackage{lineno}

\newcommand{\hp}{\hat{p}}

\newcommand{\hc}{\hat{c}}
\newcommand{\hd}{\hat{d}}

\newcommand{\hH}{\hat{H}}

\newcommand{\hn}{\hat{n}}

\newcommand{\hS}{\hat{S}}
\newcommand{\hrho}{\hat{\rho}}

\newcommand{\eqq}[1]{\begin{align} #1 \end{align}}

\begin{document}
\title{Anomalous temperature dependence of high-harmonic generation in Mott insulators}

\author{Yuta Murakami}
\affiliation{Department of Physics, Tokyo Institute of Technology, Meguro, Tokyo 152-8551, Japan}
\affiliation{Center for Emergent Matter Science, RIKEN, Wako, Saitama 351-0198, Japan}
\author{Kento Uchida}
\affiliation{Department of Physics, Graduate School of Science, Kyoto University, Sakyo-ku, Kyoto 606-8502, Japan}
\author{Akihisa Koga}
\affiliation{Department of Physics, Tokyo Institute of Technology, Meguro, Tokyo 152-8551, Japan}
\author{Koichiro Tanaka}
\affiliation{Department of Physics, Graduate School of Science, Kyoto University, Sakyo-ku, Kyoto 606-8502, Japan}
\affiliation{Institute for Integrated Cell-Material Sciences, Kyoto University, Sakyo-ku, Kyoto, 606-8501, Japan}
\author{Philipp Werner}
\affiliation{Department of Physics, University of Fribourg, 1700 Fribourg, Switzerland}
\date{\today}

\begin{abstract} 
We reveal the crucial effect of strong spin-charge coupling on high-harmonic generation (HHG) in Mott insulators. 
In a system with antiferromagnetic correlations, the HHG signal is drastically enhanced with decreasing temperature, even though the gap increases and the production of charge carriers is suppressed.
This anomalous behavior, which has also been observed in recent HHG experiments on Ca$_2$RuO$_4$,
originates from a cooperative effect between the spin-charge coupling and the thermal ensemble, and the 
strongly temperature-dependent coherence between charge carriers. 
We argue that the peculiar temperature dependence 
of HHG is a generic feature of Mott insulators, which can be controlled via the Coulomb interaction and dimensionality of the system. 
Our results demonstrate that correlations between different degrees of freedom, which are a characteristic feature of strongly correlated solids, have significant and nontrivial effects on nonlinear optical responses. 
\end{abstract}

\maketitle
High-harmonic generation (HHG) is a fundamental nonlinear optical phenomenon with potentially important technological applications.
It was first reported in atomic gases~\cite{Ferray_1988} and is utilized in attosecond laser sources as well as spectroscopies\cite{Krausz2009RMP}. 
Recently its scope is extended to condensed matters because of the observation of HHG in solids, in particular semiconductors and semimetals
\cite{Ghimire2011NatPhys,Schubert2014,Luu2015,Vampa2015Nature,Langer2016Nature,Hohenleutner2015Nature,Ndabashimiye2016,Liu2017,You2016,Yoshikawa2017Science,Hafez2018,Kaneshima2018,Yoshikawa2019,Matsunaga2020PRL,Schmid2021}.
HHG in semiconductors and semimetals can be well described by the dynamics of independent electrons (independent-particle picture)
\cite{Golde2008,Vampa2014PRL,Vampa2015PRB,Wu2015,Otobe2016,Ikemachi2017,Tancogne-Dejean2017,Luu2016,Hansen2017,Osika2017,Ikeda2018PRA,Tamaya2016,Floss2018,Markus2020SOC,Chacon2020PRB,Wilhelm2021PRB,Taya2021PRB}, 
which enables the HHG spectroscopy of band information such as dispersion relations~\cite{Vampa2015PRL,Luu2015,Li2020HHG,Luu2018Amorphas,Uchida2020PRBL}.
On the other hand, the effects of electronic correlations are often taken into account phenomenologically and a detailed understanding of their role in solid-state HHG is lacking~\cite{Kemper2013b,Kruchinin2018,Ghimire2019,Du2022PRA}.
This understanding is however essential for the exploration of HHG and the application of HHG spectroscopy in correlated materials.

 %%%%%%%%%%%%%%%%%%%%%%%%%%%%%%%%%%%%%%%%%%%%%
 \begin{figure}[t]
  \centering
    \hspace{-0.cm}
    \vspace{0.0cm}
\includegraphics[width=85mm]{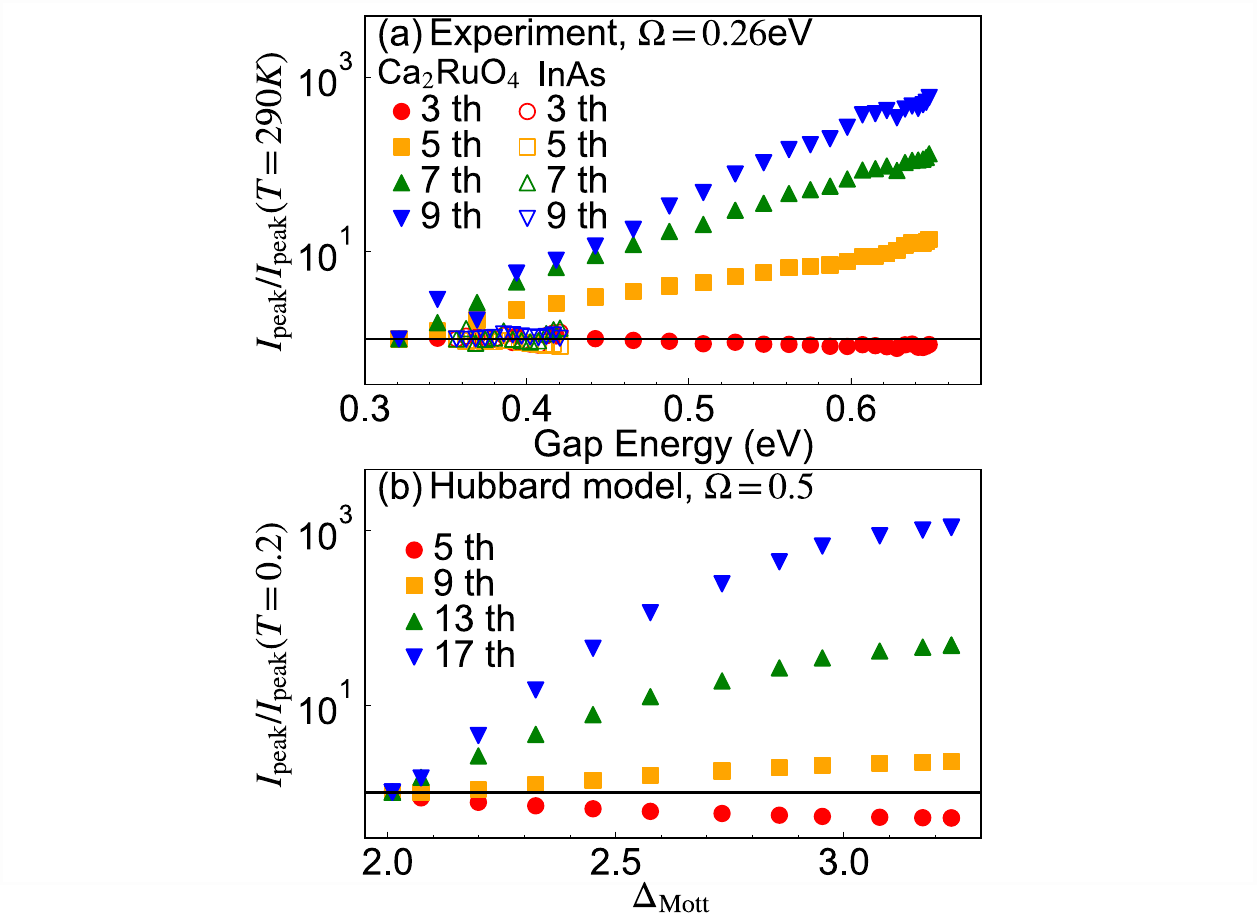} 
  \caption{(a) Experimental HHG intensity at the indicated HHG peaks as a function of the optical gap for Ca$_2$RuO$_4$ (Mott insulator) and InAs (semiconductor), reproduced from Ref.~\onlinecite{Uchida2022PRL}. The temperature $T$ is modified in the range $T\in[290$~K, $50$~K].  
 (b) DMFT results for the intensity at the indicated HHG peaks as a function of the Mott gap ($\Delta_{\rm Mott}$) for the single-band Hubbard model in the Mott insulating phase. }
%  Both in experiments and theory, the systems show a monotonic increase of the gap with decreasing temperature.}
  \label{fig:DMFT_Bethe_HHG_demo}
\end{figure}
%%%%%%%%%%%%%%%%%%%%%%%%%%%%%%%%%%%%%%%%%%%%

The new research frontier of HHG in strongly correlated systems (SCSs) has
attracted considerable interest both on the theoretical \cite{Silva2018NatPhoton,Murakami2018PRL,Murakami2018PRB,Markus2020,Tancogne-Dejean2018,Ishihara2020,Chinzei2020,Orthodoxou2021,Murakami2021PRB,Shao2022PRL,Hansen2022arXiv,Bondar2022arxiv,Udono2022PRB} and experimental \cite{vaskivskyi2020,Bionta2021PRR,Uchida2022PRL} sides. 
In contrast to semiconductors, 
which can be described in terms of electrons and holes, 
the driven state of SCSs involves
various types of many-body elemental excitations. 
This makes the mechanism and features of HHG in SCSs nontrivial. 
Previous studies revealed the direct connection between many-body excitations and HHG in SCSs~\cite{Murakami2018PRL,Ishihara2020,Murakami2021PRB}, suggesting possible spectroscopic applications of HHG to detect many-body states~\cite{Murakami2021PRB} as well as photoinduced phase transitions~\cite{Silva2018NatPhoton}.  
On the other hand, very recently, 
 an unexpected exponential enhancement of the HHG signal with increasing gap size is reported in the Mott insulator Ca$_2$RuO$_4$~\cite{Uchida2022PRL}, see Fig.~\ref{fig:DMFT_Bethe_HHG_demo} (a).
Since a larger gap should suppress the excitation of charge carriers,
this increase is opposite to the behavior expected in semiconductor HHG. 
Such a counter-intuitive result calls for a deeper theoretical understanding of HHG in SCSs.
A hallmark of SCSs is the coupling between different degrees of freedom, such as charges, orbitals and spins.
These correlations are at the origin of rich physical properties observed in equilibrium SCSs \cite{Tokura_RMP,Dagotto1994RMP}.
However, their role in highly nonlinear optical phenomena such as HHG is hardly known.

In this letter, we reveal the crucial role of spin-charge coupling on HHG in Mott insulators analyzing the Hubbard model.
Previous works showed that HHG in Mott insulators originates from the coherent dynamics of a pair of local many-body states -- a doublon (doubly occupied state) and holon (empty state) -- generated by strong fields, where the three-step model picture is applicable~\cite{Murakami2018PRL,Murakami2021PRB}. 
The kinematics of doublons and holons is strongly correlated with spins, since their hopping disturbs the spin background.
We demonstrate that this spin-charge coupling and its cooperation with thermal fluctuations produces a drastic enhancement of the HHG intensity,  accompanied with an increasing Mott 
gap, as observed in Ca$_2$RuO$_4$ (Fig.~\ref{fig:DMFT_Bethe_HHG_demo}(b)). 
These results demonstrate that strong correlations between active degrees of freedom in SCSs can result in  counter-intuitive behaviors of highly nonlinear optical phenomena such as HHG. 
%Our theoretical insights provide guiding principles for the exploration of HHG in SCSs, and suggest new ways of exploring the properties of SCSs far from equilibrium.

We focus on the single-band Hubbard model, which is a standard model for SCSs.
The Hamiltonian is
\eqq{ 
\hH(t) = -t_{\rm hop} \sum_{\langle ij\rangle} e^{i\phi_{ij}(t)} \hc^\dagger_{i\sigma} \hc_{j\sigma} + U \sum_i \hn_{i\uparrow}\hn_{i\downarrow},
}
where $\hc^\dagger_{i\sigma}$ is the creation operator for an electron with spin $\sigma$ at site $i$, $\langle ij\rangle$ indicates a pair of neighboring sites, and 
$\hn_{i\sigma} = \hc^\dagger_{i\sigma} \hc_{i\sigma}$.
$t_{\rm hop}$ is the hopping parameter and $U$ the onsite interaction. The electric field is included via a Peierls phase $\phi_{ij}$, see Supplemental Material (SM)  \footnote{Supplemental Material [url], which includes Ref. \cite{Werner2018PRB}.}.
We mainly use the nonequilibrium dynamical mean-field theory (DMFT)~\cite{Georges1996,Eckstein2010b,Aoki2013,Kilian2019PRL,Nessi2020} to solve this problem, and focus on the Bethe lattice for simplicity~\cite{Werner2017}. The qualitatively same results are obtained for the two-dimensional square lattice, see \cite{Note1}.
In the following, we use the quarter of the bandwidth at $U=0$ as the energy unit, and mainly consider $U=6$.
If our energy unit corresponds to $0.5$~eV, the Mott gap ($\Delta_{\rm Mott}\simeq 3$, see below) corresponds to 1.5~eV. %, which is a typical gap size of cuprates.
This is a typical gap size of cuprates, which are often described by the Hubbard model.

\begin{figure}[t]
  \centering
    \hspace{-0.cm}
    \vspace{0.0cm}
\includegraphics[width=87mm]{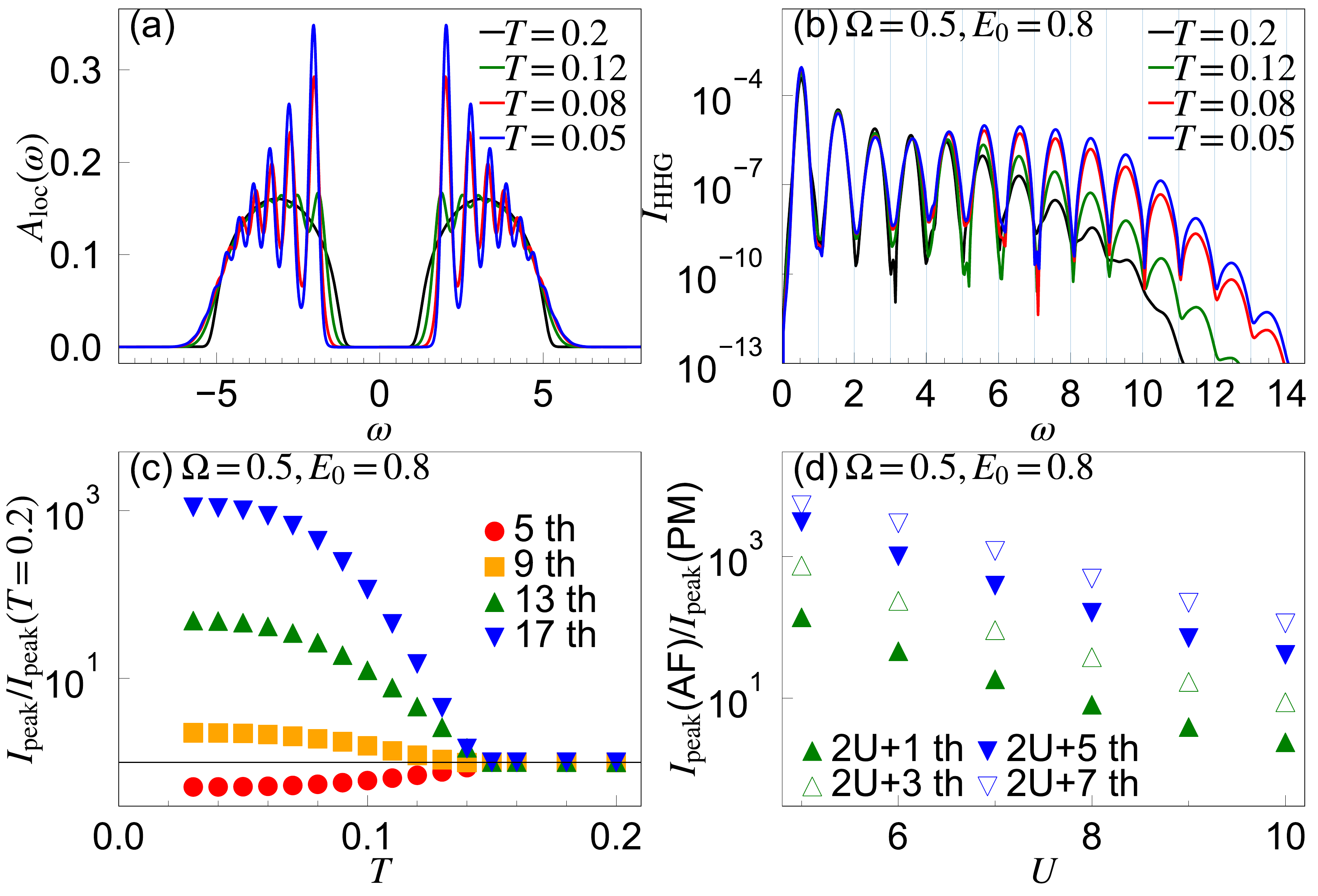} 
  \caption{(a) Local spectral functions, $A_{\rm loc}(\omega)$, in equilibrium. 
(b) HHG spectra of the Mott insulator computed with DMFT for various $T$.
  (c) The intensity at the peaks of the HHG spectra as a function of $T$.  The peak intensity is normalized by the value at $T=0.2$ (PM phase). 
  For (a-c), we use $U=6$.
 (d) $U$-dependence of the increase ratio of the HHG peaks. In order to take into account the change of the Mott gap, we compare the $(2U+n)$th HHG peaks. 
 We use $T=0.2$ for the PM phase, while we use $T=0.3/U$ for the AF state to take account of the change of $J_{\rm ex}\propto \frac{1}{U}$. 
The excitation parameters are $E_0=0.8$, $\Omega=0.5$, $t_0=75$ and $\sigma=15$. }
  \label{fig:DMFT_Bethe_HHG}
\end{figure}

We consider the half-filled system, which becomes a Mott insulator for large enough $U$ in equilibrium. 
While the Mott insulator can be realized in the paramagnetic (PM) phase, the system on the bipartite lattice exhibits an antiferromagnetic (AF) phase below the N\'eel temperature $T_c$ $(\simeq 0.15)$. 
The corresponding evolution of the single-particle spectra is shown in Fig.~\ref{fig:DMFT_Bethe_HHG}(a).
With decreasing temperature $T$, the Mott gap $\Delta_{\rm Mott}$ increases. In the PM phase, the upper and lower Hubbard bands are featureless. 
On the other hand, in the AF phase, peak structures develop within the bands, indicating the formation of spin-polarons~\cite{Martinez1991PRB,Dagotto1994RMP,
Sangiovanni2006}.
When an electron is added to (removed from) the system, a doublon (holon) is created, see \cite{Note1} for schamatics. 
When this doublon (holon) moves around, it can disturb the spin background at the cost of multiples of the exchange energy $J_{\rm ex}$  $(= \frac{4t_{\rm hop}^2}{U})$.
This results in strong spin-charge coupling, % in Mott insulators, 
of which the spin-polaron is one manifestation.

%%%%%%%%%%%%%%%%%%%%%%%%%%%%%%%%%%%%%%%%%%%%%
%%%%%%%%%%%%%%%%%%%%%%%%%%%%%%%%%%%%%%%%%%%%

Now we discuss %the role of the strong spin-charge coupling, i.e. 
the kinematics of doublons and holons accompanied by a disturbance of the spin configurations, and its effect on highly nonlinear optical phenomena.
We study the $T$-dependence of HHG in Mott insulators excited with frequency $\Omega$ smaller than the Mott gap $\Delta_{\rm Mott}$.
We mainly use $\Omega=0.5$ in the following. 
If our energy unit corresponds to $0.5$~eV, this is a mid-infrared excitation with $0.25$~eV, whose period $T_p$ is about $16$~fs.
From the $T$-dependence of the spectral functions, one would naively speculate that the HHG intensity is suppressed by lowering temperature, since the enhancement of the gap reduces the tunneling probability (see \cite{Note1}) and the formation of the spin-polarons suggests a reduced mobility of the charge carriers. 
However, the $T$-dependence turns out to be {\it opposite} to this naive expectation.

Applying a Gaussian electric field pulse $E(t)$, we evaluate the HHG intensity $I_{\rm HHG}(\omega)$ from the Fourier transformation of the current $J(t)$ as $I_{\rm HHG}(\omega)=|\omega J(\omega)|^2$.
The pulse is characterized by the standard deviation $\sigma$, the center $t_0$ and the maximum field strength $E_0$.
We show the resulting HHG spectra for various temperatures in Fig.~\ref{fig:DMFT_Bethe_HHG}(b) and plot the $T$-dependence of the relative intensity of the HHG peaks in Fig.~\ref{fig:DMFT_Bethe_HHG}(c).
$I_{\rm HHG}(\omega)$ is strongly enhanced above $\Delta_{\rm Mott}$ and the width of the HHG plateau
is enhanced with decreasing temperature.
The increase in the ratio of HHG signals is larger for the higher harmonic peaks.
Above $T_c$, the $T$-dependence becomes very weak. 	
As a function of the gap, the intensity increases almost exponentially, as illustrated in Fig.~\ref{fig:DMFT_Bethe_HHG_demo}(b).	
Importantly, the DMFT results of the simple Hubbard model reproduce the qualitative features of the HHG spectrum 
and the empirical scaling law observed in Ca$_2$RuO$_4$~\cite{Uchida2022PRL} (see Fig.~\ref{fig:DMFT_Bethe_HHG_demo}(a) and \cite{Note1}).

To reveal the origin of this $T$-dependence, we consider the $U$-dependence of the relative increase 
of the HHG signal (Fig.~\ref{fig:DMFT_Bethe_HHG}(d)).
For large $U$ the bandwidth of the upper and lower Hubbard bands is insensitive to $U$ and the Mott gap scales almost linearly with $U$.
Therefore, in order to focus on the contribution from the kinetic energy of the doublon-holon pair, we compare $I_{\rm HHG}(\omega)$ for  the same $\omega-U$. 
It turns out that the increase ratio monotonically decreases with increasing $U$.
Since $J_{\rm ex}$ is reduced with increasing $U$, the disturbance of the spin background costs less energy, and the spin-charge coupling becomes weaker. 
Hence, the $U$-dependence of the HHG increase ratio suggests that the anomalous $T$-dependence of HHG is related to the spin-charge coupling.

\begin{figure}[t]
  \centering
    \hspace{-0.cm}
    \vspace{0.0cm}
\includegraphics[width=88mm]{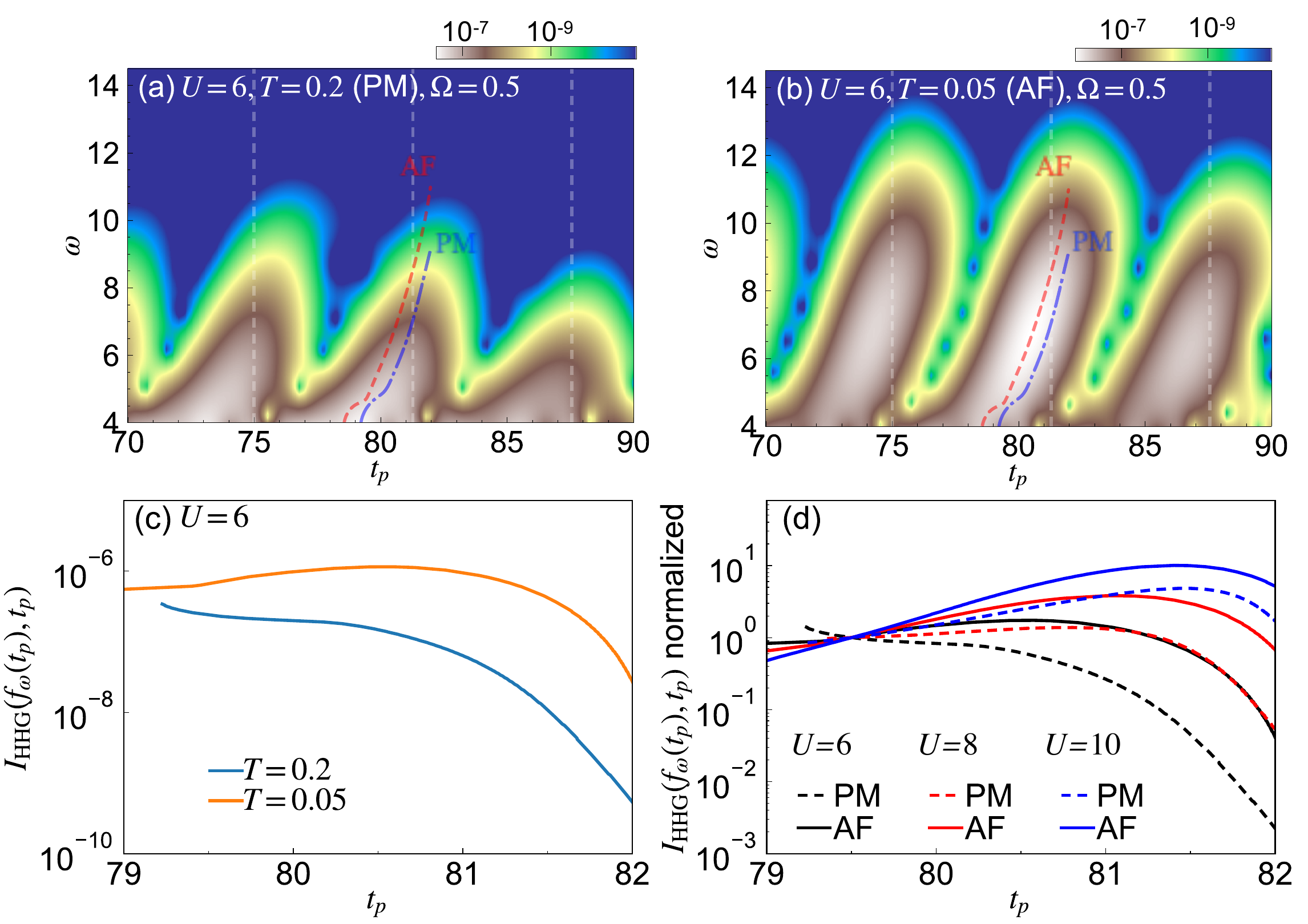} 
  \caption{(a),(b) Subcycle spectra $I_{\rm HHG}(\omega,t_p)$ for $U=6$ at (a) $T=0.2$ (PM phase) and at (b) $T=0.05$ (AF phase). A Gaussian window with standard deviation  $\sigma'=0.9$ is used. 
  The red dashed (blue dot-dashed) lines indicate the maxima of $I_{\rm HHG}(\omega,t_p)$ at $T=0.05$ ($T=0.2$) at a given $\omega$ as a function of $t_p$ around $t_p=80$, which define the function $f_\omega (t_p)$.
  The vertical dashed lines indicate the times when the electric field $E(t)=0$. 
  (c) Intensity $I_{\rm HHG}(\omega,t_p)$ along the lines $f_\omega (t_p)$ for $U=6$.  (d) Normalized intensity $I_{\rm HHG}(\omega,t_p)$ along the lines $f_\omega (t_p)$ for the indicated values of $U$. 
  We use $T=0.2$ for the PM phase and $T=0.3/U$ for the AF states to take account of the change of $J_{\rm ex}\propto \frac{1}{U}$.
  $I_{\rm HHG}(f_\omega (t_p), t_p)$ is renormalized by the value at $t_p=79.5$ in each case. 
  The excitation parameters are the same as in Fig.~\ref{fig:DMFT_Bethe_HHG}. 
  }
  \label{fig:HHG_subcycle}
\end{figure}

Next we perform a subcycle analysis considering a windowed Fourier transform $J(\omega,t_p) = \int dt e^{i\omega t}  F_{\rm window}(t-t_p)J(t)$ and evaluating $I_{\rm HHG}(\omega,t_p)\equiv|\omega J(\omega,t_p)|^2$.
$I_{\rm HHG}(\omega,t_p)$ provides the time-resolved spectral features of the emitted light around $t_p$.
Since HHG in Mott insulators mainly originates from the recombination of doublon-holon pairs~\cite{Murakami2018PRL,Murakami2021PRB}, the subcycle spectra reveal the recombination time of the pairs and their energy at that time.
In Figs.~\ref{fig:HHG_subcycle}(a),(b), we show $I_{\rm HHG}(\omega,t_p)$ in the PM and AF phases. 
In both cases, the dominant intensity appears at early times within one period, suggesting that only short trajectories of the doublon-holon pairs contribute to the HHG signal.
In other words, the coherence time of the doublon-holon pair is very short ($<T_p/4$)  compared to one cycle of the pulse field 
and to the coherence times typically considered in the analysis of semiconductors, see e.~g. Fig.~6 in Ref.~\onlinecite{Vampa2015PRB}.
The kinematics estimated from the peak position of  $I_{\rm HHG}(\omega,t_p)$ at each $\omega$ as a function of $t_p$ is represented with red dashed (blue dot-dashed) lines for the AF (PM) phase in Figs.~\ref{fig:HHG_subcycle}(a),(b). These lines define the function $f_\omega (t_p)$. 
 The difference between the blue and red lines is mostly explained by the difference in the gap size ($\simeq 1.1$), indicating that the trajectory of the doublon-holon pair is 
 almost the same in the AF and PM phases.
The main difference is the coherence time of the pair.

%%%%%%%%%%%%%%%%%%%%%%%%%%%%%%%%%%%%%%%%%%%%%
%%%%%%%%%%%%%%%%%%%%%%%%%%%%%%%%%%%%%%%%%%%%

To quantify this, we show in Fig.~\ref{fig:HHG_subcycle}(c) the intensity along the peaks, $I_{\rm HHG}(f_\omega (t_p), t_p)$. 
The results indeed show that for the higher $T$ the intensity decays faster, suggesting that the dephasing time of the doublon-holon pair is shorter.
This is in a stark contrast with the behavior of the charge distribution, where the absence of the AF spin background 
at high $T$ leads to a slower relaxation~\cite{Zala2013PRL,Denis2014PRB,Eckstein2016}.
On the other hand, with increasing $U$, 
the behavior of $I_{\rm HHG}(f_\omega (t_p), t_p)$ 
in the AF and PM phases becomes more similar, 
see Fig.~\ref{fig:HHG_subcycle}(d).
Furthermore, the peak in $I_{\rm HHG}(f_\omega (t_p), t_p)$ becomes clearer, which indicates that the intensity coming from longer-time trajectories of the doublon-holon pairs and hence the coherence time are increased. 
This feature appears counter-intuitive, because the single-particle spectrum becomes highly incoherent for large $U$~\cite{Dagotto1994RMP,Sangiovanni2006}, and demonstrates that HHG in SCSs is not directly related to the single-particle spectra, in contrast to semiconductors~\cite{Murakami2018PRL,Ishihara2020,Murakami2021PRB}.

 %%%%%%%%%%%%%%%%%%%%%%%%%%%%%%%%%%%%%%%%%%%%%
 \begin{figure}[t]
  \centering
    \hspace{-0.cm}
    \vspace{0.0cm}
\includegraphics[width=85mm]{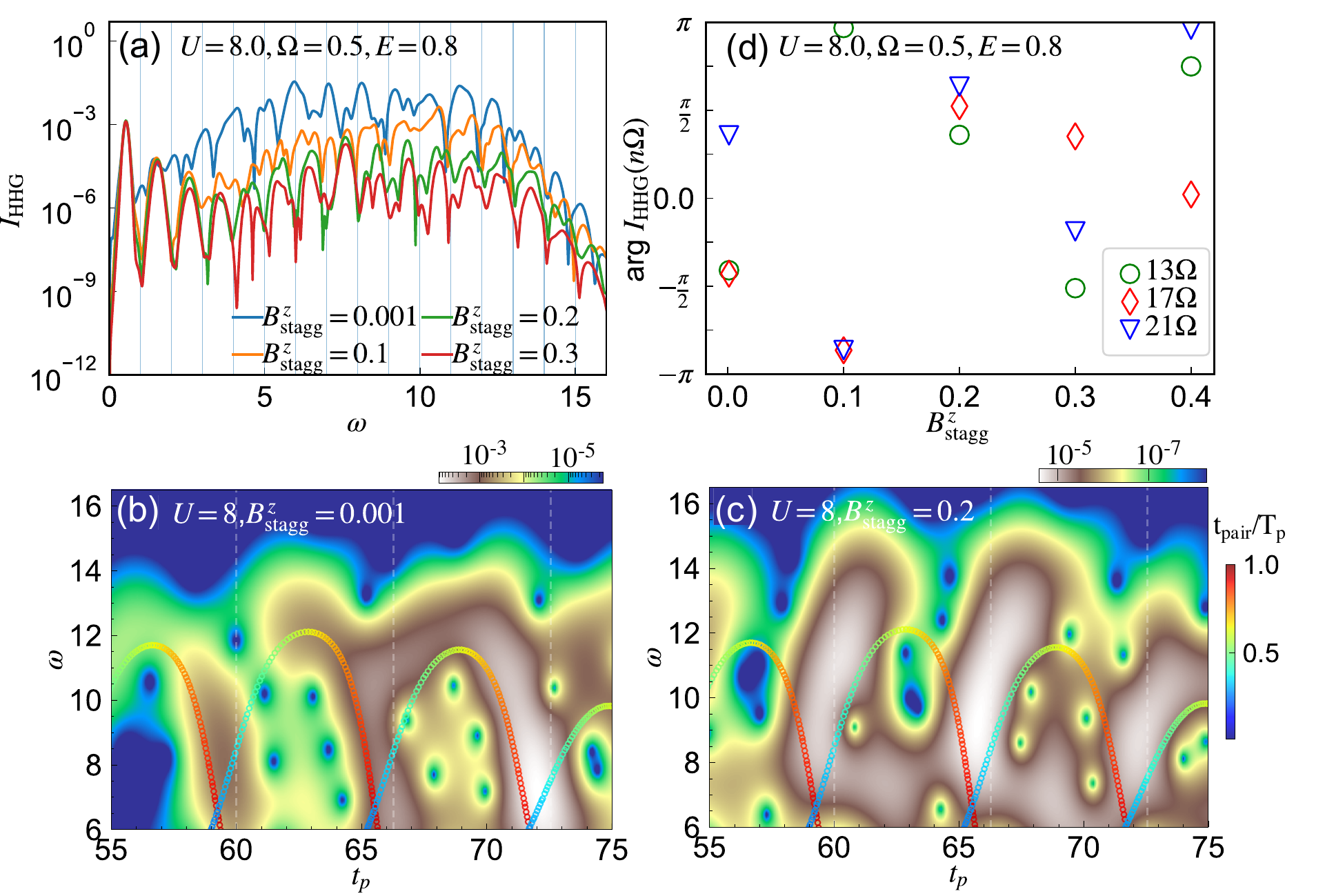} 
  \caption{(a) $I_\text{HHG}$ for different $B^z_\text{stagg}$ and subcycle analysis for $B^z_{\rm stagg}=0.001$ (b) and $B^z_{\rm stagg}=0.2$ (c). 
A Gaussian window with $\sigma'=0.9$ is used. 
  The colored markers indicate the energy emitted at $t_p$ by the recombination of a doublon-holon pair, which is predicted from the three-step model using the doublon and holon dispersions from the Bethe ansatz~\cite{Murakami2021PRB}. 
  The color indicates the time interval between the recombination and the creation of the doublon-holon pair $t_{\rm pair}$, and $T_p=\frac{2\pi}{\Omega}$.
 (d) Phase of the Fourier component of $J(\omega)$ at $\omega = n\Omega$ ($n$ is an integer). In all panels, we set $U=8$, and the excitation parameters are $\Omega=0.5,E_0=0.8$, $t_0=60$ and $\sigma=15$. }
  \label{fig:iTEBD_HHG}
\end{figure}
%%%%%%%%%%%%%%%%%%%%%%%%%%%%%%%%%%%%%%%%%%%%s

These behaviors can be consistently explained in terms of the spin-charge coupling.
To directly compare cases with and without spin-charge coupling, we switch to the one-dimensional (1D) Hubbard model with a staggered magnetic field $B^z_{\rm stagg}$.
In one dimension, without $B^z_{\rm stagg}$, the kinematics of the doublons and holons is independent of the spin-degrees of freedom (spin-charge separation), 
while for $B^z_{\rm stagg}\ne 0$, the hopping of a doublon (holon) creates a mismatch between the staggered field and the spin configuration, as it happens in higher-dimensional systems without field, see \cite{Note1} for schematics.
With this set-up, the 1D model can mimic the spin-charge coupling in higher dimensions.
The infinite time-evolving block decimation (iTEBD)~\cite{Vidal2003PRL} allows to compute accurate results for this model at $T=0$ in the thermodynamic limit.

We show the HHG spectra for various $B^z_{\rm stagg}$ in Fig.~\ref{fig:iTEBD_HHG}(a) and the corresponding subcycle analysis in Figs.~\ref{fig:iTEBD_HHG}(b),(c).
For small $B^z_{\rm stagg}$, the expected HHG peaks at $(2n+1)\Omega$ in $I_{\rm HHG}(\omega)$ are not clear, suggesting that the system is not fully time periodic during the pulse.
This is attributed to the long coherence time of the doublon-holon pair, which leads to the interference of many quasi-classical trajectories within the three-step model~\cite{Vampa2015PRB}.
Indeed, the subcycle spectra for small $B^z_{\rm stagg}$ suggest that long trajectories of doublon-holon pairs strongly contribute to the HHG signal, see Fig.~\ref{fig:iTEBD_HHG}(b)~\cite{Murakami2021PRB}. With increasing $B^z_{\rm stagg}$, the HHG intensity becomes weaker but the HHG peaks become clearer at  $(2n+1)\Omega$.  Here, $B^z_{\rm stagg}$ is chosen to be comparable to $J_{\rm ex}$.
In the subcycle spectrum, the weight is shifted to earlier times in one period, see Fig.~\ref{fig:iTEBD_HHG}(c), as it is the case in the DMFT results in Fig.~\ref{fig:HHG_subcycle}(b) at low $T$.
These results show that the coherence time of the doublon-holon pair is efficiently suppressed by the spin-charge coupling, which consistently explains the behavior of the DMFT results.
The short coherence time reduces the interference between different quasi-classical trajectories and results in clear HHG peaks both in the DMFT data and the iTEBD data for nonzero $B^z_{\rm stagg}$.

The reduction of the coherence time of the doublon-holon pair with increasing $T$ can be understood as a cooperative effect of the spin-charge coupling and the thermal ensemble.
At nonzero temperatures, the initial equilibrium state is described by an ensemble of eigenstates, represented by the density matrix $\hrho\propto e^{-\beta \hH}$.
In such a system, the total current induced by the field can be calculated as the ensemble average over the individual currents evaluated for these eigenstates.
With increasing $T$, the weight of the high-energy states increases. In our case, at higher temperatures, spin configurations different from the AF ground state are activated, see \cite{Note1}.
The dynamics of the doublon or holon is different for each configuration, since the energy transfer to the spin background during an excursion depends on the spin configuration.
This should produce emitted light with different phases for different spin configurations, resulting in phase cancellations after the ensemble average, and thus reduce the coherence between the doublon-holon pairs with increasing $T$.
Note that this effect does not rely on long-range magnetic ordering and is also relevant in the PM phase, but is absent without spin-charge coupling.
Namely, for small $J_{\rm ex}$, weaker cancellations between different spin configurations are expected, which explains the results in Fig.~\ref{fig:HHG_subcycle}(d) and the reduction of the enhancement of the HHG signal with larger $U$ in Fig.~\ref{fig:DMFT_Bethe_HHG}(d).
To exemplify that the spin-charge coupling can indeed provide such phase shifts, in Fig.~\ref{fig:iTEBD_HHG}(d), we show the $B^z_{\rm stagg}$-dependence of the phase of $J(\omega)$ for $\omega = n\Omega$ (with $n$ some integer). The result suggests that the phase is sensitive to $B^z_{\rm stagg}$, which supports the above argument.
Hence, the modification of the coherence time due to the spin-charge coupling and thermal fluctuations dominates over the reduction of the tunneling rate by the gap opening, leading to an enhancement of $I_\text{HHG}$ at lower temperatures.  

The strong $T$-dependence of the HHG spectrum observed in Mott insulators is not expected in typical semiconductors.
In the theoretical analysis of HHG in semiconductors, a short dephasing time $T_2$ of a few fs for an electron-hole pair is often used.
The main origin of the fast dephasing is the experimental setup, i.e. the dephasing by the propagation of light and the inhomogeneity of the field strength~\cite{Floss2018,Kilen2020PRL},
which is insensitive to temperature.
Another relevant factor is the electron-electron scattering among excited carriers in semiconductors~\cite{Becker1988PRL,Nagai2022}. 
Still, this is also expected to be insensitive to temperature, since thermal fluctuations cannot efficiently excite carriers across the gap.
These considerations are supported by the experimental HHG spectrum for the semiconductor InAs shown in Fig.~\ref{fig:DMFT_Bethe_HHG_demo}(a).

In summary, our theoretical study revealed important effects of strong spin-charge coupling on the coherent carrier dynamics in Mott insulators, which lead to the counter-intuitive enhancement of HHG accompanied by a gap enhancement.
Spin-charge coupling is inevitable in Mott insulators in dimensions larger than one, so that this peculiar behavior should be a generic feature of HHG in SCSs (see \cite{Note1}).
In addition, in multiorbital systems like Ca$_2$RuO$_4$, the orbital-charge coupling should have a similar effect as the spin-charge coupling (see \cite{Note1})~\cite{Hugo2017PRB}.
These insights demonstrate the important role of correlations in highly nonlinear optical responses and provide useful guidance for the future exploration of HHG in SCSs.
On the one hand, our results suggest that the $T$-dependence of $I_\text{HHG}$ can be controlled by changing the ratio $\frac{U}{t_{\rm hop}}$,
which is feasible with the application of chemical or physical pressure. 
On the other hand, to realize a strong HHG signal, 1D Mott systems are more favorable than higher-dimensional ones due to the absence of spin-charge coupling.
The recovery of coherence and the possible increase of the HHG intensity due to the reduction of the dimensionally could be systematically analyzed by exploiting the dimensional crossover in ladder-type compounds such as  Sr$_{n-1}$Cu$_{n+1}$O$_{2n}$~\cite{Tokura_RMP}. 
Furthermore, the sensitivity of HHG to the temperature and spin-charge coupling suggests possible HHG-based techniques for detecting and characterizing thermal and non-thermal phases, and for measuring the strength of the spin-charge coupling. In the future, it will also be interesting to study HHG with more sophisticated methods such as cluster DMFT to reveal the role of magnetic fluctuations.

\begin{acknowledgments}
The calculations have been performed on the Beo05 cluster at the University of Fribourg. 
This work is supported by Grant-in-Aid for Scientific Research from JSPS, KAKENHI Grant Nos. JP20K14412(Y.M.), JP21H05017(Y.M.,K.U.,K.T.), JP19H05821, JP18K04678, JP17K05536 (A.K.), JST CREST Grant No. JPMJCR1901 (Y.M.), and ERC Consolidator Grant No.~724103 (P.W.). The nonequilibrium DMFT calculations have been implemented using the open source library Nessi~\cite{Nessi2020}.
\end{acknowledgments}

\clearpage 

\section{Non-equilibrium dynamical mean-field theory}
Dynamical mean-field theory (DMFT) is a powerful theoretical framework that can deal with strongly correlated systems~\cite{Georges1996}, and in particular Mott physics.  Nonequilibrium DMFT is the extension of DMFT to nonequilibrium problems~\cite{Aoki2013}.
The DMFT approach gives reliable results for high-dimensional systems. This has been confirmed also for the nonequilibrium dynamics by 
a recent ab-initio comparison between nonequilibrium DMFT and cold atom quantum simulators for a three dimensional system.\cite{Kilian2019PRL}
DMFT is based on the Green's function formalism, and in nonequilibrium DMFT the Green's functions $G(t,t')$ are defined on the so-called L-shape contour, which includes the Matsubara (imaginary time) branch and a real-time contour.\cite{Aoki2013}
The information on the temperature enters through the Matsubara branch, which describes the initial equilibrium system (at $t=0$) at a given temperature.

To evaluate the Green's functions, in DMFT, we map the lattice system to an effective impurity model with a self-consistently determined time-dependent noninteracting bath~\cite{Aoki2013}.
In the present case, we consider the single-band Hubbard model with a possible antiferromagnetic (AF) order, and assume that the spins can be polarized along the $z$ axis.
In the AF phase, the two sublattices A and B show opposite magnetizations and effective impurity models are introduced for each sublattice.
The action of the impurity model for the sublattice $\alpha$ can be expressed as 
\small
\eqq{
\mathcal{S}^\alpha_{\rm imp} = -i\int dt dt' \sum_\sigma \hd^\dagger_{\sigma}(t) \Delta^\alpha_\sigma (t,t') \hd_\sigma(t')-i\int dt \hH_{\rm loc}(t),
}
\normalsize
where $\Delta$ is the hybridization function and $\hH_{\rm loc}(t)=-\mu\sum_{\sigma} \hd^\dagger_\sigma(t) \hd_\sigma(t) + U  \hd^\dagger_\uparrow(t) \hd_\uparrow(t) \hd^\dagger_\downarrow(t) \hd_\downarrow(t)$.
We note that here the times $t,t'$ are defined on the L-shaped contour.
In DMFT, $\Delta$ is self-consistently determined such that the local Green's functions and the self-energies of the lattice are the same as those of the impurity model.
We solve the impurity model using the non-crossing approximation (NCA)~\cite{Eckstein2010b}, which yields reliable results in the strong coupling regime.

The lattice self-consistency condition for the hybridization function used in the main text is $\Delta^\alpha_\sigma(t,t')= \sum_{\xi=\pm} \Delta^\alpha_{\sigma,\xi}(t,t')$,
where $\xi=\pm$ corresponds to the positive/negative bond direction (relative to the polarization of the field)
and $\Delta^\alpha_{\sigma,\pm}(t,t')=  \frac{t_{\rm hop}^2}{2} e^{\pm iA(t)} G_{\rm imp,\sigma}^{\bar{\alpha}}(t,t')e^{\mp iA(t')}$,
with $A$ the vector potential of the field pulse \cite{Werner2017}. Here, the bond length $a$ and the electron charge are set to unity.
The vector potential is related to the electric field $E(t)$ by $E(t)=-\partial_t A(t)$. 
We choose the vector potential $A(t) = \frac{E_0}{\Omega} F_{\rm G}(t,t_0,\sigma) \sin(\Omega(t-t_0))$ with $F_{\rm G}(t,t_0,\sigma) = \exp[-\frac{(t-t_0)^2}{2\sigma^2}]$.
$E_0$ indicates the maximum value of the electric field.
The corresponding current (per site) can be computed as $J(t) = {\rm Im}[\sum_{\sigma,\xi=\pm} \xi \Gamma^\alpha_{\sigma,\xi}(t)]$, where $\Gamma^\alpha_{\sigma,\pm}(t) \equiv -i [G_{\rm imp,\sigma}^\alpha*\Delta_{\sigma,\pm}^\alpha]^<(t,t)$.
The self-consistency condition represents a Bethe lattice with $d$ bonds connected to each lattice site, where we take the limit of $d\rightarrow \infty$ with a rescaled hopping parameter $t_{\rm hop}/\sqrt{d}$.
In the free system ($U=0$), the full bandwidth becomes $W=4t_{\rm hop}$. For the Bethe lattice, the self-consistency condition is simplified, compared to other lattices, which reduces the numerical cost and enables a systematic analysis.
The qualitative features of the HHG spectrum are expected to be insensitive to the choice of the lattice. Below we confirm this point with simulations for the two-dimensional square lattice, although the scope of this analysis is limited.

In equilibrium, we define the momentum-averaged single-particle spectral function as 
\eqq{
& A_{\rm loc}(\omega) = -\frac{1}{\pi} {\rm Im} \; G^R_{\rm loc}(\omega).
} 
Here, we introduced the retarded Green's function $G^R_{ij,\sigma}(t-t') = -i\theta(t-t')\langle  \{ \hc_{i\sigma}(t),\hc^\dagger_{j\sigma}(t')\}\rangle$, and 
$G_{\rm loc}=\frac{1}{N}\sum_iG_{ii,\sigma}$. $G^{R}(\omega)$ is defined as $\int d t e^{i\omega t} G^{R}(t)$.
In practice, since the simulation is limited to finite $t$, we use a Gaussian window function with standard deviation $\sigma'=8$ in the Fourier transformation.
The Mott gap $\Delta_{\rm Mott}$ is determined by the criterion $A_{\rm loc}(\Delta_{\rm Mott}/2)=\delta$ with $\delta =0.005$. We checked that the choice of $\delta$ has no qualitative effect on the results.

 %%%%%%%%%%%%%%%%%%%%%%%%%%%%%%%%%%%%%%%%%%%%%
 \begin{figure}[t]
  \centering
    \hspace{-0.cm}
    \vspace{0.0cm}
\includegraphics[width=58mm]{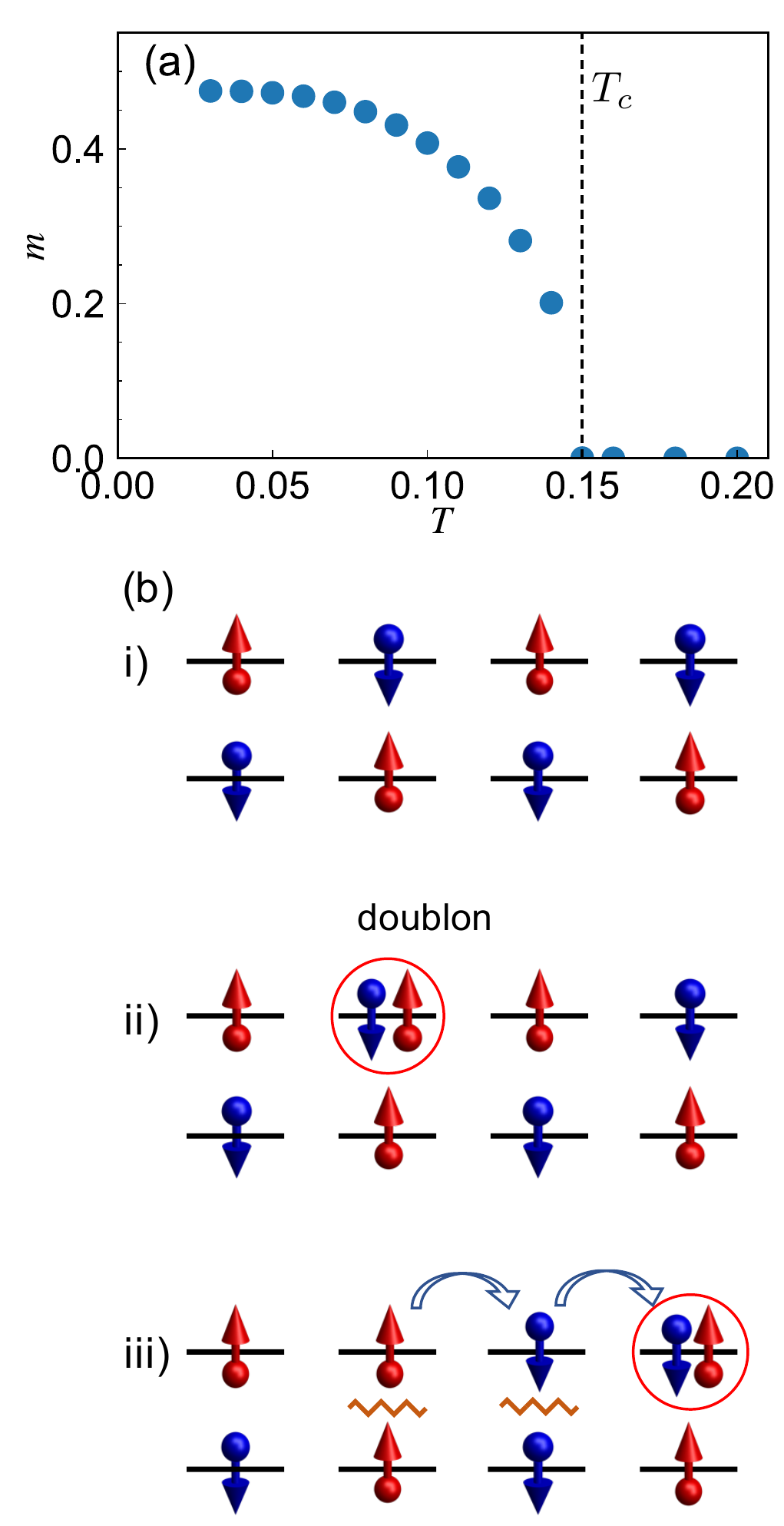} 
  \caption{(a) Temperature dependence of the magnetization,  $m=|\langle \hn_{i\uparrow} \rangle - \langle \hn_{i\downarrow} \rangle|/2$. 
  The vertical dashed line indicates the transition temperature $T_c$.   We use $U=6$ and consider the Bethe lattice.
  (b) Schematic pictures of the spin-charge coupling accompanying the kinematics of a doublon (circle). Panel i) shows the spin configuration of an antiferromagnetic state, ii) shows a doublon added to this state, and iii) shows the dynamics of the doublon, which disturbs the spin configuration (zigzag lines) at the cost of multiples of the exchange energy $J_{\rm ex}$.}
  \label{fig:DMFT_U6_eq_tot}
\end{figure}
%%%%%%%%%%%%%%%%%%%%%%%%%%%%%%%%%%%%%%%%%%%%

\section{Infinite time-evolving block decimation}
 The Hamiltonian of the one-dimensional Hubbard model considered here is
 \begin{align}
\hH(t) =& -t_{\rm hop}\sum_{i,\sigma} [e^{-iA(t)} \hc^\dagger_{i,\sigma} \hc_{i+1,\sigma} + h.c.]\\ \nonumber 
&+ U\sum_i  \hn_{i,\uparrow}\hn_{i,\downarrow}  +B^z_{\rm stagg} \sum_i (-1)^i \hS_{z,i}, \label{eq:Hubbard}
\end{align}
where $\hS_{z,i}=\frac{1}{2}( \hn_{i,\uparrow} - \hn_{i,\downarrow})$ and $t_{\rm hop}$ is set to unity.
We analyze this model with  the infinite time-evolving block decimation (iTEBD) method~\cite{Vidal2003PRL}.
In iTEBD, assuming translational invariance, we express the wave function of the system as a matrix product state (MPS).
iTEBD directly treats the thermodynamic limit and we use the cut-off dimension $D=2000$ for the MPS to obtain converged results.
In the implementation, we use the conservation laws for the numbers of spin-up and spin-down electrons to improve the numerical efficiency.

 %%%%%%%%%%%%%%%%%%%%%%%%%%%%%%%%%%%%%%%%%%%%%
 \begin{figure}[t]
  \centering
    \hspace{-0.cm}
    \vspace{0.0cm}
\includegraphics[width=60mm]{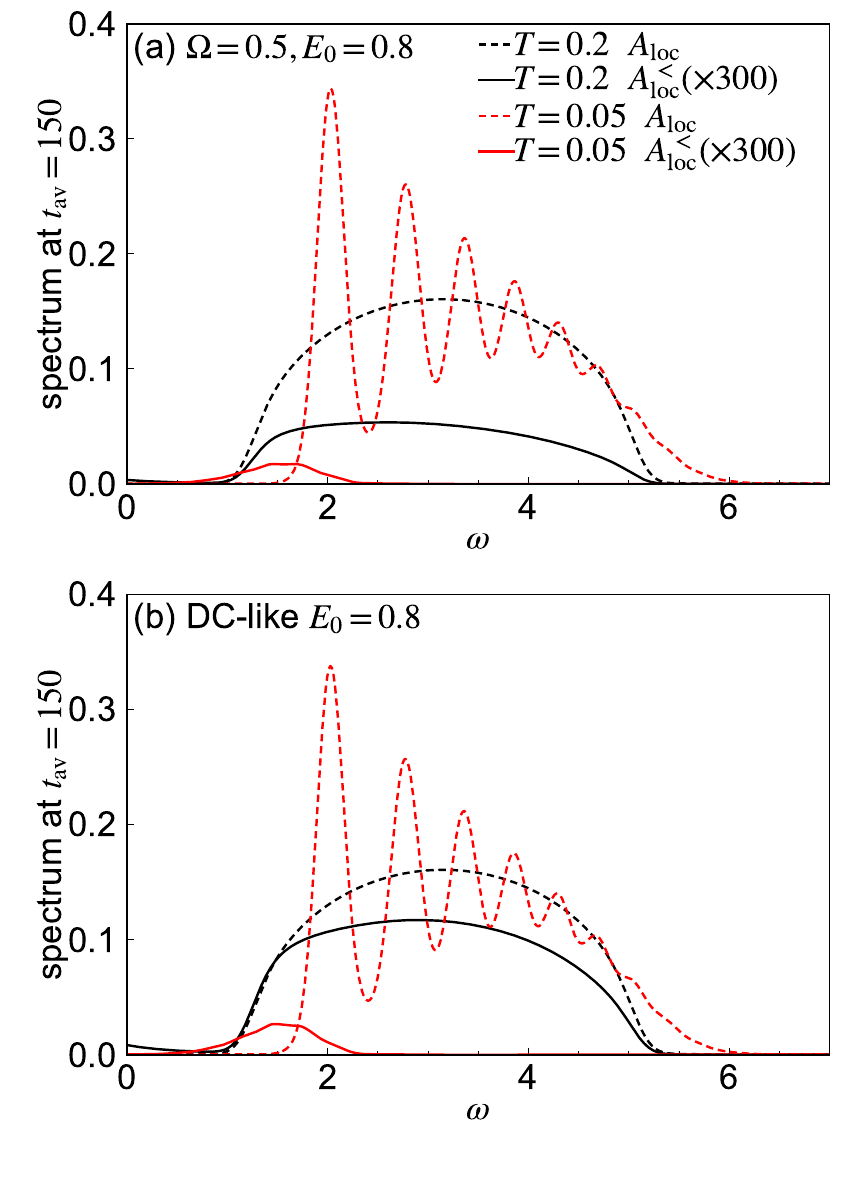} 
  \caption{Single-particle spectral functions $A_{\rm loc}(\omega,t_{\rm av})$ and $A^<_{\rm loc}(\omega,t_{\rm av})$ after the application of the electric field. $A^<_{\rm loc}(\omega,t_{\rm av})$ measures the density of carriers excited by the field. We set $U=6$ and
 the pulse parameters are (a) $\Omega=0.5,E_0=0.8, t_0=75$ and $\sigma=15$, and (b) $E_0=0.8, t_0=75$ and $\sigma=15$. In all cases, we take $t_{\rm av}=150$. }
  \label{fig:DMFT_bethe_tunnel}
\end{figure}
%%%%%%%%%%%%%%%%%%%%%%%%%%%%%%%%%%%%%%%%%%%%

\section{Supplementary results for the Bethe lattice}

In this section, we present supplementary results obtained with DMFT for the Hubbard model on the Bethe lattice.
First, we show the magnetization in equilibrium in Fig.~\ref{fig:DMFT_U6_eq_tot}(a). As explained in the main text, 
it shows the appearance of an AF phase below the transition temperature $T_\text{N\'eel}\approx 0.15$. 
Figure~\ref{fig:DMFT_U6_eq_tot}(b) illustrates the kinematics of a doublon added to the AF state.
It explains how the doublon dynamics couples to the modification of the spin background. 
Panel i) shows the spin configuration of an antiferromagnetic state, ii) shows a doublon added to this state, and iii) shows the dynamics of the doublon, which disturbs the spin configuration (zigzag lines) at the cost of multiples of the exchange energy $J_{\rm ex}$.

 %%%%%%%%%%%%%%%%%%%%%%%%%%%%%%%%%%%%%%%%%%%%%
 \begin{figure}[t]
  \centering
    \hspace{-0.cm}
    \vspace{0.0cm}
\includegraphics[width=60mm]{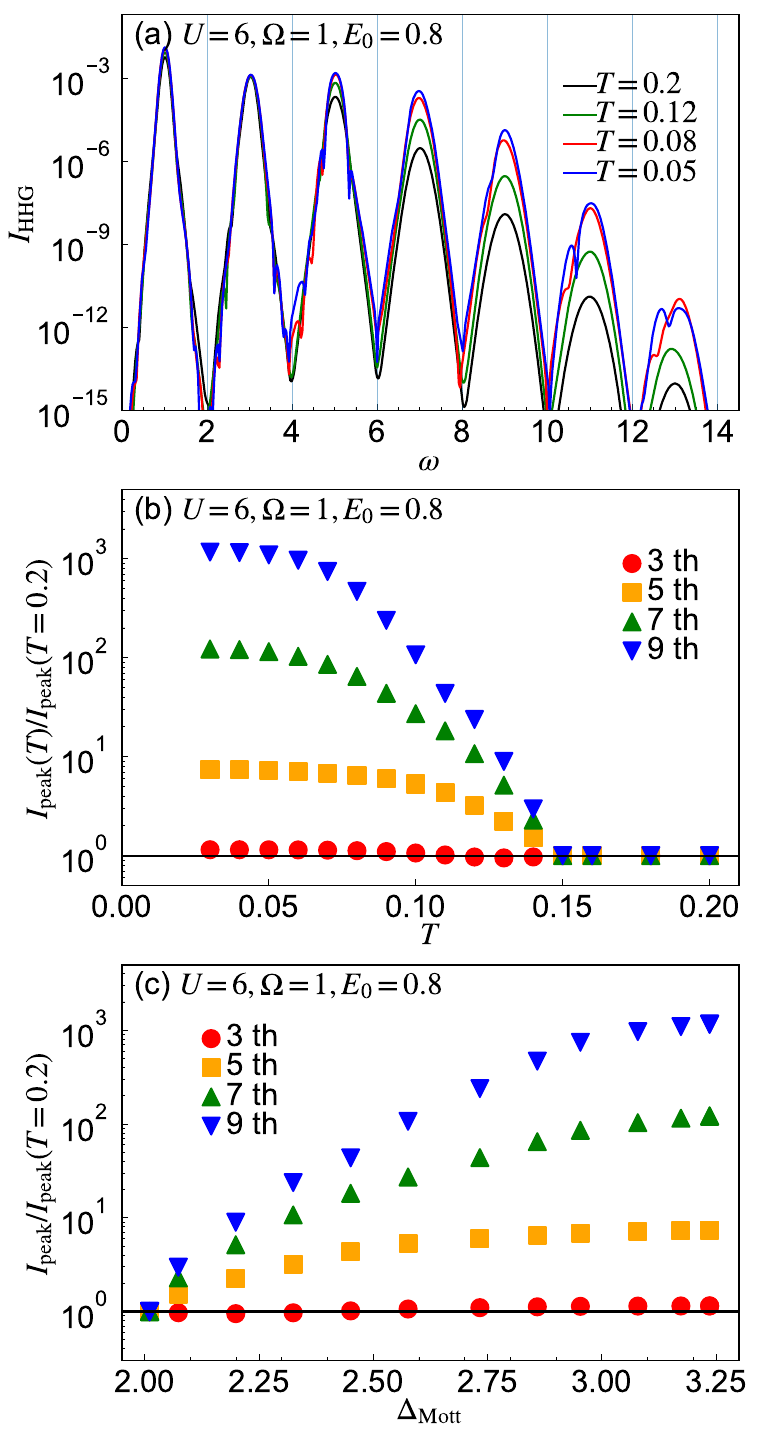} 
  \caption{(a) HHG spectra $I_{\rm HHG}(\omega)$ of the Mott insulator evaluated with DMFT for various temperatures.
  (b) The intensity at the peaks of the HHG spectra $I_{\rm peak}$ as a function of the temperature.
  (c) The intensity at the peaks of the HHG spectra as a function of the Mott gap size $\Delta_\text{Mott}$. In panels (b) and (c), the peak intensity is normalized by the value at $T=0.2$.
  Here, we set $U=6$ and consider the Bethe lattice.  The pulse parameters are  $\Omega=1,E_0=0.8, t_0=75$ and $\sigma=15$.}
  \label{fig:DMFT_bethe_HHG_Ome1}
\end{figure}
%%%%%%%%%%%%%%%%%%%%%%%%%%%%%%%%%%%%%%%%%%%%

Next, we show the temperature dependence of the number of charge carriers created by the field. 
To this end, we compute the time-dependent single-particle spectral functions~\cite{Aoki2013}
\eqq{
& A_{\rm loc}(\omega,t_{\rm av}) = -\frac{1}{\pi} {\rm Im} \; G^R_{\rm loc}(\omega,t_{\rm av}), \\ \nonumber
& A^<_{\rm loc}(\omega,t_{\rm av}) = \frac{1}{2\pi}  {\rm Im} \; G^<_{\rm loc}(\omega,t_{\rm av}).
} 
Here, we introduced the Green's functions $G_{ij,\sigma}(t,t') = -i\langle T_{\mathcal C} \hc_{i\sigma}(t)\hc^\dagger_{j\sigma}(t')\rangle$ with $T_{\mathcal C}$ being the contour ordering operator,
$G_{\rm loc}=\frac{1}{N}\sum_iG_{ii,\sigma}$, and $G^R$ and $G^<$ are the retarded and lesser parts of the Green's function. 
$G^{R/<}(\omega,t_{\rm av})$ is defined as $\int d t_{\rm rel} e^{i\omega t_{\rm rel}} G^{R/<}(t_{\rm rel},t_{\rm av})$, where $G^{R/<}(t_{\rm rel},t_{\rm av}) = G^{R/<}(t,t')$, $t_{\rm rel}=t-t'$ and $t_{\rm av}=\frac{t+t'}{2}$.
$A_{\rm loc}(\omega,t_{\rm av})$ yields the energy spectrum at time $t_{\rm av}$ and $A^<_{\rm loc}(\omega,t_{\rm av})$ the occupied states.
Figure~\ref{fig:DMFT_bethe_tunnel} shows how many charge carriers are created after the pulse field.
Panel (a) plots the results for the same condition as in the main text, while panel (b) shows the results for a pulse without oscillations ($E(t)=E_0 F_{\rm G}(t,t_0,\sigma)$), i.e. close to a DC excitation.
Both panels demonstrate that a much smaller amount of charge carriers is produced when the temperature is lowered and the Mott 
gap is increased, as  expected.

Next, we show the supplemental data for the HHG spectra for $\Omega=1$. 
This excitation frequency is twice larger than what is used in the main text, and, in terms of the value of $\Delta_{\rm Mott}/\Omega$, this choice is closer to the situation in the experiments
on Ca$_2$RuO$_4$~\cite{Uchida2022PRL}.
In Fig.~\ref{fig:DMFT_bethe_HHG_Ome1} (a), we show the HHG spectra for $U=6$ for different temperatures. 
The corresponding temperature dependence of the intensity of the HHG peaks is shown in Fig.~\ref{fig:DMFT_bethe_HHG_Ome1}(b), while the HHG peak intensity is shown as a function of the Mott gap in Fig.~\ref{fig:DMFT_bethe_HHG_Ome1}(c).
The intensity of the $n$th HHG peak $I_{\rm peak}$ is defined as the maximum value of $I_{\rm HHG}(\omega)$ for $\omega\in[(n-1)\Omega,(n+1)\Omega]$.
Since now the gap size is two to three times the excitation frequency, one may expect that multi-photon excitation processes play an important role.
Still, the results shown in Fig.~\ref{fig:DMFT_bethe_HHG_Ome1} indicate that the qualitative temperature dependence of the HHG spectrum is qualitatively the same as that for $\Omega=0.5$
and that the characteristic increase of the HHG intensity with decreasing temperature is insensitive to the excitation condition.
Furthermore, the match between the experiment for Ca$_2$RuO$_4$ and the theory is better for $\Omega=1$ than for $\Omega=0.5$, compare Fig.~\ref{fig:DMFT_bethe_HHG_Ome1}(c) with Fig.~1 in the main text.

 %%%%%%%%%%%%%%%%%%%%%%%%%%%%%%%%%%%%%%%%%%%%%
 \begin{figure}[tb]
  \centering
    \hspace{-0.cm}
    \vspace{0.0cm}
\includegraphics[width=55mm]{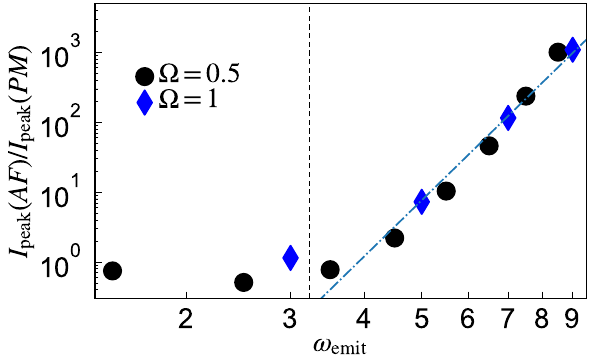} 
  \caption{The increase ratio of the HHG peak intensity for the $n$th order harmonics as a function of the emitted energy $\omega_\textrm{emit}=n\Omega$. Note that both axes are plotted on a log scale.
  The vertical dashed line indicates the gap at low temperatures ($\Delta(T\rightarrow0)$). The dot-dashed line shows the fit to the data for $\omega_\textrm{emit}>\Delta(T\rightarrow0)$, which supports the feature (iii).
   Here, we set $U=6$ and consider the Bethe lattice.  The pulse parameters are $E_0=0.8$, $t_0=75$ and $\sigma=15$. }
  \label{fig:DMFT_Bethe_w_emit}
\end{figure}
%%%%%%%%%%%%%%%%%%%%%%%%%%%%%%%%%%%%%%%%%%%%

Now we discuss to what extent the empirical scaling relation for the HHG intensity introduced in Ref.~\onlinecite{Uchida2022PRL} for Ca$_2$RuO$_4$ is reproduced by our theoretical results.
The empirical relation has the form
\eqq{
I_{\rm peak}(n,T)/I_{\rm peak}(n, T_0)\simeq \Bigl(\frac{n\Omega}{\Omega_\text{th}}\Bigl) ^{\frac{\Delta_{\rm Mott}(T)-\Delta_{\rm Mott}(T_0)}{\Delta_\text{th}}}, \label{eq:eq1}
}
where $I_{\rm peak}(n,T)$ is the peak intensity of the $n$th HHG peak at temperature $T$, $T_0$ is a reference temperature, and $\Omega_{\rm th}$ and $\Delta_{\rm th}$ are fitting parameters.
%\textcolor{blue}{[probably we should explain the parameters here]}
This equation implies the following three features of HHG:
\begin{enumerate}[(i)]
\item The HHG intensity for a given order of harmonics exponentially increases with $\Delta(T)-\Delta(T_0)$.
\item The increase ratio of the HHG peak intensity  is determined only by the emission energy $\omega_\text{emit} (=n\Omega)$.
\item At fixed temperature, the increase ratio exhibits a power law as a function of $\omega_\text{emit}$ with positive exponent.
\end{enumerate}
These three features are the necessary and sufficient conditions for Eq.~\eqref{eq:eq1}.
The present theory does not perfectly, but well reproduce these features in the following sense.
Property (i) is almost satisfied as pointed out in the main text. A close inspection reveals that the theoretical results show a slight saturation at large $\Delta$, which makes the fitting by Eq.~\eqref{eq:eq1} not perfect, but the main trend is consistent. 
Property (ii) is nicely reproduced by the present analysis as shown in Fig.~\ref{fig:DMFT_Bethe_w_emit}, where we compare the increase ratio for different excitation frequencies.
Property (iii) is also consistently reproduced as long as we focus on $\omega_\text{emit}\geq \Delta_{\rm Mott}(T\rightarrow0)$, see Fig.~\ref{fig:DMFT_Bethe_w_emit}. 
Note that in Ref.~\onlinecite{Uchida2022PRL} the empirical law has been proposed for the harmonics above the maximum gap of the system. 
In the theoretical results, the ratio for the harmonics below the gap does not follow the scaling law \eqref{eq:eq1}. This qualitatively different behavior below the gap likely originates from %the fact that the mechanism of HHG is subtle in this regime.
a different HHG mechanism in this regime. 
Above the gap, HHG mainly originates from the recombination of doublon-holon pairs. 
On the other hand, around or below the gap, the contribution from the hopping of doublons or holons becomes relevant, as in the case of the intraband current in  semiconductors.
The resulting change in the HHG mechanism is the likely origin of the deviation from the formula.

 %%%%%%%%%%%%%%%%%%%%%%%%%%%%%%%%%%%%%%%%%%%%%
 \begin{figure}[t]
  \centering
    \hspace{-0.cm}
    \vspace{0.0cm}
\includegraphics[width=48mm]{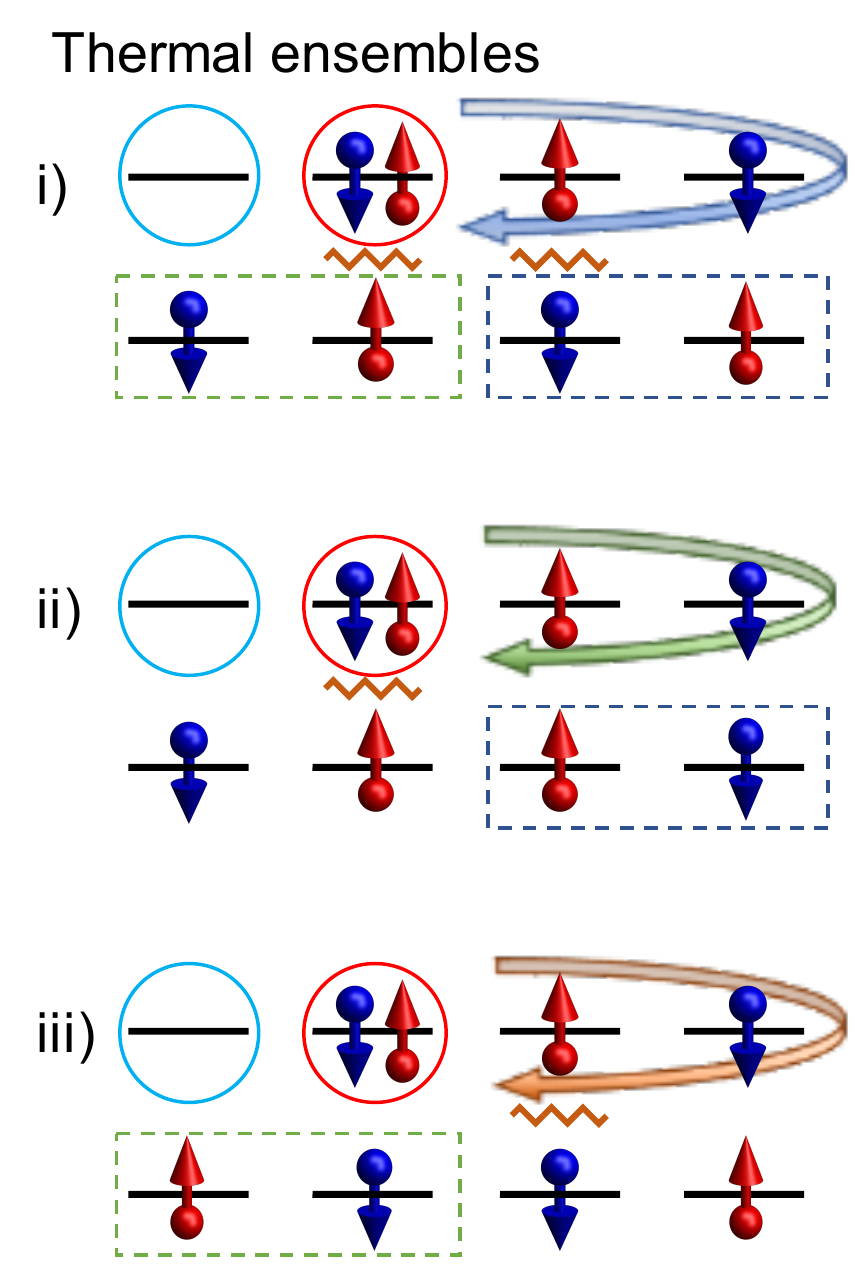} 
  \caption{Schematic pictures of the interplay between thermal ensembles and spin-charge coupling. Panels i)-iii) show cases with different spin configurations activated by thermal fluctuations.  
The difference between case i) and case ii) is indicated by the blue dashed rectangle, while that between case i) and case iii) is indicated by the green dashed rectangle. 
The zigzag lines in each panel show the sites where the mismatch in the spin configuration occurs as the doublon moves around. The horizontal arrows with different colors indicate the difference in the kinematics of the doublon due to the spin-charge coupling. }
  \label{fig:schematic3}
\end{figure}
%%%%%%%%%%%%%%%%%%%%%%%%%%%%%%%%%%%%%%%%%%%%

We note that a systematic derivation and explanation of the empirical formula is difficult and beyond the scope of this paper. Here we simply propose plausible scenarios for some of the above features and leave a detailed analysis to some future work. 
As for (i), the strong increase of the HHG intensity originates from the large change in the coherence time of the doublon-holon pairs. This is indeed an important message of this paper.
The exponential enhancement of the HHG intensity as a function of the gap size can be explained in the following manner. It is natural to assume that the factor in the intensity originating from the dephasing can be expressed as 
$\exp(-t_p/T_2)$. Here $t_p$ is the time interval between the creation and recombination of a relevant doublon-holon pair, and $T_2$ is the coherence time.
This $T_2$ decreases with increasing temperature, which is accompanies by a decrease of the gap.
In this situation, if the inverse of the coherence time behaves as $\frac{1}{T_2} = C_1 -C_2 \Delta_\text{Mott}$, with $C_1$ and $C_2$ some positive constants,
an exponential enhancement of the HHG signal as a function of the gap size is expected.
As for (ii), we need to remember that the strength of the electric field ($E_0$) is fixed. Therefore, when the excitation frequency ($\Omega$) is increased, the amplitude of the vector potential ($A_0=E_0/\Omega$) is reduced, and vice versa. 
The increase of the excitation frequency tends to reduce the time interval $t_p$, while the decrease of the vector potential tends to increase it.
These two effects may compensate each other such that the interval $t_p$ for a given emission energy ($\omega_{\rm emit}$) becomes insensitive to a change in the excitation frequency. 
The ratio of the interval $t_p$ and the coherence time determines how much the pairs contribute to the emission. Since this ratio is almost unchanged, the effects of  the dephasing for a given $\omega_\text{emit}$ should be almost independent of the excitation frequency. 
Thus, given that the change of the HHG intensity mainly originates from the change of the dephasing time, the increase ratio should also be insensitive to the excitation frequency.
As for (iii), we can only explain why the exponent should be positive, i.e. that 
the increase ratio is larger for a larger emission energy. For a larger emission energy, the corresponding interval $t_p$ increases, and therefore the effect of the change in the coherence time becomes larger.
Further systematic analyses are needed to reveal the origin of the power law.
Finally, we note that, for a detailed discussion, the efficiency of the creation of the doublon-holon pair should also be taken into account. 
However, this effect should be subdominant for the analysis of the temperature dependence of a given harmonic, and the above discussion should cover the main effects.
We note that the creation becomes more efficient as we increase the temperature and suppress the gap, 
as shown in Fig.~2 of the Supplemental Material. However, the HHG intensity is enhanced if temperature is decreased, 
which suggests that the dephasing effect dominates.

 %%%%%%%%%%%%%%%%%%%%%%%%%%%%%%%%%%%%%%%%%%%%%
 \begin{figure}[tb]
  \centering
    \hspace{-0.cm}
    \vspace{0.0cm}
\includegraphics[width=85mm]{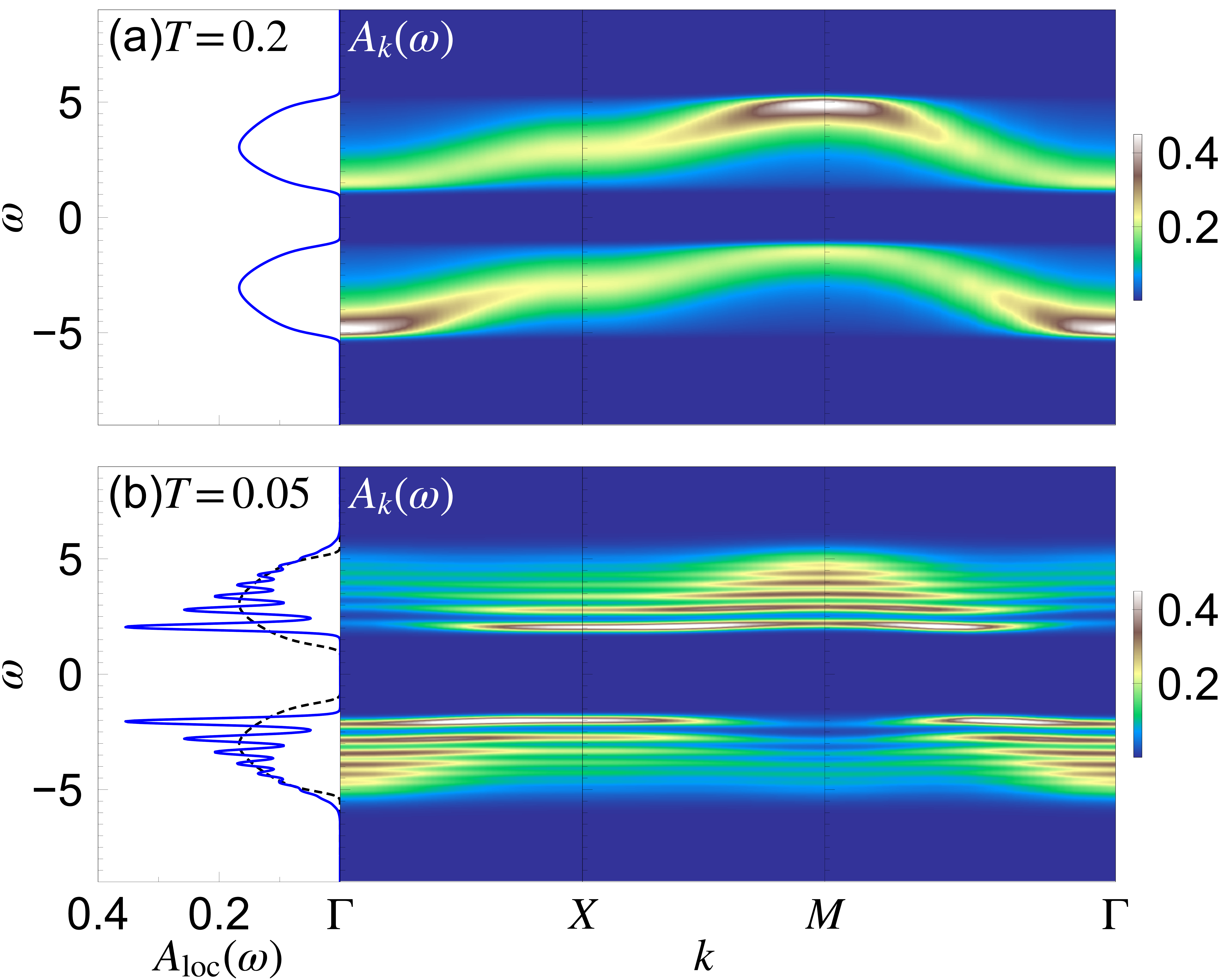} 
  \caption{Local single-particle spectral functions $A_{\rm loc}(\omega)$ and momentum-resolved single-particle spectral functions $A_{\bf k}(\omega)$ for the Hubbard model on the two-dimensional square lattice.
 Panels (a) is for $T=0.2$ in the PM phase and (b) is for $T=0.05$ in the AF phase. Here, $U=6$, $t_{\rm hop}=0.5$ and $(L_x,L_y)=(16,16)$. In the momentum-resolved spectra, $\Gamma, X$ and $M$ follow the canonical notation for the Brillouin zone of the PM phase. }
  \label{fig:DMFT_2dSQ_spectrum}
\end{figure}
%%%%%%%%%%%%%%%%%%%%%%%%%%%%%%%%%%%%%%%%%%%%

 %%%%%%%%%%%%%%%%%%%%%%%%%%%%%%%%%%%%%%%%%%%%%
 \begin{figure*}[tb]
  \centering
    \hspace{-0.cm}
    \vspace{0.0cm}
\includegraphics[width=150mm]{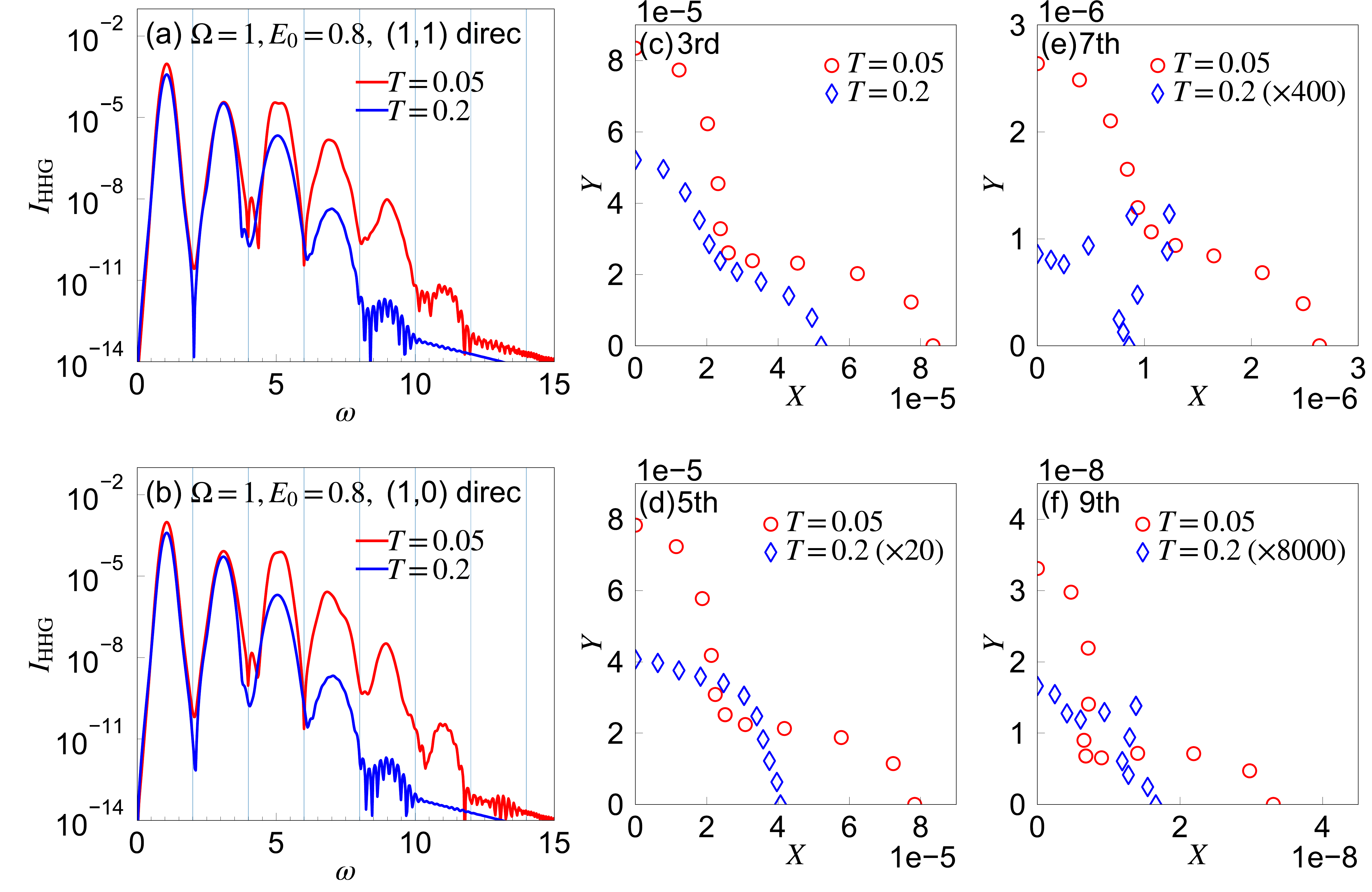} 
  \caption{(a)(b) HHG spectra of the Mott insulator evaluated with DMFT for the indicated temperatures. Here, we consider the Hubbard model on the two-dimensional square lattice and set $U=6$ and $t_{\rm hop}=0.5$.
  The pulse parameters are (a) $\Omega=1,E_0=0.8,t_0=30,\sigma=7.5$ and $\theta=0$, and (b) $\Omega=1,E_0=0.8,t_0=30,\sigma=7.5$ and $\theta=\pi/4$. 
  (c-f) The corresponding polarization dependence of the HHG peak intensity. In all cases, we evaluate the HHG intensity along the direction of the applied electric field.  }
  \label{fig:2dim_HHG}
\end{figure*}
%%%%%%%%%%%%%%%%%%%%%%%%%%%%%%%%%%%%%%%%%%%%

In Fig.~\ref{fig:schematic3}, to support the explanations about the effect of thermal fluctuations in the main text, we show schematic pictures of 
the dynamics of a doublon-holon pair in different spin backgrounds.
Panels i)-iii) show cases with different spin configurations activated by thermal fluctuations.  
The difference between case i) and case ii) is indicated by the blue dashed rectangle, while that between case i) and case iii) is indicated by the green dashed rectangle. 
The zigzag lines in each panel show the sites where the mismatch in the spin configuration occurs as the doublon moves around. The curved arrows with different colors indicate the difference in the kinematics of the doublon due to the spin-charge coupling.

Finally, we comment on the relevance of the physics of the single-band Hubbard model for Ca$_2$RuO$_4$. 
A more accurate modeling of Ca$_2$RuO$_4$ should involve a three-band Hubbard model. In multi-orbital strongly correlated systems, spin as well as orbital degrees of freedom become important.
Still, we believe that similar physics as discussed for the single-band Hubbard model is relevant in this case.
Firstly, the spin degrees of freedom should play a similar role also in the three band case, since the dynamics of excited local multiplets disturbs the spin background.
Secondly, in multi-orbital systems, the orbital-charge coupling can play a similar role as the spin-charge coupling, as has been discussed for the charge relaxation in a previous study.\cite{Hugo2017PRB} More specifically, the dynamics of the excited multiples disturbs the orbital configurations at the cost of the Hund energy.
We expect that, in Ca$_2$RuO$_4$, both the spin-charge coupling and the orbital-charge coupling should contribute to the peculiar behavior of HHG, while the physical mechanism is essentially the one discussed for the single-band Hubbard model.

\section{Supplementary results for the square lattice}

In this section, we present supplementary DMFT results for the Hubbard model on the two-dimensional square lattice at half filling.
We show that the equilibrium features and the characteristic temperature dependence of the HHG are essentially the same as in the case of the Bethe lattice, 
although a systematic analysis is difficult since the DMFT calculation becomes numerically more demanding. 
We assume that the system only has nearest neighbor hopping, $t_{\rm hop}$.
We set  $t_{\rm hop}=0.5$ so that the bandwidth of the free system becomes $4$ as in the case of the Bethe lattice.
The lattice constant $a$ is set unity.
In the following the system size is $(L_x,L_y)=(16,16)$, where $L_x$ ($L_y$) is the number of sites along the $x$ ($y$) axis, and we use periodic boundary conditions.

In Fig.~\ref{fig:DMFT_2dSQ_spectrum}, we show the local single-particle spectral function $A_{\rm loc}(\omega)$ and the momentum-resolved single-particle spectral function $A_{\bf k}(\omega)$ in equilibrium.
The latter is defined as $A_{\bf k}(\omega)=-\frac{1}{\pi}{\rm Im} G^R_{\bf k,\sigma} (\omega)$, where $G^R_{\bf k,\sigma} (\omega)$ is the Fourier component of the retarded part of the Green's function $G_{\bf k,\sigma} (t)$.
Here, $G_{\bf k,\sigma} (t) = -i\langle T_{\mathcal C} \hc_{\bf k\sigma}(t) \hc^\dagger_{\bf k\sigma}(0) \rangle$ and $\hc^\dagger_{\bf k\sigma}=\frac{1}{\sqrt{N}} \sum_l e^{i{\bf k}\cdot {\bf r}_l} \hc_{l\sigma}^\dagger$.
As in the case of the Bethe lattice, the local spectral function $A_{\rm loc}(\omega)$ is featureless above the transition temperature $T_c$. Below $T_c$ the Mott gap is enhanced and peaks corresponding to spin polarons emerge.
In the PM phase, the dispersions of the upper and lower Hubbard bands are parallel to each other, similar to the prediction of the Hubbard I approximation, see Fig.~\ref{fig:DMFT_2dSQ_spectrum}(a).
On the other hand, in the AF phase, the momentum-resolved spectral function consists of many almost-flat spin-polaron bands, see Fig.~\ref{fig:DMFT_2dSQ_spectrum}(b).
Also, at each momentum, the spectral function covers a wider frequency range than in the PM phase, and as a whole the dispersion relation is less well-defined than in the PM phase, i.e. the spectra look more incoherent.
Based on the experience from HHG in semiconductors, these observations lead to the natural expectation that the HHG intensity should decrease with decreasing temperature.

 %%%%%%%%%%%%%%%%%%%%%%%%%%%%%%%%%%%%%%%%%%%%%
 \begin{figure}[t]
  \centering
    \hspace{-0.cm}
    \vspace{0.0cm}
\includegraphics[width=88mm]{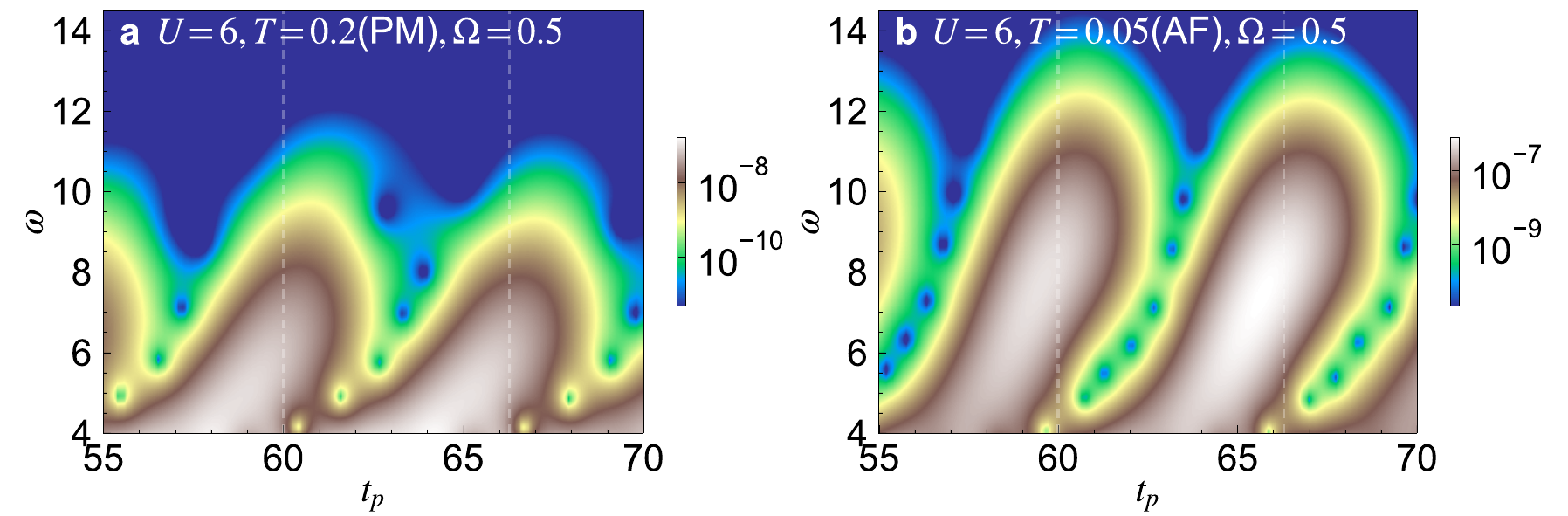} 
  \caption{(a)(b) Subcycle analysis of the HHG signal, $I_{\rm HHG}(\omega,t_p)$, for the field along the $(1,1)$ direction ($\theta=\frac{\pi}{4}$). Panel (a) is for $T=0.2$ (PM phase) and (b) is  for $T=0.05$ (AF phase).    The pulse parameters are $\Omega=0.5$, $E_0=0.9\sqrt{2}$, $t_0=60$, $\sigma=15$. The vertical dashed lines indicate the times when $E(t)=0$. }
  \label{fig:2dim_subcycle}
\end{figure}
%%%%%%%%%%%%%%%%%%%%%%%%%%%%%%%%%%%%%%%%%%%%

Next, we discuss the HHG spectrum of this system.
We apply linearly polarized light along the ${\bf e}_\theta = [ \cos(\theta),\sin(\theta)]$ direction:
\eqq{
A_x(t) &= \frac{E_0}{\Omega} \cos(\theta)  F_{\rm G}(t,t_0,\sigma) \sin(\Omega(t-t_0)),  \\
A_y(t) &= \frac{E_0}{\Omega} \sin(\theta)  F_{\rm G}(t,t_0,\sigma)  \sin(\Omega(t-t_0)).
}
We extract the HHG signal along the ${\bf e}_\theta$ direction from the Fourier transformation of $J_\theta(t)$, where $J_\theta(t)={\bf e}_\theta\cdot {\bf J}(t)$ and ${\bf J}(t)=[J_x(t),J_y(t)]$.
The resulting HHG spectra are shown in Figs.~\ref{fig:2dim_HHG}(a)(b) for $\Omega=1$ and $U=6$.
These data show that the temperature dependence of the HHG intensity is qualitatively the same as that obtained for the Bethe lattice, and
the enhancement of the HHG signal with decreasing temperature is independent of the polarization.
In Figs.~\ref{fig:2dim_HHG}(c-f), we show the polarization dependence of the intensity of the HHG peaks for a given order.
In the AF phase, the intensity becomes largest when the field is polarized along the bond direction, which is consistent with the experiment on Ca$_2$RuO$_4$~\cite{Uchida2022PRL}.
However, interestingly, in the PM phase, the polarization dependence changes qualitatively for the higher harmonic components ($n\geq 5$).

For lower frequencies, like $\Omega=0.5$ used in the main text, the DMFT self-consistency becomes too expensive for the full simulation of the time evolution.
Still, one can simulate halfway, and analyze the subcycle features around the peak of the pulse, see Fig.~\ref{fig:2dim_subcycle}.
Again, we find qualitatively the same behavior as in the Bethe lattice, which supports the generality of the physics discussed in the main tex.

\section{Supplemental results for other models}

In the main text, we revealed  the important effect of the correlations between different degrees of freedom on HHG in strongly correlated systems, focusing of the single-band Hubbard model, where the spin-charge coupling is strong. We showed that this simple model can explain many peculiar HHG features reported in an experimental study of Ca$_2$RuO$_4$.\cite{Uchida2022PRL}
The underlying physics should be applicable to a wide range of Mott insulators.
In particular, orbital-charge coupling in addition to spin-charge coupling can yield similar phenomena in multi-orbital Mott insulators, since a previous study~\cite{Hugo2017PRB} showed that orbital-charge coupling can have a similar effect on nonequilibrium charge carriers as spin-charge coupling.
More specifically, in multi-orbital systems with Hund coupling, the dynamics of local multiplets disturbs the orbital configurations in a similar way as the doublon dynamics disturbs the spin background in the single-band Hubbard model.
Here, we briefly show some results for two other models to support the above statements.

 %%%%%%%%%%%%%%%%%%%%%%%%%%%%%%%%%%%%%%%%%%%%%%
 \begin{figure}[t]
  \centering
    \hspace{-0.cm}
    \vspace{0.0cm}
\includegraphics[width=85mm]{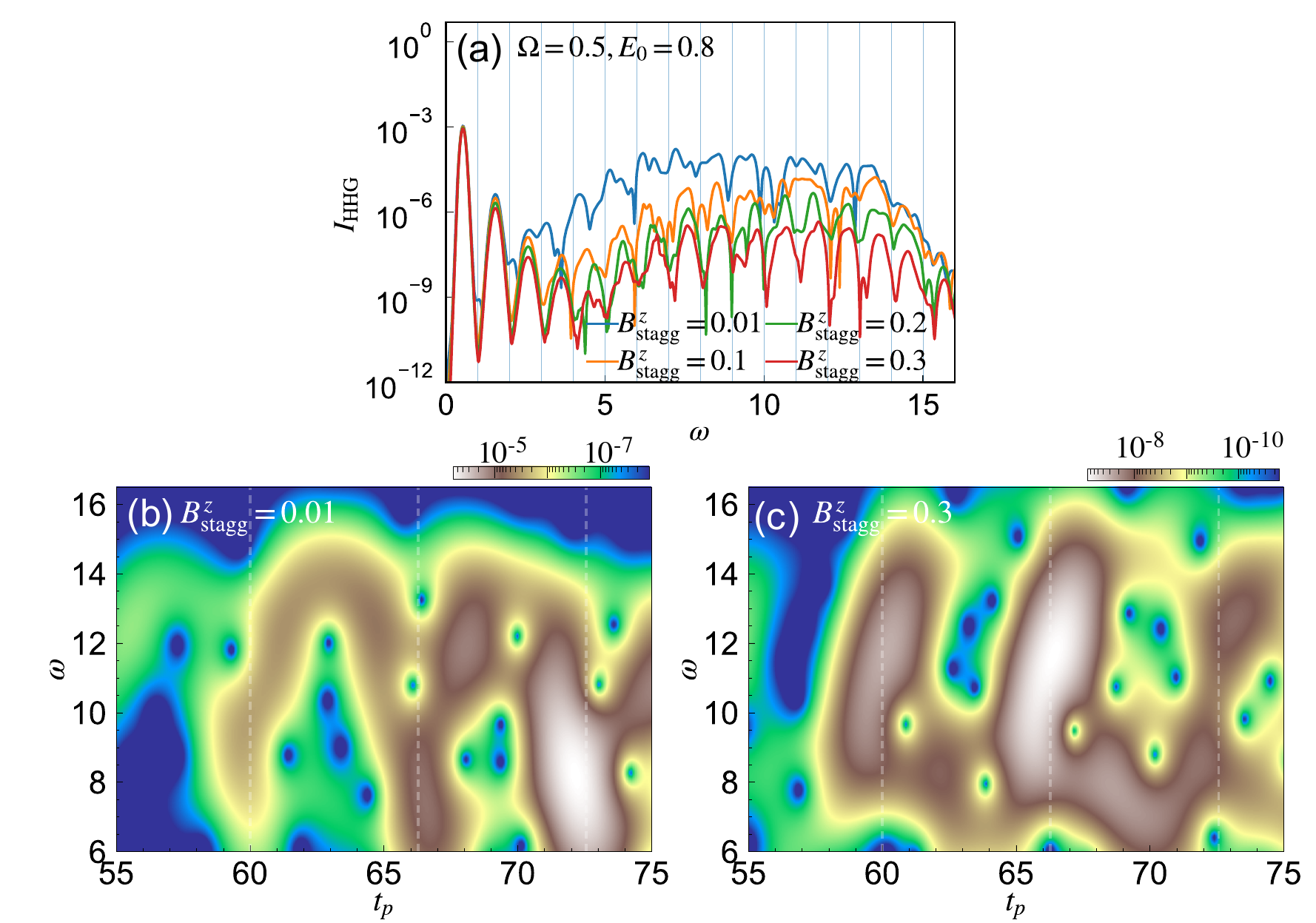} 
  \caption{ (a) $I_\text{HHG}$ for different $B^z_\text{stagg}$ for the charge transfer insulator described by the one-dimensional $d$-$p$ model. (b)(c) The corresponding subcycle analysis for $B^z_\text{stagg}=0.01$ (b) and $B^z_\text{stagg}=0.3$ (c). 
A Gaussian window with $\sigma'=0.9$ is used. In all panels, we set $v_{dp}=3, \epsilon_d-\epsilon_p= -10,U_{dd}=20,U_{pp}=0,V_{dp}=0$ and consider a three-quarter filled system. The excitation parameters are $\Omega=0.5,E_0=0.8$, $t_0=60$ and $\sigma=15$. }
    \label{fig:fig2_response}
\end{figure}
%%%%%%%%%%%%%%%%%%%%%%%%%%%%%%%%%%%%%%%%%%%

\subsection{$d$-$p$ model for charge transfer insulators}
In this section we analyze the $d$-$p$ model in one dimension using iTEBD.
The Hamiltonian consists of the two parts 
\eqq{
\hH_{dp}= \hH_{dp,\text{kin}} + \hH_{dp,\text{int}}.
}
If we assume that the $d$ orbital corresponds to $d_{x^2-y^2}$ and the $p$ orbital to $p_x$, the explicit for of these terms is
\eqq{
\hH_{dp,\text{kin}} =&  -\sum_{i,\sigma} [v_{dp}(t)\hp^\dagger_{i,\sigma} \hd_{i,\sigma}+ h.c.] \nonumber\\
 &+  \sum_{i,\sigma} [v_{dp}(t)\hd^\dagger_{i,\sigma}\hp_{i-1,\sigma} + h.c.]   \\
& + \epsilon_d \sum_{i} n^d_i + \epsilon_p \sum_i n^p_i + B^z_\text{stagg}\sum_i (-1)^i \hS^d_{z,i}, \nonumber}
\eqq{
 \hH_{dp,\text{int}} &= U_{dd}\sum_i n^d_{i\uparrow}n^d_{i\downarrow} + U_{pp}\sum_i n^p_{i\uparrow}n^p_{i\downarrow}   \\
 & \;\;\;\; + V_{dp}\sum_{i}n^d_{i}(n^p_{i}+n^p_{i-1}). \nonumber
}
Here, $\hd^\dagger$ represents the creation operator of the $d$-orbital electron, while $\hp^\dagger$ represents the creation operator of the $p$-orbital electron.
$\hn_{i\sigma}^d = \hd^\dagger_{i\sigma} \hd_{i\sigma}$, $\hn_{i\sigma}^p = \hp^\dagger_{i\sigma} \hp_{i\sigma}$, $\hn_i^d = \sum_\sigma \hn_{i\sigma}^d $, $\hn_i^p = \sum_\sigma \hn_{i\sigma}^p $ and $\hS^d_{z,i} = \frac{1}{2}[\hd^\dagger_{i\uparrow} \hd_{i\uparrow}-\hd^\dagger_{i\downarrow} \hd_{i\downarrow}] $.
$v_{dp}$ is the hopping parameter between the neighboring $d$ and $p$ orbitals, $\epsilon_a$ indicates the energy level of the orbital $a$, and $B^z_\text{stagg}$ is the staggered magnetic field applied to the $d$ orbitals.
$U_{dd}$ and $U_{pp}$ are the on-site interactions for the $d$ and $p$ orbitals, respectively, and $V_{dp}$ is the nearest neighbor interaction.
We assume that the $d$ and $p$ orbitals are equally distanced, and set the lattice constant and the charge to unity. The effect of the electric field is considered through the Peierls substitution as in the Hubbard model, i.e., $v_{dp}(t)= v_{dp}e^{iA(t)/2}$. The current is defined as $\hat{J}(t)=-\frac{\delta \hH_{dp}}{\delta A(t)}$.
We focus on the three-quarter filled system, which corresponds to a Mott or charge transfer (CT) insulator, and simulate the time evolution by iTEBD.
The HHG intensity is evaluated by the field induced current as in the case of the Hubbard model.
We note again that the staggered magnetic field mimics the effect of the spin-charge coupling expected in higher dimensions.

In Fig.~\ref{fig:fig2_response}, we show the results for a parameter set corresponding to a CT insulator. 
We choose the parameters such that the bandwidth of the upper Hubbard band and the Zhang-Rice singlet band, as well as the band gap, become similar to those in the Hubbard model studied in the main text.
One can identify essentially the same HHG features as in the results for the single-band Hubbard model discussed in the main text.
Namely, when $B^z_\text{stagg}$ is small, the HHG spectrum does not show clear HHG peaks, while with larger $B^z_\text{stagg}$ the HHG peaks become clearer, see Fig.~\ref{fig:fig2_response}(a).
Also the subcycle analysis shows that the coherence time of the pairs of excited local multiplets is efficiently reduced by $B^z_\text{stagg}$, see Fig.~\ref{fig:fig2_response}(b)(c).
These results suggest that also in CT insulators, the spin-charge coupling plays a crucial role in the HHG mechanism and that the physics discussed for the single-band Hubbard model is relevant. 

 %%%%%%%%%%%%%%%%%%%%%%%%%%%%%%%%%%%%%%%%%%%%%%
 \begin{figure}[t]
  \centering
    \hspace{-0.cm}
    \vspace{0.0cm}
\includegraphics[width=85mm]{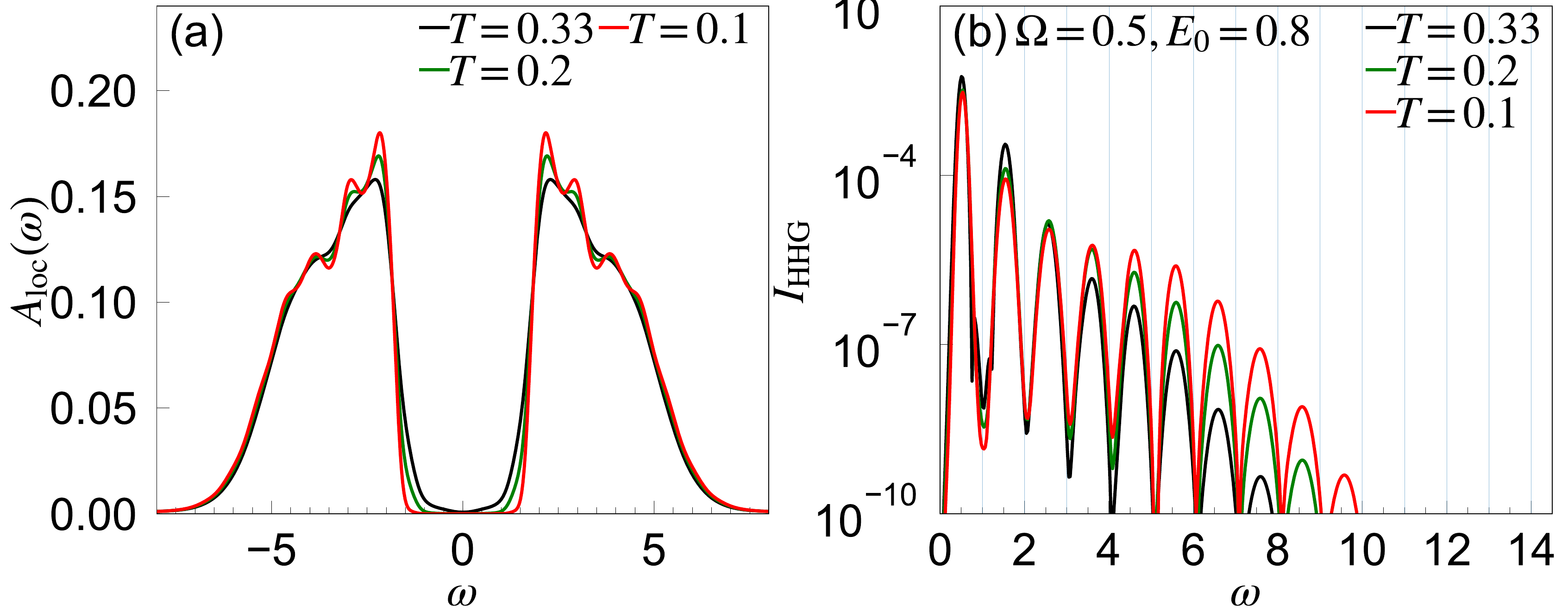} 
  \caption{Results for paramagnetic Mott insulators described by the two band Hubbard model. (a) Equilibrium local spectral functions $A_\text{loc}(\omega)$ computed with DMFT for various $T$.
(b) HHG spectra of the Mott insulator computed with DMFT. In all panels, we set $U=6,J=0.5$ and consider half filling. The excitation parameters are $\Omega=0.5,E_0=0.8$, $t_0=75$ and $\sigma=15$. }
    \label{fig:fig3_response}
\end{figure}
%%%%%%%%%%%%%%%%%%%%%%%%%%%%%%%%%%%%%%%%%%%

\subsection{Two-band Hubbard model}
As a second example, we consider a two-orbital Hubbard model with Hamiltonian
\eqq{
  &\hH_\text{2bH} = \sum_{\langle ij\rangle}\sum_{\alpha = 1,2} \sum_\sigma v_{ij}^\alpha(t) \hc^\dagger_{i,\alpha\sigma} \hc_{j,\alpha\sigma}  \nonumber\\
  & + \sum_i \sum_{\alpha = 1,2 } \Bigl[ U \hn_{i,\alpha\uparrow}  \hn_{i,\alpha\downarrow} - \mu(\hn_{i,\alpha\uparrow} + n_{i,\alpha\downarrow})\Bigl] \\
&+ \sum_i \sum_{\sigma } \Bigl[ (U-2J) \hn_{i,\alpha\sigma}  \hn_{i,\alpha\bar{\sigma}} + (U-3J) \hn_{i,\alpha\sigma}  \hn_{i,\alpha \sigma}\Bigl]  \nonumber.
}
Here $\hc^\dagger_{\alpha\sigma}$  is the creation operator of the electron in orbital $\alpha$ with spin $\sigma$ and $\hn_{i,\alpha\sigma} = \hc^\dagger_{i,\alpha\sigma}\hc_{i,\alpha\sigma}$.
 $v_{ij}^\alpha$ is the hopping parameter between site $i$ and $j$ for orbital $\alpha$, $\mu$ the chemical potential, $U$ the intra-orbital interaction and $J$ the Hund coupling.
For simplicity, we neglect the spin-flip and pair-hopping terms.
In the following, we put $v_{ij}^1 = v_{ij}^2$, and consider the Bethe lattice as in the Hubbard model case, following Refs.~\onlinecite{Werner2018PRB, Markus2020}.
We set the quarter of the free electron bandwidth to unity.
Focusing on half filling, we simulate the time evolution of the system under an electric field pulse with DMFT.

 %%%%%%%%%%%%%%%%%%%%%%%%%%%%%%%%%%%%%%%%%%%%%
 \begin{figure}[t]
  \centering
    \hspace{-0.cm}
    \vspace{0.0cm}
\includegraphics[width=48mm]{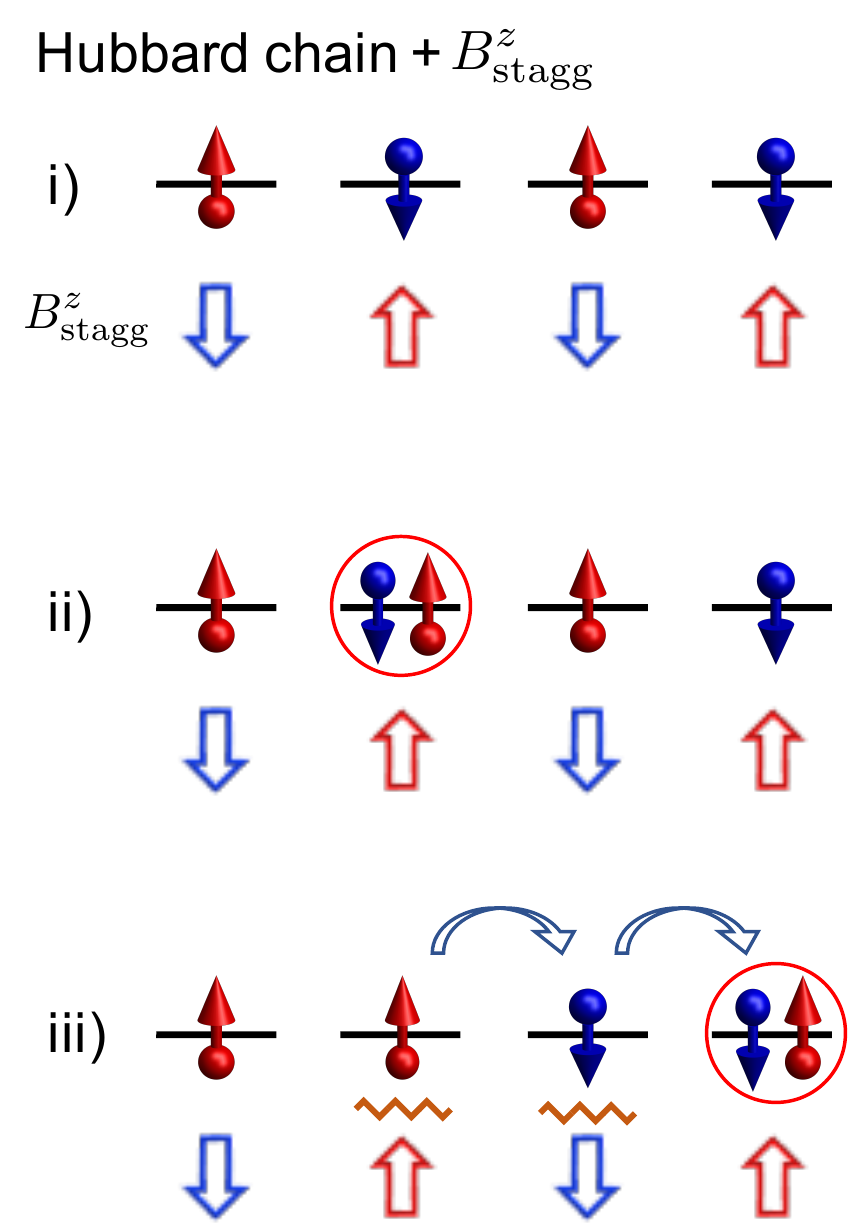} 
  \caption{Schematic pictures of the effects of the staggered magnetic field $B^z_{\rm stagg}$ applied to the one-dimensional Hubbard chain, which can be directly compared to Fig.~\ref{fig:DMFT_U6_eq_tot}~(b).  i) In equilibrium, each site is occupied by one electron, and the spin is aligned anti-parallel to the staggered field. ii), iii) As the doublon moves, it produces a mismatch between the spin configuration and the staggered magnetic field at the cost of the Zeeman energy, about $B^z_{\rm stagg}$ (zigzag lines).}
  \label{fig:schematic2}
\end{figure}
%%%%%%%%%%%%%%%%%%%%%%%%%%%%%%%%%%%%%%%%%%%%

 %%%%%%%%%%%%%%%%%%%%%%%%%%%%%%%%%%%%%%%%%%%%%
 \begin{figure}[t]
  \centering
    \hspace{-0.cm}
    \vspace{0.0cm}
\includegraphics[width=85mm]{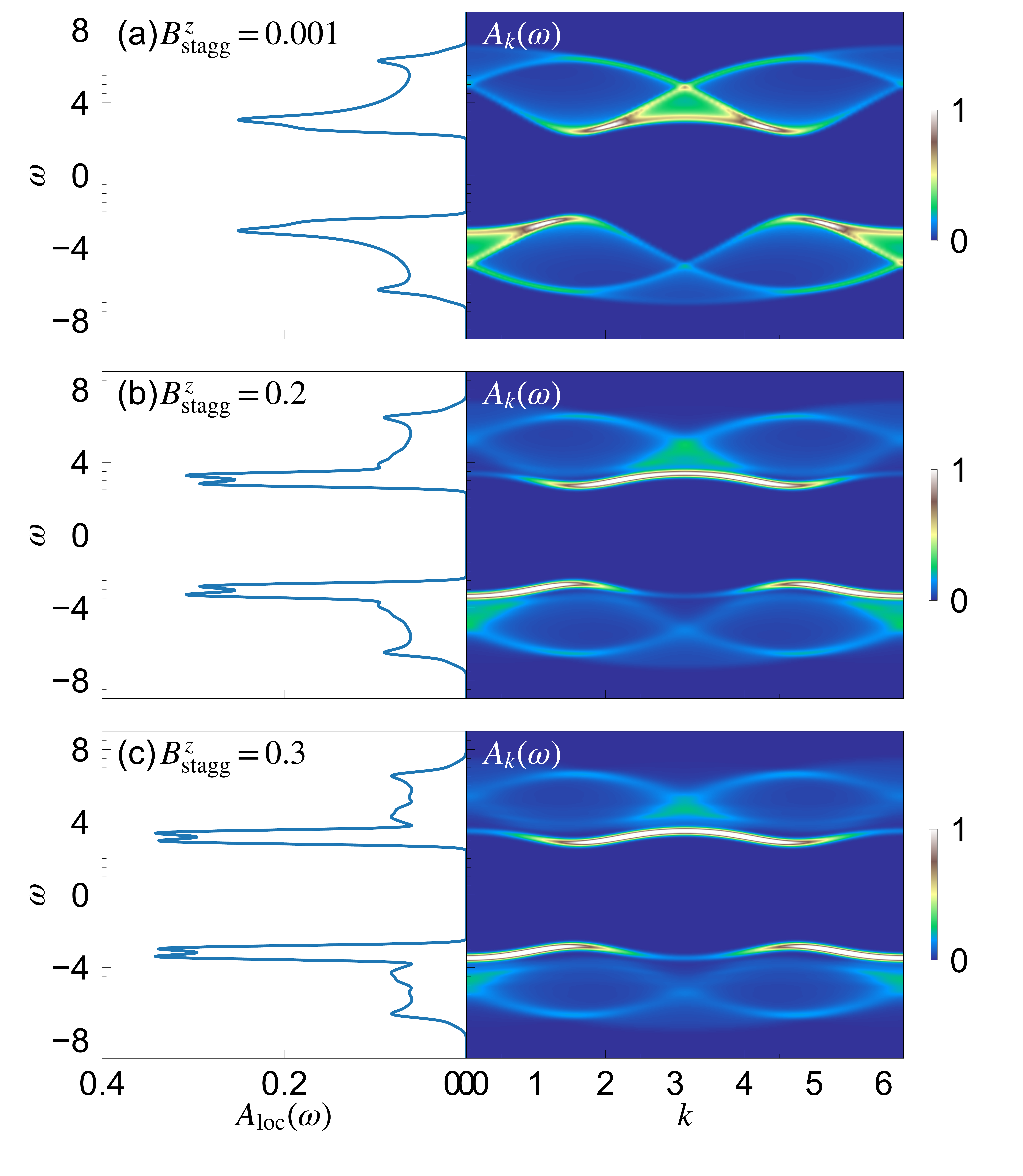} 
  \caption{ Momentum-integrated spectrum $A_{\rm loc}(\omega)$ and momentum-resolved spectrum $A_{k}(\omega)$ for the one-dimensional Hubbard model with staggered magnetic field at half filling.
Panel (a) is for $U=8$, $B_{\rm stagg}^z=0.001$, (b) is for $U=8$, $B_{\rm stagg}^z = 0.2$ and (c) is for $U=8$, $B_{\rm stagg}^z= 0.3$. Here, we use the iTEBD method.}
  \label{fig:1d_spectrum_string}
\end{figure}
%%%%%%%%%%%%%%%%%%%%%%%%%%%%%%%%%%%%%%%%%%%%

In Fig.~\ref{fig:fig3_response}, we show the local spectrum averaged over spins and orbitals ($A_\text{loc}(\omega)$) and the HHG spectra for different temperatures. 
Here, we simulate the system in the paramagnetic (PM) phase. As we decrease the temperature, the band gap increases and peak structures develop within the bands [see Fig.~\ref{fig:fig3_response}(a)]. 
This is in contrast to the single-band case, where the local spectrum in the PM phase is featureless. The difference originates from the Hund coupling and the resultant orbital-charge coupling.
When a test charge is injected into the half-filled system, a triplon is created and its dynamics disturbs the orbital background at the cost of the Hund energy.
This results in a strong orbital-charge coupling, which plays a similar role as the spin-charge coupling in the single-band Hubbard model.
Figure~\ref{fig:fig3_response}(b) shows that the HHG intensity increases as the temperature decreases (the gap increases), as in the single-band Hubbard model.
These results suggest that indeed the orbital-charge coupling in multi-orbital systems has a similar effect on HHG as the spin-charge coupling in the single-band Hubbard model,
and thus the physics discussed in the paper should also be applicable for this case.
We note that, if we go into the antiferromagnetic phase in the two-orbital model, the spin-charge coupling also starts to play a role and a further increase of the HHG signal is expected.

\section{Supplementary data for the one-dimensional system}
In this section, we present supplementary results obtained with iTEBD for the one-dimensional Hubbard model with staggered magnetic field.
First, we show a schematic picture to illustrate the analogy between this model and the Hubbard model in higher dimensions, which is discussed in the main text, see Fig.~\ref{fig:schematic2}.
 i) In equilibrium, each site is occupied by one electron, and the spin is aligned anti-parallel to the staggered field. ii), iii) As the doublon moves, it produces a mismatch between the spin configuration and the staggered magnetic field at the cost of the Zeeman energy, about $B^z_{\rm stagg}$ (zigzag lines).

In Fig.~\ref{fig:1d_spectrum_string}, we show how the single-particle spectra change with the magnetic field $B_{\rm stagg}^z$.
One can see that, with increasing field strength, there emerges a separated band at the bottom (top) of the upper (lower) Hubbard band in the momentum-resolved spectrum $A_{k}(\omega)$.
Furthermore, replicas of these separated bands can be identified. In the momentum-integrated spectrum $A_{\rm loc}(\omega)$, these sub-bands correspond to the peaks that emerge with increasing field strength.
These features are qualitatively the same as those of the DMFT spectra, and originate from the effective spin-charge coupling and the resulting spin-polarons that are induced by the staggered magnetic field, as mentioned in the main text.

\bibliography{HHG_Ref}

\begin{thebibliography}{78}%
\makeatletter
\providecommand \@ifxundefined [1]{%
 \@ifx{#1\undefined}
}%
\providecommand \@ifnum [1]{%
 \ifnum #1\expandafter \@firstoftwo
 \else \expandafter \@secondoftwo
 \fi
}%
\providecommand \@ifx [1]{%
 \ifx #1\expandafter \@firstoftwo
 \else \expandafter \@secondoftwo
 \fi
}%
\providecommand \natexlab [1]{#1}%
\providecommand \enquote  [1]{``#1''}%
\providecommand \bibnamefont  [1]{#1}%
\providecommand \bibfnamefont [1]{#1}%
\providecommand \citenamefont [1]{#1}%
\providecommand \href@noop [0]{\@secondoftwo}%
\providecommand \href [0]{\begingroup \@sanitize@url \@href}%
\providecommand \@href[1]{\@@startlink{#1}\@@href}%
\providecommand \@@href[1]{\endgroup#1\@@endlink}%
\providecommand \@sanitize@url [0]{\catcode `\\12\catcode `\$12\catcode
  `\&12\catcode `\#12\catcode `\^12\catcode `\_12\catcode `\%12\relax}%
\providecommand \@@startlink[1]{}%
\providecommand \@@endlink[0]{}%
\providecommand \url  [0]{\begingroup\@sanitize@url \@url }%
\providecommand \@url [1]{\endgroup\@href {#1}{\urlprefix }}%
\providecommand \urlprefix  [0]{URL }%
\providecommand \Eprint [0]{\href }%
\providecommand \doibase [0]{http://dx.doi.org/}%
\providecommand \selectlanguage [0]{\@gobble}%
\providecommand \bibinfo  [0]{\@secondoftwo}%
\providecommand \bibfield  [0]{\@secondoftwo}%
\providecommand \translation [1]{[#1]}%
\providecommand \BibitemOpen [0]{}%
\providecommand \bibitemStop [0]{}%
\providecommand \bibitemNoStop [0]{.\EOS\space}%
\providecommand \EOS [0]{\spacefactor3000\relax}%
\providecommand \BibitemShut  [1]{\csname bibitem#1\endcsname}%
\let\auto@bib@innerbib\@empty
%</preamble>
\bibitem [{\citenamefont {Ferray}\ \emph {et~al.}(1988)\citenamefont {Ferray},
  \citenamefont {L'Huillier}, \citenamefont {Li}, \citenamefont {Lompre},
  \citenamefont {Mainfray},\ and\ \citenamefont {Manus}}]{Ferray_1988}%
  \BibitemOpen
  \bibfield  {author} {\bibinfo {author} {\bibfnamefont {M.}~\bibnamefont
  {Ferray}}, \bibinfo {author} {\bibfnamefont {A.}~\bibnamefont {L'Huillier}},
  \bibinfo {author} {\bibfnamefont {X.~F.}\ \bibnamefont {Li}}, \bibinfo
  {author} {\bibfnamefont {L.~A.}\ \bibnamefont {Lompre}}, \bibinfo {author}
  {\bibfnamefont {G.}~\bibnamefont {Mainfray}}, \ and\ \bibinfo {author}
  {\bibfnamefont {C.}~\bibnamefont {Manus}},\ }\href {\doibase
  10.1088/0953-4075/21/3/001} {\bibfield  {journal} {\bibinfo  {journal}
  {Journal of Physics B: Atomic, Molecular and Optical Physics}\ }\textbf
  {\bibinfo {volume} {21}},\ \bibinfo {pages} {L31} (\bibinfo {year}
  {1988})}\BibitemShut {NoStop}%
\bibitem [{\citenamefont {Krausz}\ and\ \citenamefont
  {Ivanov}(2009)}]{Krausz2009RMP}%
  \BibitemOpen
  \bibfield  {author} {\bibinfo {author} {\bibfnamefont {F.}~\bibnamefont
  {Krausz}}\ and\ \bibinfo {author} {\bibfnamefont {M.}~\bibnamefont
  {Ivanov}},\ }\href {\doibase 10.1103/RevModPhys.81.163} {\bibfield  {journal}
  {\bibinfo  {journal} {Rev. Mod. Phys.}\ }\textbf {\bibinfo {volume} {81}},\
  \bibinfo {pages} {163} (\bibinfo {year} {2009})}\BibitemShut {NoStop}%
\bibitem [{\citenamefont {Ghimire}\ \emph {et~al.}(2011)\citenamefont
  {Ghimire}, \citenamefont {DiChiara}, \citenamefont {Sistrunk}, \citenamefont
  {Agostini}, \citenamefont {DiMauro},\ and\ \citenamefont
  {Reis}}]{Ghimire2011NatPhys}%
  \BibitemOpen
  \bibfield  {author} {\bibinfo {author} {\bibfnamefont {S.}~\bibnamefont
  {Ghimire}}, \bibinfo {author} {\bibfnamefont {A.~D.}\ \bibnamefont
  {DiChiara}}, \bibinfo {author} {\bibfnamefont {E.}~\bibnamefont {Sistrunk}},
  \bibinfo {author} {\bibfnamefont {P.}~\bibnamefont {Agostini}}, \bibinfo
  {author} {\bibfnamefont {L.~F.}\ \bibnamefont {DiMauro}}, \ and\ \bibinfo
  {author} {\bibfnamefont {D.~A.}\ \bibnamefont {Reis}},\ }\href
  {https://doi.org/10.1038/nphys1847} {\bibfield  {journal} {\bibinfo
  {journal} {Nat. Phys.}\ }\textbf {\bibinfo {volume} {7}},\ \bibinfo {pages}
  {138} (\bibinfo {year} {2011})}\BibitemShut {NoStop}%
\bibitem [{\citenamefont {Schubert}\ \emph {et~al.}(2014)\citenamefont
  {Schubert}, \citenamefont {Hohenleutner}, \citenamefont {Langer},
  \citenamefont {Urbanek}, \citenamefont {Lange}, \citenamefont {Huttner},
  \citenamefont {Golde}, \citenamefont {Meier}, \citenamefont {Kira},
  \citenamefont {Koch},\ and\ \citenamefont {Huber}}]{Schubert2014}%
  \BibitemOpen
  \bibfield  {author} {\bibinfo {author} {\bibfnamefont {O.}~\bibnamefont
  {Schubert}}, \bibinfo {author} {\bibfnamefont {M.}~\bibnamefont
  {Hohenleutner}}, \bibinfo {author} {\bibfnamefont {F.}~\bibnamefont
  {Langer}}, \bibinfo {author} {\bibfnamefont {B.}~\bibnamefont {Urbanek}},
  \bibinfo {author} {\bibfnamefont {C.}~\bibnamefont {Lange}}, \bibinfo
  {author} {\bibfnamefont {U.}~\bibnamefont {Huttner}}, \bibinfo {author}
  {\bibfnamefont {D.}~\bibnamefont {Golde}}, \bibinfo {author} {\bibfnamefont
  {T.}~\bibnamefont {Meier}}, \bibinfo {author} {\bibfnamefont
  {M.}~\bibnamefont {Kira}}, \bibinfo {author} {\bibfnamefont {S.~W.}\
  \bibnamefont {Koch}}, \ and\ \bibinfo {author} {\bibfnamefont
  {R.}~\bibnamefont {Huber}},\ }\href
  {http://dx.doi.org/10.1038/nphoton.2013.349} {\bibfield  {journal} {\bibinfo
  {journal} {Nat. Photon.}\ }\textbf {\bibinfo {volume} {8}},\ \bibinfo {pages}
  {119} (\bibinfo {year} {2014})}\BibitemShut {NoStop}%
\bibitem [{\citenamefont {Luu}\ \emph {et~al.}(2015)\citenamefont {Luu},
  \citenamefont {Garg}, \citenamefont {Kruchinin}, \citenamefont {Moulet},
  \citenamefont {Hassan},\ and\ \citenamefont {Goulielmakis}}]{Luu2015}%
  \BibitemOpen
  \bibfield  {author} {\bibinfo {author} {\bibfnamefont {T.~T.}\ \bibnamefont
  {Luu}}, \bibinfo {author} {\bibfnamefont {M.}~\bibnamefont {Garg}}, \bibinfo
  {author} {\bibfnamefont {S.~Y.}\ \bibnamefont {Kruchinin}}, \bibinfo {author}
  {\bibfnamefont {A.}~\bibnamefont {Moulet}}, \bibinfo {author} {\bibfnamefont
  {M.~T.}\ \bibnamefont {Hassan}}, \ and\ \bibinfo {author} {\bibfnamefont
  {E.}~\bibnamefont {Goulielmakis}},\ }\href
  {http://www.nature.com/doifinder/10.1038/nature14456} {\bibfield  {journal}
  {\bibinfo  {journal} {Nature (London)}\ }\textbf {\bibinfo {volume} {521}},\
  \bibinfo {pages} {498} (\bibinfo {year} {2015})}\BibitemShut {NoStop}%
\bibitem [{\citenamefont {Vampa}\ \emph
  {et~al.}(2015{\natexlab{a}})\citenamefont {Vampa}, \citenamefont {Hammond},
  \citenamefont {Thire}, \citenamefont {Schmidt}, \citenamefont {Legare},
  \citenamefont {McDonald}, \citenamefont {Brabec},\ and\ \citenamefont
  {Corkum}}]{Vampa2015Nature}%
  \BibitemOpen
  \bibfield  {author} {\bibinfo {author} {\bibfnamefont {G.}~\bibnamefont
  {Vampa}}, \bibinfo {author} {\bibfnamefont {T.~J.}\ \bibnamefont {Hammond}},
  \bibinfo {author} {\bibfnamefont {N.}~\bibnamefont {Thire}}, \bibinfo
  {author} {\bibfnamefont {B.~E.}\ \bibnamefont {Schmidt}}, \bibinfo {author}
  {\bibfnamefont {F.}~\bibnamefont {Legare}}, \bibinfo {author} {\bibfnamefont
  {C.~R.}\ \bibnamefont {McDonald}}, \bibinfo {author} {\bibfnamefont
  {T.}~\bibnamefont {Brabec}}, \ and\ \bibinfo {author} {\bibfnamefont {P.~B.}\
  \bibnamefont {Corkum}},\ }\href {http://dx.doi.org/10.1038/nature14517}
  {\bibfield  {journal} {\bibinfo  {journal} {Nature (London)}\ }\textbf
  {\bibinfo {volume} {522}},\ \bibinfo {pages} {462} (\bibinfo {year}
  {2015}{\natexlab{a}})}\BibitemShut {NoStop}%
\bibitem [{\citenamefont {Langer}\ \emph {et~al.}(2016)\citenamefont {Langer},
  \citenamefont {Hohenleutner}, \citenamefont {Schmid}, \citenamefont
  {P{\"o}llmann}, \citenamefont {Nagler}, \citenamefont {Korn}, \citenamefont
  {Sch{\"u}ller}, \citenamefont {Sherwin}, \citenamefont {Huttner},
  \citenamefont {Steiner}, \citenamefont {Koch}, \citenamefont {Kira},\ and\
  \citenamefont {Huber}}]{Langer2016Nature}%
  \BibitemOpen
  \bibfield  {author} {\bibinfo {author} {\bibfnamefont {F.}~\bibnamefont
  {Langer}}, \bibinfo {author} {\bibfnamefont {M.}~\bibnamefont
  {Hohenleutner}}, \bibinfo {author} {\bibfnamefont {C.~P.}\ \bibnamefont
  {Schmid}}, \bibinfo {author} {\bibfnamefont {C.}~\bibnamefont
  {P{\"o}llmann}}, \bibinfo {author} {\bibfnamefont {P.}~\bibnamefont
  {Nagler}}, \bibinfo {author} {\bibfnamefont {T.}~\bibnamefont {Korn}},
  \bibinfo {author} {\bibfnamefont {C.}~\bibnamefont {Sch{\"u}ller}}, \bibinfo
  {author} {\bibfnamefont {M.}~\bibnamefont {Sherwin}}, \bibinfo {author}
  {\bibfnamefont {U.}~\bibnamefont {Huttner}}, \bibinfo {author} {\bibfnamefont
  {J.}~\bibnamefont {Steiner}}, \bibinfo {author} {\bibfnamefont
  {S.}~\bibnamefont {Koch}}, \bibinfo {author} {\bibfnamefont {M.}~\bibnamefont
  {Kira}}, \ and\ \bibinfo {author} {\bibfnamefont {R.}~\bibnamefont {Huber}},\
  }\href {http://www.nature.com/doifinder/10.1038/nature17958} {\bibfield
  {journal} {\bibinfo  {journal} {Nature (London)}\ }\textbf {\bibinfo {volume}
  {533}},\ \bibinfo {pages} {225} (\bibinfo {year} {2016})}\BibitemShut
  {NoStop}%
\bibitem [{\citenamefont {Hohenleutner}\ \emph {et~al.}(2015)\citenamefont
  {Hohenleutner}, \citenamefont {Langer}, \citenamefont {Schubert},
  \citenamefont {Knorr}, \citenamefont {Huttner}, \citenamefont {Koch},
  \citenamefont {Kira},\ and\ \citenamefont {Huber}}]{Hohenleutner2015Nature}%
  \BibitemOpen
  \bibfield  {author} {\bibinfo {author} {\bibfnamefont {M.}~\bibnamefont
  {Hohenleutner}}, \bibinfo {author} {\bibfnamefont {F.}~\bibnamefont
  {Langer}}, \bibinfo {author} {\bibfnamefont {O.}~\bibnamefont {Schubert}},
  \bibinfo {author} {\bibfnamefont {M.}~\bibnamefont {Knorr}}, \bibinfo
  {author} {\bibfnamefont {U.}~\bibnamefont {Huttner}}, \bibinfo {author}
  {\bibfnamefont {S.}~\bibnamefont {Koch}}, \bibinfo {author} {\bibfnamefont
  {M.}~\bibnamefont {Kira}}, \ and\ \bibinfo {author} {\bibfnamefont
  {R.}~\bibnamefont {Huber}},\ }\href {https://doi.org/10.1038/nature14652}
  {\bibfield  {journal} {\bibinfo  {journal} {Nature (London)}\ }\textbf
  {\bibinfo {volume} {523}},\ \bibinfo {pages} {572} (\bibinfo {year}
  {2015})}\BibitemShut {NoStop}%
\bibitem [{\citenamefont {Ndabashimiye}\ \emph {et~al.}(2016)\citenamefont
  {Ndabashimiye}, \citenamefont {Ghimire}, \citenamefont {Wu}, \citenamefont
  {Browne}, \citenamefont {Schafer}, \citenamefont {Gaarde},\ and\
  \citenamefont {Reis}}]{Ndabashimiye2016}%
  \BibitemOpen
  \bibfield  {author} {\bibinfo {author} {\bibfnamefont {G.}~\bibnamefont
  {Ndabashimiye}}, \bibinfo {author} {\bibfnamefont {S.}~\bibnamefont
  {Ghimire}}, \bibinfo {author} {\bibfnamefont {M.}~\bibnamefont {Wu}},
  \bibinfo {author} {\bibfnamefont {D.~A.}\ \bibnamefont {Browne}}, \bibinfo
  {author} {\bibfnamefont {K.~J.}\ \bibnamefont {Schafer}}, \bibinfo {author}
  {\bibfnamefont {M.~B.}\ \bibnamefont {Gaarde}}, \ and\ \bibinfo {author}
  {\bibfnamefont {D.~A.}\ \bibnamefont {Reis}},\ }\href
  {http://www.nature.com/doifinder/10.1038/nature17660} {\bibfield  {journal}
  {\bibinfo  {journal} {Nature (London)}\ }\textbf {\bibinfo {volume} {534}},\
  \bibinfo {pages} {520} (\bibinfo {year} {2016})}\BibitemShut {NoStop}%
\bibitem [{\citenamefont {Liu}\ \emph {et~al.}(2017)\citenamefont {Liu},
  \citenamefont {Li}, \citenamefont {You}, \citenamefont {Ghimire},
  \citenamefont {Heinz},\ and\ \citenamefont {Reis}}]{Liu2017}%
  \BibitemOpen
  \bibfield  {author} {\bibinfo {author} {\bibfnamefont {H.}~\bibnamefont
  {Liu}}, \bibinfo {author} {\bibfnamefont {Y.}~\bibnamefont {Li}}, \bibinfo
  {author} {\bibfnamefont {Y.~S.}\ \bibnamefont {You}}, \bibinfo {author}
  {\bibfnamefont {S.}~\bibnamefont {Ghimire}}, \bibinfo {author} {\bibfnamefont
  {T.~F.}\ \bibnamefont {Heinz}}, \ and\ \bibinfo {author} {\bibfnamefont
  {D.~A.}\ \bibnamefont {Reis}},\ }\href
  {https://www.nature.com/articles/nphys3946} {\bibfield  {journal} {\bibinfo
  {journal} {Nat. Phys.}\ }\textbf {\bibinfo {volume} {13}},\ \bibinfo {pages}
  {262} (\bibinfo {year} {2017})}\BibitemShut {NoStop}%
\bibitem [{\citenamefont {You}\ \emph {et~al.}(2017)\citenamefont {You},
  \citenamefont {Reis},\ and\ \citenamefont {Ghimire}}]{You2016}%
  \BibitemOpen
  \bibfield  {author} {\bibinfo {author} {\bibfnamefont {Y.~S.}\ \bibnamefont
  {You}}, \bibinfo {author} {\bibfnamefont {D.~A.}\ \bibnamefont {Reis}}, \
  and\ \bibinfo {author} {\bibfnamefont {S.}~\bibnamefont {Ghimire}},\ }\href
  {http://www.nature.com/doifinder/10.1038/nphys3955} {\bibfield  {journal}
  {\bibinfo  {journal} {Nat. Phys.}\ }\textbf {\bibinfo {volume} {13}},\
  \bibinfo {pages} {345} (\bibinfo {year} {2017})}\BibitemShut {NoStop}%
\bibitem [{\citenamefont {Yoshikawa}\ \emph {et~al.}(2017)\citenamefont
  {Yoshikawa}, \citenamefont {Tamaya},\ and\ \citenamefont
  {Tanaka}}]{Yoshikawa2017Science}%
  \BibitemOpen
  \bibfield  {author} {\bibinfo {author} {\bibfnamefont {N.}~\bibnamefont
  {Yoshikawa}}, \bibinfo {author} {\bibfnamefont {T.}~\bibnamefont {Tamaya}}, \
  and\ \bibinfo {author} {\bibfnamefont {K.}~\bibnamefont {Tanaka}},\ }\href
  {\doibase 10.1126/science.aam8861} {\bibfield  {journal} {\bibinfo  {journal}
  {Science}\ }\textbf {\bibinfo {volume} {356}},\ \bibinfo {pages} {736}
  (\bibinfo {year} {2017})}\BibitemShut {NoStop}%
\bibitem [{\citenamefont {Hafez}\ \emph {et~al.}(2018)\citenamefont {Hafez},
  \citenamefont {Kovalev}, \citenamefont {Deinert}, \citenamefont {Mics},
  \citenamefont {Green}, \citenamefont {Awari}, \citenamefont {Chen},
  \citenamefont {Germanskiy}, \citenamefont {Lehnert}, \citenamefont
  {Teichert}, \citenamefont {Wang}, \citenamefont {Tielrooij}, \citenamefont
  {Liu}, \citenamefont {Chen}, \citenamefont {Narita}, \citenamefont {Mullen},
  \citenamefont {Bonn}, \citenamefont {Gensch},\ and\ \citenamefont
  {Turchinovich}}]{Hafez2018}%
  \BibitemOpen
  \bibfield  {author} {\bibinfo {author} {\bibfnamefont {H.~A.}\ \bibnamefont
  {Hafez}}, \bibinfo {author} {\bibfnamefont {S.}~\bibnamefont {Kovalev}},
  \bibinfo {author} {\bibfnamefont {J.-C.}\ \bibnamefont {Deinert}}, \bibinfo
  {author} {\bibfnamefont {Z.}~\bibnamefont {Mics}}, \bibinfo {author}
  {\bibfnamefont {B.}~\bibnamefont {Green}}, \bibinfo {author} {\bibfnamefont
  {N.}~\bibnamefont {Awari}}, \bibinfo {author} {\bibfnamefont
  {M.}~\bibnamefont {Chen}}, \bibinfo {author} {\bibfnamefont {S.}~\bibnamefont
  {Germanskiy}}, \bibinfo {author} {\bibfnamefont {U.}~\bibnamefont {Lehnert}},
  \bibinfo {author} {\bibfnamefont {J.}~\bibnamefont {Teichert}}, \bibinfo
  {author} {\bibfnamefont {Z.}~\bibnamefont {Wang}}, \bibinfo {author}
  {\bibfnamefont {K.-J.}\ \bibnamefont {Tielrooij}}, \bibinfo {author}
  {\bibfnamefont {Z.}~\bibnamefont {Liu}}, \bibinfo {author} {\bibfnamefont
  {Z.}~\bibnamefont {Chen}}, \bibinfo {author} {\bibfnamefont {A.}~\bibnamefont
  {Narita}}, \bibinfo {author} {\bibfnamefont {K.}~\bibnamefont {Mullen}},
  \bibinfo {author} {\bibfnamefont {M.}~\bibnamefont {Bonn}}, \bibinfo {author}
  {\bibfnamefont {M.}~\bibnamefont {Gensch}}, \ and\ \bibinfo {author}
  {\bibfnamefont {D.}~\bibnamefont {Turchinovich}},\ }\href
  {http://dx.doi.org/10.1038/s41586-018-0508-1} {\bibfield  {journal} {\bibinfo
   {journal} {Nature (London)}\ }\textbf {\bibinfo {volume} {561}},\ \bibinfo
  {pages} {507} (\bibinfo {year} {2018})}\BibitemShut {NoStop}%
\bibitem [{\citenamefont {Kaneshima}\ \emph {et~al.}(2018)\citenamefont
  {Kaneshima}, \citenamefont {Shinohara}, \citenamefont {Takeuchi},
  \citenamefont {Ishii}, \citenamefont {Imasaka}, \citenamefont {Kaji},
  \citenamefont {Ashihara}, \citenamefont {Ishikawa},\ and\ \citenamefont
  {Itatani}}]{Kaneshima2018}%
  \BibitemOpen
  \bibfield  {author} {\bibinfo {author} {\bibfnamefont {K.}~\bibnamefont
  {Kaneshima}}, \bibinfo {author} {\bibfnamefont {Y.}~\bibnamefont
  {Shinohara}}, \bibinfo {author} {\bibfnamefont {K.}~\bibnamefont {Takeuchi}},
  \bibinfo {author} {\bibfnamefont {N.}~\bibnamefont {Ishii}}, \bibinfo
  {author} {\bibfnamefont {K.}~\bibnamefont {Imasaka}}, \bibinfo {author}
  {\bibfnamefont {T.}~\bibnamefont {Kaji}}, \bibinfo {author} {\bibfnamefont
  {S.}~\bibnamefont {Ashihara}}, \bibinfo {author} {\bibfnamefont {K.~L.}\
  \bibnamefont {Ishikawa}}, \ and\ \bibinfo {author} {\bibfnamefont
  {J.}~\bibnamefont {Itatani}},\ }\href {\doibase
  10.1103/PhysRevLett.120.243903} {\bibfield  {journal} {\bibinfo  {journal}
  {Phys. Rev. Lett.}\ }\textbf {\bibinfo {volume} {120}},\ \bibinfo {pages}
  {243903} (\bibinfo {year} {2018})}\BibitemShut {NoStop}%
\bibitem [{\citenamefont {Yoshikawa}\ \emph {et~al.}(2019)\citenamefont
  {Yoshikawa}, \citenamefont {Nagai}, \citenamefont {Uchida}, \citenamefont
  {Takaguchi}, \citenamefont {Sasaki}, \citenamefont {Miyata},\ and\
  \citenamefont {Tanaka}}]{Yoshikawa2019}%
  \BibitemOpen
  \bibfield  {author} {\bibinfo {author} {\bibfnamefont {N.}~\bibnamefont
  {Yoshikawa}}, \bibinfo {author} {\bibfnamefont {K.}~\bibnamefont {Nagai}},
  \bibinfo {author} {\bibfnamefont {K.}~\bibnamefont {Uchida}}, \bibinfo
  {author} {\bibfnamefont {Y.}~\bibnamefont {Takaguchi}}, \bibinfo {author}
  {\bibfnamefont {S.}~\bibnamefont {Sasaki}}, \bibinfo {author} {\bibfnamefont
  {Y.}~\bibnamefont {Miyata}}, \ and\ \bibinfo {author} {\bibfnamefont
  {K.}~\bibnamefont {Tanaka}},\ }\href
  {https://doi.org/10.1038/s41467-019-11697-6} {\bibfield  {journal} {\bibinfo
  {journal} {Nat. Comm.}\ }\textbf {\bibinfo {volume} {10}},\ \bibinfo {pages}
  {1} (\bibinfo {year} {2019})}\BibitemShut {NoStop}%
\bibitem [{\citenamefont {Cheng}\ \emph {et~al.}(2020)\citenamefont {Cheng},
  \citenamefont {Kanda}, \citenamefont {Ikeda}, \citenamefont {Matsuda},
  \citenamefont {Xia}, \citenamefont {Schumann}, \citenamefont {Stemmer},
  \citenamefont {Itatani}, \citenamefont {Armitage},\ and\ \citenamefont
  {Matsunaga}}]{Matsunaga2020PRL}%
  \BibitemOpen
  \bibfield  {author} {\bibinfo {author} {\bibfnamefont {B.}~\bibnamefont
  {Cheng}}, \bibinfo {author} {\bibfnamefont {N.}~\bibnamefont {Kanda}},
  \bibinfo {author} {\bibfnamefont {T.~N.}\ \bibnamefont {Ikeda}}, \bibinfo
  {author} {\bibfnamefont {T.}~\bibnamefont {Matsuda}}, \bibinfo {author}
  {\bibfnamefont {P.}~\bibnamefont {Xia}}, \bibinfo {author} {\bibfnamefont
  {T.}~\bibnamefont {Schumann}}, \bibinfo {author} {\bibfnamefont
  {S.}~\bibnamefont {Stemmer}}, \bibinfo {author} {\bibfnamefont
  {J.}~\bibnamefont {Itatani}}, \bibinfo {author} {\bibfnamefont {N.~P.}\
  \bibnamefont {Armitage}}, \ and\ \bibinfo {author} {\bibfnamefont
  {R.}~\bibnamefont {Matsunaga}},\ }\href {\doibase
  10.1103/PhysRevLett.124.117402} {\bibfield  {journal} {\bibinfo  {journal}
  {Phys. Rev. Lett.}\ }\textbf {\bibinfo {volume} {124}},\ \bibinfo {pages}
  {117402} (\bibinfo {year} {2020})}\BibitemShut {NoStop}%
\bibitem [{\citenamefont {Schmid}\ \emph {et~al.}(2021)\citenamefont {Schmid},
  \citenamefont {Weigl}, \citenamefont {Gr{\"o}ssing}, \citenamefont {Junk},
  \citenamefont {Gorini}, \citenamefont {Schlauderer}, \citenamefont {Ito},
  \citenamefont {Meierhofer}, \citenamefont {Hofmann}, \citenamefont
  {Afanasiev}, \citenamefont {Crewse}, \citenamefont {Kokh}, \citenamefont
  {Tereshchenko}, \citenamefont {G{\"u}dde}, \citenamefont {Evers},
  \citenamefont {Wilhelm}, \citenamefont {Richter}, \citenamefont {H{\"o}fer},\
  and\ \citenamefont {Huber}}]{Schmid2021}%
  \BibitemOpen
  \bibfield  {author} {\bibinfo {author} {\bibfnamefont {C.~P.}\ \bibnamefont
  {Schmid}}, \bibinfo {author} {\bibfnamefont {L.}~\bibnamefont {Weigl}},
  \bibinfo {author} {\bibfnamefont {P.}~\bibnamefont {Gr{\"o}ssing}}, \bibinfo
  {author} {\bibfnamefont {V.}~\bibnamefont {Junk}}, \bibinfo {author}
  {\bibfnamefont {C.}~\bibnamefont {Gorini}}, \bibinfo {author} {\bibfnamefont
  {S.}~\bibnamefont {Schlauderer}}, \bibinfo {author} {\bibfnamefont
  {S.}~\bibnamefont {Ito}}, \bibinfo {author} {\bibfnamefont {M.}~\bibnamefont
  {Meierhofer}}, \bibinfo {author} {\bibfnamefont {N.}~\bibnamefont {Hofmann}},
  \bibinfo {author} {\bibfnamefont {D.}~\bibnamefont {Afanasiev}}, \bibinfo
  {author} {\bibfnamefont {J.}~\bibnamefont {Crewse}}, \bibinfo {author}
  {\bibfnamefont {K.~A.}\ \bibnamefont {Kokh}}, \bibinfo {author}
  {\bibfnamefont {O.~E.}\ \bibnamefont {Tereshchenko}}, \bibinfo {author}
  {\bibfnamefont {J.}~\bibnamefont {G{\"u}dde}}, \bibinfo {author}
  {\bibfnamefont {F.}~\bibnamefont {Evers}}, \bibinfo {author} {\bibfnamefont
  {J.}~\bibnamefont {Wilhelm}}, \bibinfo {author} {\bibfnamefont
  {K.}~\bibnamefont {Richter}}, \bibinfo {author} {\bibfnamefont
  {U.}~\bibnamefont {H{\"o}fer}}, \ and\ \bibinfo {author} {\bibfnamefont
  {R.}~\bibnamefont {Huber}},\ }\href {\doibase 10.1038/s41586-021-03466-7}
  {\bibfield  {journal} {\bibinfo  {journal} {Nature}\ }\textbf {\bibinfo
  {volume} {593}},\ \bibinfo {pages} {385} (\bibinfo {year}
  {2021})}\BibitemShut {NoStop}%
\bibitem [{\citenamefont {Golde}\ \emph {et~al.}(2008)\citenamefont {Golde},
  \citenamefont {Meier},\ and\ \citenamefont {Koch}}]{Golde2008}%
  \BibitemOpen
  \bibfield  {author} {\bibinfo {author} {\bibfnamefont {D.}~\bibnamefont
  {Golde}}, \bibinfo {author} {\bibfnamefont {T.}~\bibnamefont {Meier}}, \ and\
  \bibinfo {author} {\bibfnamefont {S.~W.}\ \bibnamefont {Koch}},\ }\href
  {\doibase 10.1103/PhysRevB.77.075330} {\bibfield  {journal} {\bibinfo
  {journal} {Phys. Rev. B}\ }\textbf {\bibinfo {volume} {77}},\ \bibinfo
  {pages} {075330} (\bibinfo {year} {2008})}\BibitemShut {NoStop}%
\bibitem [{\citenamefont {Vampa}\ \emph {et~al.}(2014)\citenamefont {Vampa},
  \citenamefont {McDonald}, \citenamefont {Orlando}, \citenamefont {Klug},
  \citenamefont {Corkum},\ and\ \citenamefont {Brabec}}]{Vampa2014PRL}%
  \BibitemOpen
  \bibfield  {author} {\bibinfo {author} {\bibfnamefont {G.}~\bibnamefont
  {Vampa}}, \bibinfo {author} {\bibfnamefont {C.~R.}\ \bibnamefont {McDonald}},
  \bibinfo {author} {\bibfnamefont {G.}~\bibnamefont {Orlando}}, \bibinfo
  {author} {\bibfnamefont {D.~D.}\ \bibnamefont {Klug}}, \bibinfo {author}
  {\bibfnamefont {P.~B.}\ \bibnamefont {Corkum}}, \ and\ \bibinfo {author}
  {\bibfnamefont {T.}~\bibnamefont {Brabec}},\ }\href {\doibase
  10.1103/PhysRevLett.113.073901} {\bibfield  {journal} {\bibinfo  {journal}
  {Phys. Rev. Lett.}\ }\textbf {\bibinfo {volume} {113}},\ \bibinfo {pages}
  {073901} (\bibinfo {year} {2014})}\BibitemShut {NoStop}%
\bibitem [{\citenamefont {Vampa}\ \emph
  {et~al.}(2015{\natexlab{b}})\citenamefont {Vampa}, \citenamefont {McDonald},
  \citenamefont {Orlando}, \citenamefont {Corkum},\ and\ \citenamefont
  {Brabec}}]{Vampa2015PRB}%
  \BibitemOpen
  \bibfield  {author} {\bibinfo {author} {\bibfnamefont {G.}~\bibnamefont
  {Vampa}}, \bibinfo {author} {\bibfnamefont {C.~R.}\ \bibnamefont {McDonald}},
  \bibinfo {author} {\bibfnamefont {G.}~\bibnamefont {Orlando}}, \bibinfo
  {author} {\bibfnamefont {P.~B.}\ \bibnamefont {Corkum}}, \ and\ \bibinfo
  {author} {\bibfnamefont {T.}~\bibnamefont {Brabec}},\ }\href {\doibase
  10.1103/PhysRevB.91.064302} {\bibfield  {journal} {\bibinfo  {journal} {Phys.
  Rev. B}\ }\textbf {\bibinfo {volume} {91}},\ \bibinfo {pages} {064302}
  (\bibinfo {year} {2015}{\natexlab{b}})}\BibitemShut {NoStop}%
\bibitem [{\citenamefont {Wu}\ \emph {et~al.}(2015)\citenamefont {Wu},
  \citenamefont {Ghimire}, \citenamefont {Reis}, \citenamefont {Schafer},\ and\
  \citenamefont {Gaarde}}]{Wu2015}%
  \BibitemOpen
  \bibfield  {author} {\bibinfo {author} {\bibfnamefont {M.}~\bibnamefont
  {Wu}}, \bibinfo {author} {\bibfnamefont {S.}~\bibnamefont {Ghimire}},
  \bibinfo {author} {\bibfnamefont {D.~A.}\ \bibnamefont {Reis}}, \bibinfo
  {author} {\bibfnamefont {K.~J.}\ \bibnamefont {Schafer}}, \ and\ \bibinfo
  {author} {\bibfnamefont {M.~B.}\ \bibnamefont {Gaarde}},\ }\href {\doibase
  10.1103/PhysRevA.91.043839} {\bibfield  {journal} {\bibinfo  {journal} {Phys.
  Rev. A}\ }\textbf {\bibinfo {volume} {91}},\ \bibinfo {pages} {043839}
  (\bibinfo {year} {2015})}\BibitemShut {NoStop}%
\bibitem [{\citenamefont {Otobe}(2016)}]{Otobe2016}%
  \BibitemOpen
  \bibfield  {author} {\bibinfo {author} {\bibfnamefont {T.}~\bibnamefont
  {Otobe}},\ }\href {\doibase 10.1103/PhysRevB.94.235152} {\bibfield  {journal}
  {\bibinfo  {journal} {Phys. Rev. B}\ }\textbf {\bibinfo {volume} {94}},\
  \bibinfo {pages} {235152} (\bibinfo {year} {2016})}\BibitemShut {NoStop}%
\bibitem [{\citenamefont {Ikemachi}\ \emph {et~al.}(2017)\citenamefont
  {Ikemachi}, \citenamefont {Shinohara}, \citenamefont {Sato}, \citenamefont
  {Yumoto}, \citenamefont {Kuwata-Gonokami},\ and\ \citenamefont
  {Ishikawa}}]{Ikemachi2017}%
  \BibitemOpen
  \bibfield  {author} {\bibinfo {author} {\bibfnamefont {T.}~\bibnamefont
  {Ikemachi}}, \bibinfo {author} {\bibfnamefont {Y.}~\bibnamefont {Shinohara}},
  \bibinfo {author} {\bibfnamefont {T.}~\bibnamefont {Sato}}, \bibinfo {author}
  {\bibfnamefont {J.}~\bibnamefont {Yumoto}}, \bibinfo {author} {\bibfnamefont
  {M.}~\bibnamefont {Kuwata-Gonokami}}, \ and\ \bibinfo {author} {\bibfnamefont
  {K.~L.}\ \bibnamefont {Ishikawa}},\ }\href {\doibase
  10.1103/PhysRevA.95.043416} {\bibfield  {journal} {\bibinfo  {journal} {Phys.
  Rev. A}\ }\textbf {\bibinfo {volume} {95}},\ \bibinfo {pages} {043416}
  (\bibinfo {year} {2017})}\BibitemShut {NoStop}%
\bibitem [{\citenamefont {Tancogne-Dejean}\ \emph {et~al.}(2017)\citenamefont
  {Tancogne-Dejean}, \citenamefont {M{\"u}cke}, \citenamefont {K{\"a}rtner},\
  and\ \citenamefont {Rubio}}]{Tancogne-Dejean2017}%
  \BibitemOpen
  \bibfield  {author} {\bibinfo {author} {\bibfnamefont {N.}~\bibnamefont
  {Tancogne-Dejean}}, \bibinfo {author} {\bibfnamefont {O.~D.}\ \bibnamefont
  {M{\"u}cke}}, \bibinfo {author} {\bibfnamefont {F.~X.}\ \bibnamefont
  {K{\"a}rtner}}, \ and\ \bibinfo {author} {\bibfnamefont {A.}~\bibnamefont
  {Rubio}},\ }\href {\doibase 10.1038/s41467-017-00764-5} {\bibfield  {journal}
  {\bibinfo  {journal} {Nat. Comm.}\ }\textbf {\bibinfo {volume} {8}},\
  \bibinfo {pages} {745} (\bibinfo {year} {2017})}\BibitemShut {NoStop}%
\bibitem [{\citenamefont {Luu}\ and\ \citenamefont {W\"orner}(2016)}]{Luu2016}%
  \BibitemOpen
  \bibfield  {author} {\bibinfo {author} {\bibfnamefont {T.~T.}\ \bibnamefont
  {Luu}}\ and\ \bibinfo {author} {\bibfnamefont {H.~J.}\ \bibnamefont
  {W\"orner}},\ }\href {\doibase 10.1103/PhysRevB.94.115164} {\bibfield
  {journal} {\bibinfo  {journal} {Phys. Rev. B}\ }\textbf {\bibinfo {volume}
  {94}},\ \bibinfo {pages} {115164} (\bibinfo {year} {2016})}\BibitemShut
  {NoStop}%
\bibitem [{\citenamefont {Hansen}\ \emph {et~al.}(2017)\citenamefont {Hansen},
  \citenamefont {Deffge},\ and\ \citenamefont {Bauer}}]{Hansen2017}%
  \BibitemOpen
  \bibfield  {author} {\bibinfo {author} {\bibfnamefont {K.~K.}\ \bibnamefont
  {Hansen}}, \bibinfo {author} {\bibfnamefont {T.}~\bibnamefont {Deffge}}, \
  and\ \bibinfo {author} {\bibfnamefont {D.}~\bibnamefont {Bauer}},\ }\href
  {\doibase 10.1103/PhysRevA.96.053418} {\bibfield  {journal} {\bibinfo
  {journal} {Phys. Rev. A}\ }\textbf {\bibinfo {volume} {96}},\ \bibinfo
  {pages} {053418} (\bibinfo {year} {2017})}\BibitemShut {NoStop}%
\bibitem [{\citenamefont {Osika}\ \emph {et~al.}(2017)\citenamefont {Osika},
  \citenamefont {Chac\'on}, \citenamefont {Ortmann}, \citenamefont {Su\'arez},
  \citenamefont {P\'erez-Hern\'andez}, \citenamefont {Szafran}, \citenamefont
  {Ciappina}, \citenamefont {Sols}, \citenamefont {Landsman},\ and\
  \citenamefont {Lewenstein}}]{Osika2017}%
  \BibitemOpen
  \bibfield  {author} {\bibinfo {author} {\bibfnamefont {E.~N.}\ \bibnamefont
  {Osika}}, \bibinfo {author} {\bibfnamefont {A.}~\bibnamefont {Chac\'on}},
  \bibinfo {author} {\bibfnamefont {L.}~\bibnamefont {Ortmann}}, \bibinfo
  {author} {\bibfnamefont {N.}~\bibnamefont {Su\'arez}}, \bibinfo {author}
  {\bibfnamefont {J.~A.}\ \bibnamefont {P\'erez-Hern\'andez}}, \bibinfo
  {author} {\bibfnamefont {B.}~\bibnamefont {Szafran}}, \bibinfo {author}
  {\bibfnamefont {M.~F.}\ \bibnamefont {Ciappina}}, \bibinfo {author}
  {\bibfnamefont {F.}~\bibnamefont {Sols}}, \bibinfo {author} {\bibfnamefont
  {A.~S.}\ \bibnamefont {Landsman}}, \ and\ \bibinfo {author} {\bibfnamefont
  {M.}~\bibnamefont {Lewenstein}},\ }\href {\doibase 10.1103/PhysRevX.7.021017}
  {\bibfield  {journal} {\bibinfo  {journal} {Phys. Rev. X}\ }\textbf {\bibinfo
  {volume} {7}},\ \bibinfo {pages} {021017} (\bibinfo {year}
  {2017})}\BibitemShut {NoStop}%
\bibitem [{\citenamefont {Ikeda}\ \emph {et~al.}(2018)\citenamefont {Ikeda},
  \citenamefont {Chinzei},\ and\ \citenamefont {Tsunetsugu}}]{Ikeda2018PRA}%
  \BibitemOpen
  \bibfield  {author} {\bibinfo {author} {\bibfnamefont {T.~N.}\ \bibnamefont
  {Ikeda}}, \bibinfo {author} {\bibfnamefont {K.}~\bibnamefont {Chinzei}}, \
  and\ \bibinfo {author} {\bibfnamefont {H.}~\bibnamefont {Tsunetsugu}},\
  }\href {\doibase 10.1103/PhysRevA.98.063426} {\bibfield  {journal} {\bibinfo
  {journal} {Phys. Rev. A}\ }\textbf {\bibinfo {volume} {98}},\ \bibinfo
  {pages} {063426} (\bibinfo {year} {2018})}\BibitemShut {NoStop}%
\bibitem [{\citenamefont {Tamaya}\ \emph {et~al.}(2016)\citenamefont {Tamaya},
  \citenamefont {Ishikawa}, \citenamefont {Ogawa},\ and\ \citenamefont
  {Tanaka}}]{Tamaya2016}%
  \BibitemOpen
  \bibfield  {author} {\bibinfo {author} {\bibfnamefont {T.}~\bibnamefont
  {Tamaya}}, \bibinfo {author} {\bibfnamefont {A.}~\bibnamefont {Ishikawa}},
  \bibinfo {author} {\bibfnamefont {T.}~\bibnamefont {Ogawa}}, \ and\ \bibinfo
  {author} {\bibfnamefont {K.}~\bibnamefont {Tanaka}},\ }\href {\doibase
  10.1103/PhysRevLett.116.016601} {\bibfield  {journal} {\bibinfo  {journal}
  {Phys. Rev. Lett.}\ }\textbf {\bibinfo {volume} {116}},\ \bibinfo {pages}
  {016601} (\bibinfo {year} {2016})}\BibitemShut {NoStop}%
\bibitem [{\citenamefont {Floss}\ \emph {et~al.}(2018)\citenamefont {Floss},
  \citenamefont {Lemell}, \citenamefont {Wachter}, \citenamefont {Smejkal},
  \citenamefont {Sato}, \citenamefont {Tong}, \citenamefont {Yabana},\ and\
  \citenamefont {Burgd\"orfer}}]{Floss2018}%
  \BibitemOpen
  \bibfield  {author} {\bibinfo {author} {\bibfnamefont {I.}~\bibnamefont
  {Floss}}, \bibinfo {author} {\bibfnamefont {C.}~\bibnamefont {Lemell}},
  \bibinfo {author} {\bibfnamefont {G.}~\bibnamefont {Wachter}}, \bibinfo
  {author} {\bibfnamefont {V.}~\bibnamefont {Smejkal}}, \bibinfo {author}
  {\bibfnamefont {S.~A.}\ \bibnamefont {Sato}}, \bibinfo {author}
  {\bibfnamefont {X.-M.}\ \bibnamefont {Tong}}, \bibinfo {author}
  {\bibfnamefont {K.}~\bibnamefont {Yabana}}, \ and\ \bibinfo {author}
  {\bibfnamefont {J.}~\bibnamefont {Burgd\"orfer}},\ }\href {\doibase
  10.1103/PhysRevA.97.011401} {\bibfield  {journal} {\bibinfo  {journal} {Phys.
  Rev. A}\ }\textbf {\bibinfo {volume} {97}},\ \bibinfo {pages} {011401}
  (\bibinfo {year} {2018})}\BibitemShut {NoStop}%
\bibitem [{\citenamefont {Lysne}\ \emph
  {et~al.}(2020{\natexlab{a}})\citenamefont {Lysne}, \citenamefont {Murakami},
  \citenamefont {Sch\"uler},\ and\ \citenamefont {Werner}}]{Markus2020SOC}%
  \BibitemOpen
  \bibfield  {author} {\bibinfo {author} {\bibfnamefont {M.}~\bibnamefont
  {Lysne}}, \bibinfo {author} {\bibfnamefont {Y.}~\bibnamefont {Murakami}},
  \bibinfo {author} {\bibfnamefont {M.}~\bibnamefont {Sch\"uler}}, \ and\
  \bibinfo {author} {\bibfnamefont {P.}~\bibnamefont {Werner}},\ }\href
  {\doibase 10.1103/PhysRevB.102.081121} {\bibfield  {journal} {\bibinfo
  {journal} {Phys. Rev. B}\ }\textbf {\bibinfo {volume} {102}},\ \bibinfo
  {pages} {081121} (\bibinfo {year} {2020}{\natexlab{a}})}\BibitemShut
  {NoStop}%
\bibitem [{\citenamefont {Chac{\'o}n}\ \emph {et~al.}(2020)\citenamefont
  {Chac{\'o}n}, \citenamefont {Kim}, \citenamefont {Zhu}, \citenamefont
  {Kelly}, \citenamefont {Dauphin}, \citenamefont {Pisanty}, \citenamefont
  {Maxwell}, \citenamefont {Pic\'on}, \citenamefont {Ciappina}, \citenamefont
  {Kim}, \citenamefont {Ticknor}, \citenamefont {Saxena},\ and\ \citenamefont
  {Lewenstein}}]{Chacon2020PRB}%
  \BibitemOpen
  \bibfield  {author} {\bibinfo {author} {\bibfnamefont {A.}~\bibnamefont
  {Chac{\'o}n}}, \bibinfo {author} {\bibfnamefont {D.}~\bibnamefont {Kim}},
  \bibinfo {author} {\bibfnamefont {W.}~\bibnamefont {Zhu}}, \bibinfo {author}
  {\bibfnamefont {S.~P.}\ \bibnamefont {Kelly}}, \bibinfo {author}
  {\bibfnamefont {A.}~\bibnamefont {Dauphin}}, \bibinfo {author} {\bibfnamefont
  {E.}~\bibnamefont {Pisanty}}, \bibinfo {author} {\bibfnamefont {A.~S.}\
  \bibnamefont {Maxwell}}, \bibinfo {author} {\bibfnamefont {A.}~\bibnamefont
  {Pic\'on}}, \bibinfo {author} {\bibfnamefont {M.~F.}\ \bibnamefont
  {Ciappina}}, \bibinfo {author} {\bibfnamefont {D.~E.}\ \bibnamefont {Kim}},
  \bibinfo {author} {\bibfnamefont {C.}~\bibnamefont {Ticknor}}, \bibinfo
  {author} {\bibfnamefont {A.}~\bibnamefont {Saxena}}, \ and\ \bibinfo {author}
  {\bibfnamefont {M.}~\bibnamefont {Lewenstein}},\ }\href {\doibase
  10.1103/PhysRevB.102.134115} {\bibfield  {journal} {\bibinfo  {journal}
  {Phys. Rev. B}\ }\textbf {\bibinfo {volume} {102}},\ \bibinfo {pages}
  {134115} (\bibinfo {year} {2020})}\BibitemShut {NoStop}%
\bibitem [{\citenamefont {Wilhelm}\ \emph {et~al.}(2021)\citenamefont
  {Wilhelm}, \citenamefont {Gr\"ossing}, \citenamefont {Seith}, \citenamefont
  {Crewse}, \citenamefont {Nitsch}, \citenamefont {Weigl}, \citenamefont
  {Schmid},\ and\ \citenamefont {Evers}}]{Wilhelm2021PRB}%
  \BibitemOpen
  \bibfield  {author} {\bibinfo {author} {\bibfnamefont {J.}~\bibnamefont
  {Wilhelm}}, \bibinfo {author} {\bibfnamefont {P.}~\bibnamefont {Gr\"ossing}},
  \bibinfo {author} {\bibfnamefont {A.}~\bibnamefont {Seith}}, \bibinfo
  {author} {\bibfnamefont {J.}~\bibnamefont {Crewse}}, \bibinfo {author}
  {\bibfnamefont {M.}~\bibnamefont {Nitsch}}, \bibinfo {author} {\bibfnamefont
  {L.}~\bibnamefont {Weigl}}, \bibinfo {author} {\bibfnamefont
  {C.}~\bibnamefont {Schmid}}, \ and\ \bibinfo {author} {\bibfnamefont
  {F.}~\bibnamefont {Evers}},\ }\href {\doibase 10.1103/PhysRevB.103.125419}
  {\bibfield  {journal} {\bibinfo  {journal} {Phys. Rev. B}\ }\textbf {\bibinfo
  {volume} {103}},\ \bibinfo {pages} {125419} (\bibinfo {year}
  {2021})}\BibitemShut {NoStop}%
\bibitem [{\citenamefont {Taya}\ \emph {et~al.}(2021)\citenamefont {Taya},
  \citenamefont {Hongo},\ and\ \citenamefont {Ikeda}}]{Taya2021PRB}%
  \BibitemOpen
  \bibfield  {author} {\bibinfo {author} {\bibfnamefont {H.}~\bibnamefont
  {Taya}}, \bibinfo {author} {\bibfnamefont {M.}~\bibnamefont {Hongo}}, \ and\
  \bibinfo {author} {\bibfnamefont {T.~N.}\ \bibnamefont {Ikeda}},\ }\href
  {\doibase 10.1103/PhysRevB.104.L140305} {\bibfield  {journal} {\bibinfo
  {journal} {Phys. Rev. B}\ }\textbf {\bibinfo {volume} {104}},\ \bibinfo
  {pages} {L140305} (\bibinfo {year} {2021})}\BibitemShut {NoStop}%
\bibitem [{\citenamefont {Vampa}\ \emph
  {et~al.}(2015{\natexlab{c}})\citenamefont {Vampa}, \citenamefont {Hammond},
  \citenamefont {Thir\'e}, \citenamefont {Schmidt}, \citenamefont {L\'egar\'e},
  \citenamefont {McDonald}, \citenamefont {Brabec}, \citenamefont {Klug},\ and\
  \citenamefont {Corkum}}]{Vampa2015PRL}%
  \BibitemOpen
  \bibfield  {author} {\bibinfo {author} {\bibfnamefont {G.}~\bibnamefont
  {Vampa}}, \bibinfo {author} {\bibfnamefont {T.~J.}\ \bibnamefont {Hammond}},
  \bibinfo {author} {\bibfnamefont {N.}~\bibnamefont {Thir\'e}}, \bibinfo
  {author} {\bibfnamefont {B.~E.}\ \bibnamefont {Schmidt}}, \bibinfo {author}
  {\bibfnamefont {F.}~\bibnamefont {L\'egar\'e}}, \bibinfo {author}
  {\bibfnamefont {C.~R.}\ \bibnamefont {McDonald}}, \bibinfo {author}
  {\bibfnamefont {T.}~\bibnamefont {Brabec}}, \bibinfo {author} {\bibfnamefont
  {D.~D.}\ \bibnamefont {Klug}}, \ and\ \bibinfo {author} {\bibfnamefont
  {P.~B.}\ \bibnamefont {Corkum}},\ }\href {\doibase
  10.1103/PhysRevLett.115.193603} {\bibfield  {journal} {\bibinfo  {journal}
  {Phys. Rev. Lett.}\ }\textbf {\bibinfo {volume} {115}},\ \bibinfo {pages}
  {193603} (\bibinfo {year} {2015}{\natexlab{c}})}\BibitemShut {NoStop}%
\bibitem [{\citenamefont {Li}\ \emph {et~al.}(2020)\citenamefont {Li},
  \citenamefont {Lan}, \citenamefont {He}, \citenamefont {Cao}, \citenamefont
  {Zhang},\ and\ \citenamefont {Lu}}]{Li2020HHG}%
  \BibitemOpen
  \bibfield  {author} {\bibinfo {author} {\bibfnamefont {L.}~\bibnamefont
  {Li}}, \bibinfo {author} {\bibfnamefont {P.}~\bibnamefont {Lan}}, \bibinfo
  {author} {\bibfnamefont {L.}~\bibnamefont {He}}, \bibinfo {author}
  {\bibfnamefont {W.}~\bibnamefont {Cao}}, \bibinfo {author} {\bibfnamefont
  {Q.}~\bibnamefont {Zhang}}, \ and\ \bibinfo {author} {\bibfnamefont
  {P.}~\bibnamefont {Lu}},\ }\href {\doibase 10.1103/PhysRevLett.124.157403}
  {\bibfield  {journal} {\bibinfo  {journal} {Phys. Rev. Lett.}\ }\textbf
  {\bibinfo {volume} {124}},\ \bibinfo {pages} {157403} (\bibinfo {year}
  {2020})}\BibitemShut {NoStop}%
\bibitem [{\citenamefont {Luu}\ and\ \citenamefont
  {W{\"o}rner}(2018)}]{Luu2018Amorphas}%
  \BibitemOpen
  \bibfield  {author} {\bibinfo {author} {\bibfnamefont {T.~T.}\ \bibnamefont
  {Luu}}\ and\ \bibinfo {author} {\bibfnamefont {H.~J.}\ \bibnamefont
  {W{\"o}rner}},\ }\href {https://www.nature.com/articles/s41467-018-03397-4}
  {\bibfield  {journal} {\bibinfo  {journal} {Nat. Comm.}\ }\textbf {\bibinfo
  {volume} {9}},\ \bibinfo {pages} {916} (\bibinfo {year} {2018})}\BibitemShut
  {NoStop}%
\bibitem [{\citenamefont {Uchida}\ \emph {et~al.}(2021)\citenamefont {Uchida},
  \citenamefont {Pareek}, \citenamefont {Nagai}, \citenamefont {Dani},\ and\
  \citenamefont {Tanaka}}]{Uchida2020PRBL}%
  \BibitemOpen
  \bibfield  {author} {\bibinfo {author} {\bibfnamefont {K.}~\bibnamefont
  {Uchida}}, \bibinfo {author} {\bibfnamefont {V.}~\bibnamefont {Pareek}},
  \bibinfo {author} {\bibfnamefont {K.}~\bibnamefont {Nagai}}, \bibinfo
  {author} {\bibfnamefont {K.~M.}\ \bibnamefont {Dani}}, \ and\ \bibinfo
  {author} {\bibfnamefont {K.}~\bibnamefont {Tanaka}},\ }\href {\doibase
  10.1103/PhysRevB.103.L161406} {\bibfield  {journal} {\bibinfo  {journal}
  {Phys. Rev. B}\ }\textbf {\bibinfo {volume} {103}},\ \bibinfo {pages}
  {L161406} (\bibinfo {year} {2021})}\BibitemShut {NoStop}%
\bibitem [{\citenamefont {Kemper}\ \emph {et~al.}(2013)\citenamefont {Kemper},
  \citenamefont {Moritz}, \citenamefont {Freericks},\ and\ \citenamefont
  {Devereaux}}]{Kemper2013b}%
  \BibitemOpen
  \bibfield  {author} {\bibinfo {author} {\bibfnamefont {A.~F.}\ \bibnamefont
  {Kemper}}, \bibinfo {author} {\bibfnamefont {B.}~\bibnamefont {Moritz}},
  \bibinfo {author} {\bibfnamefont {J.~K.}\ \bibnamefont {Freericks}}, \ and\
  \bibinfo {author} {\bibfnamefont {T.~P.}\ \bibnamefont {Devereaux}},\ }\href
  {http://stacks.iop.org/1367-2630/15/i=2/a=023003} {\bibfield  {journal}
  {\bibinfo  {journal} {New J. Phys.}\ }\textbf {\bibinfo {volume} {15}},\
  \bibinfo {pages} {023003} (\bibinfo {year} {2013})}\BibitemShut {NoStop}%
\bibitem [{\citenamefont {Kruchinin}\ \emph {et~al.}(2018)\citenamefont
  {Kruchinin}, \citenamefont {Krausz},\ and\ \citenamefont
  {Yakovlev}}]{Kruchinin2018}%
  \BibitemOpen
  \bibfield  {author} {\bibinfo {author} {\bibfnamefont {S.~Y.}\ \bibnamefont
  {Kruchinin}}, \bibinfo {author} {\bibfnamefont {F.}~\bibnamefont {Krausz}}, \
  and\ \bibinfo {author} {\bibfnamefont {V.~S.}\ \bibnamefont {Yakovlev}},\
  }\href {\doibase 10.1103/RevModPhys.90.021002} {\bibfield  {journal}
  {\bibinfo  {journal} {Rev. Mod. Phys.}\ }\textbf {\bibinfo {volume} {90}},\
  \bibinfo {pages} {021002} (\bibinfo {year} {2018})}\BibitemShut {NoStop}%
\bibitem [{\citenamefont {Ghimire}\ and\ \citenamefont
  {Reis}(2019)}]{Ghimire2019}%
  \BibitemOpen
  \bibfield  {author} {\bibinfo {author} {\bibfnamefont {S.}~\bibnamefont
  {Ghimire}}\ and\ \bibinfo {author} {\bibfnamefont {D.~A.}\ \bibnamefont
  {Reis}},\ }\href {\doibase 10.1038/s41567-018-0315-5} {\bibfield  {journal}
  {\bibinfo  {journal} {Nat. Phys.}\ }\textbf {\bibinfo {volume} {15}},\
  \bibinfo {pages} {10} (\bibinfo {year} {2019})}\BibitemShut {NoStop}%
\bibitem [{\citenamefont {Du}\ and\ \citenamefont {Ma}(2022)}]{Du2022PRA}%
  \BibitemOpen
  \bibfield  {author} {\bibinfo {author} {\bibfnamefont {T.-Y.}\ \bibnamefont
  {Du}}\ and\ \bibinfo {author} {\bibfnamefont {C.}~\bibnamefont {Ma}},\ }\href
  {\doibase 10.1103/PhysRevA.105.053125} {\bibfield  {journal} {\bibinfo
  {journal} {Phys. Rev. A}\ }\textbf {\bibinfo {volume} {105}},\ \bibinfo
  {pages} {053125} (\bibinfo {year} {2022})}\BibitemShut {NoStop}%
\bibitem [{\citenamefont {Uchida}\ \emph {et~al.}(2022)\citenamefont {Uchida},
  \citenamefont {Mattoni}, \citenamefont {Yonezawa}, \citenamefont {Nakamura},
  \citenamefont {Maeno},\ and\ \citenamefont {Tanaka}}]{Uchida2022PRL}%
  \BibitemOpen
  \bibfield  {author} {\bibinfo {author} {\bibfnamefont {K.}~\bibnamefont
  {Uchida}}, \bibinfo {author} {\bibfnamefont {G.}~\bibnamefont {Mattoni}},
  \bibinfo {author} {\bibfnamefont {S.}~\bibnamefont {Yonezawa}}, \bibinfo
  {author} {\bibfnamefont {F.}~\bibnamefont {Nakamura}}, \bibinfo {author}
  {\bibfnamefont {Y.}~\bibnamefont {Maeno}}, \ and\ \bibinfo {author}
  {\bibfnamefont {K.}~\bibnamefont {Tanaka}},\ }\href {\doibase
  10.1103/PhysRevLett.128.127401} {\bibfield  {journal} {\bibinfo  {journal}
  {Phys. Rev. Lett.}\ }\textbf {\bibinfo {volume} {128}},\ \bibinfo {pages}
  {127401} (\bibinfo {year} {2022})}\BibitemShut {NoStop}%
\bibitem [{\citenamefont {Silva}\ \emph {et~al.}(2018)\citenamefont {Silva},
  \citenamefont {Blinov}, \citenamefont {Rubtsov}, \citenamefont {Smirnova},\
  and\ \citenamefont {Ivanov}}]{Silva2018NatPhoton}%
  \BibitemOpen
  \bibfield  {author} {\bibinfo {author} {\bibfnamefont {R.~E.~F.}\
  \bibnamefont {Silva}}, \bibinfo {author} {\bibfnamefont {I.~V.}\ \bibnamefont
  {Blinov}}, \bibinfo {author} {\bibfnamefont {A.~N.}\ \bibnamefont {Rubtsov}},
  \bibinfo {author} {\bibfnamefont {O.}~\bibnamefont {Smirnova}}, \ and\
  \bibinfo {author} {\bibfnamefont {M.}~\bibnamefont {Ivanov}},\ }\href
  {https://doi.org/10.1038/s41566-018-0129-0} {\bibfield  {journal} {\bibinfo
  {journal} {Nat. Photon.}\ }\textbf {\bibinfo {volume} {12}},\ \bibinfo
  {pages} {266} (\bibinfo {year} {2018})}\BibitemShut {NoStop}%
\bibitem [{\citenamefont {Murakami}\ \emph {et~al.}(2018)\citenamefont
  {Murakami}, \citenamefont {Eckstein},\ and\ \citenamefont
  {Werner}}]{Murakami2018PRL}%
  \BibitemOpen
  \bibfield  {author} {\bibinfo {author} {\bibfnamefont {Y.}~\bibnamefont
  {Murakami}}, \bibinfo {author} {\bibfnamefont {M.}~\bibnamefont {Eckstein}},
  \ and\ \bibinfo {author} {\bibfnamefont {P.}~\bibnamefont {Werner}},\ }\href
  {\doibase 10.1103/PhysRevLett.121.057405} {\bibfield  {journal} {\bibinfo
  {journal} {Phys. Rev. Lett.}\ }\textbf {\bibinfo {volume} {121}},\ \bibinfo
  {pages} {057405} (\bibinfo {year} {2018})}\BibitemShut {NoStop}%
\bibitem [{\citenamefont {Murakami}\ and\ \citenamefont
  {Werner}(2018)}]{Murakami2018PRB}%
  \BibitemOpen
  \bibfield  {author} {\bibinfo {author} {\bibfnamefont {Y.}~\bibnamefont
  {Murakami}}\ and\ \bibinfo {author} {\bibfnamefont {P.}~\bibnamefont
  {Werner}},\ }\href {\doibase 10.1103/PhysRevB.98.075102} {\bibfield
  {journal} {\bibinfo  {journal} {Phys. Rev. B}\ }\textbf {\bibinfo {volume}
  {98}},\ \bibinfo {pages} {075102} (\bibinfo {year} {2018})}\BibitemShut
  {NoStop}%
\bibitem [{\citenamefont {Lysne}\ \emph
  {et~al.}(2020{\natexlab{b}})\citenamefont {Lysne}, \citenamefont {Murakami},\
  and\ \citenamefont {Werner}}]{Markus2020}%
  \BibitemOpen
  \bibfield  {author} {\bibinfo {author} {\bibfnamefont {M.}~\bibnamefont
  {Lysne}}, \bibinfo {author} {\bibfnamefont {Y.}~\bibnamefont {Murakami}}, \
  and\ \bibinfo {author} {\bibfnamefont {P.}~\bibnamefont {Werner}},\ }\href
  {\doibase 10.1103/PhysRevB.101.195139} {\bibfield  {journal} {\bibinfo
  {journal} {Phys. Rev. B}\ }\textbf {\bibinfo {volume} {101}},\ \bibinfo
  {pages} {195139} (\bibinfo {year} {2020}{\natexlab{b}})}\BibitemShut
  {NoStop}%
\bibitem [{\citenamefont {Tancogne-Dejean}\ \emph {et~al.}(2018)\citenamefont
  {Tancogne-Dejean}, \citenamefont {Sentef},\ and\ \citenamefont
  {Rubio}}]{Tancogne-Dejean2018}%
  \BibitemOpen
  \bibfield  {author} {\bibinfo {author} {\bibfnamefont {N.}~\bibnamefont
  {Tancogne-Dejean}}, \bibinfo {author} {\bibfnamefont {M.~A.}\ \bibnamefont
  {Sentef}}, \ and\ \bibinfo {author} {\bibfnamefont {A.}~\bibnamefont
  {Rubio}},\ }\href {\doibase 10.1103/PhysRevLett.121.097402} {\bibfield
  {journal} {\bibinfo  {journal} {Phys. Rev. Lett.}\ }\textbf {\bibinfo
  {volume} {121}},\ \bibinfo {pages} {097402} (\bibinfo {year}
  {2018})}\BibitemShut {NoStop}%
\bibitem [{\citenamefont {Imai}\ \emph {et~al.}(2020)\citenamefont {Imai},
  \citenamefont {Ono},\ and\ \citenamefont {Ishihara}}]{Ishihara2020}%
  \BibitemOpen
  \bibfield  {author} {\bibinfo {author} {\bibfnamefont {S.}~\bibnamefont
  {Imai}}, \bibinfo {author} {\bibfnamefont {A.}~\bibnamefont {Ono}}, \ and\
  \bibinfo {author} {\bibfnamefont {S.}~\bibnamefont {Ishihara}},\ }\href
  {\doibase 10.1103/PhysRevLett.124.157404} {\bibfield  {journal} {\bibinfo
  {journal} {Phys. Rev. Lett.}\ }\textbf {\bibinfo {volume} {124}},\ \bibinfo
  {pages} {157404} (\bibinfo {year} {2020})}\BibitemShut {NoStop}%
\bibitem [{\citenamefont {Chinzei}\ and\ \citenamefont
  {Ikeda}(2020)}]{Chinzei2020}%
  \BibitemOpen
  \bibfield  {author} {\bibinfo {author} {\bibfnamefont {K.}~\bibnamefont
  {Chinzei}}\ and\ \bibinfo {author} {\bibfnamefont {T.~N.}\ \bibnamefont
  {Ikeda}},\ }\href {\doibase 10.1103/PhysRevResearch.2.013033} {\bibfield
  {journal} {\bibinfo  {journal} {Phys. Rev. Research}\ }\textbf {\bibinfo
  {volume} {2}},\ \bibinfo {pages} {013033} (\bibinfo {year}
  {2020})}\BibitemShut {NoStop}%
\bibitem [{\citenamefont {Orthodoxou}\ \emph {et~al.}(2021)\citenamefont
  {Orthodoxou}, \citenamefont {Za{\"i}r},\ and\ \citenamefont
  {Booth}}]{Orthodoxou2021}%
  \BibitemOpen
  \bibfield  {author} {\bibinfo {author} {\bibfnamefont {C.}~\bibnamefont
  {Orthodoxou}}, \bibinfo {author} {\bibfnamefont {A.}~\bibnamefont
  {Za{\"i}r}}, \ and\ \bibinfo {author} {\bibfnamefont {G.~H.}\ \bibnamefont
  {Booth}},\ }\href {\doibase 10.1038/s41535-021-00377-8} {\bibfield  {journal}
  {\bibinfo  {journal} {npj Quantum Materials}\ }\textbf {\bibinfo {volume}
  {6}},\ \bibinfo {pages} {76} (\bibinfo {year} {2021})}\BibitemShut {NoStop}%
\bibitem [{\citenamefont {Murakami}\ \emph {et~al.}(2021)\citenamefont
  {Murakami}, \citenamefont {Takayoshi}, \citenamefont {Koga},\ and\
  \citenamefont {Werner}}]{Murakami2021PRB}%
  \BibitemOpen
  \bibfield  {author} {\bibinfo {author} {\bibfnamefont {Y.}~\bibnamefont
  {Murakami}}, \bibinfo {author} {\bibfnamefont {S.}~\bibnamefont {Takayoshi}},
  \bibinfo {author} {\bibfnamefont {A.}~\bibnamefont {Koga}}, \ and\ \bibinfo
  {author} {\bibfnamefont {P.}~\bibnamefont {Werner}},\ }\href {\doibase
  10.1103/PhysRevB.103.035110} {\bibfield  {journal} {\bibinfo  {journal}
  {Phys. Rev. B}\ }\textbf {\bibinfo {volume} {103}},\ \bibinfo {pages}
  {035110} (\bibinfo {year} {2021})}\BibitemShut {NoStop}%
\bibitem [{\citenamefont {Shao}\ \emph {et~al.}(2022)\citenamefont {Shao},
  \citenamefont {Lu}, \citenamefont {Zhang}, \citenamefont {Yu}, \citenamefont
  {Tohyama},\ and\ \citenamefont {Lu}}]{Shao2022PRL}%
  \BibitemOpen
  \bibfield  {author} {\bibinfo {author} {\bibfnamefont {C.}~\bibnamefont
  {Shao}}, \bibinfo {author} {\bibfnamefont {H.}~\bibnamefont {Lu}}, \bibinfo
  {author} {\bibfnamefont {X.}~\bibnamefont {Zhang}}, \bibinfo {author}
  {\bibfnamefont {C.}~\bibnamefont {Yu}}, \bibinfo {author} {\bibfnamefont
  {T.}~\bibnamefont {Tohyama}}, \ and\ \bibinfo {author} {\bibfnamefont
  {R.}~\bibnamefont {Lu}},\ }\href {\doibase 10.1103/PhysRevLett.128.047401}
  {\bibfield  {journal} {\bibinfo  {journal} {Phys. Rev. Lett.}\ }\textbf
  {\bibinfo {volume} {128}},\ \bibinfo {pages} {047401} (\bibinfo {year}
  {2022})}\BibitemShut {NoStop}%
\bibitem [{\citenamefont {Hansen}\ \emph {et~al.}(2022)\citenamefont {Hansen},
  \citenamefont {Jensen},\ and\ \citenamefont {Madsen}}]{Hansen2022arXiv}%
  \BibitemOpen
  \bibfield  {author} {\bibinfo {author} {\bibfnamefont {T.}~\bibnamefont
  {Hansen}}, \bibinfo {author} {\bibfnamefont {S.~V.~B.}\ \bibnamefont
  {Jensen}}, \ and\ \bibinfo {author} {\bibfnamefont {L.~B.}\ \bibnamefont
  {Madsen}},\ }\href@noop {} {\bibfield  {journal} {\bibinfo  {journal}
  {arXiv:2201.06879}\ } (\bibinfo {year} {2022})}\BibitemShut {NoStop}%
\bibitem [{\citenamefont {Masur}\ \emph {et~al.}(2022)\citenamefont {Masur},
  \citenamefont {Bondar},\ and\ \citenamefont {McCaul}}]{Bondar2022arxiv}%
  \BibitemOpen
  \bibfield  {author} {\bibinfo {author} {\bibfnamefont {J.}~\bibnamefont
  {Masur}}, \bibinfo {author} {\bibfnamefont {D.~I.}\ \bibnamefont {Bondar}}, \
  and\ \bibinfo {author} {\bibfnamefont {G.}~\bibnamefont {McCaul}},\
  }\href@noop {} {\bibfield  {journal} {\bibinfo  {journal} {arXiv:2202.06895}\
  } (\bibinfo {year} {2022})}\BibitemShut {NoStop}%
\bibitem [{\citenamefont {Udono}\ \emph {et~al.}(2022)\citenamefont {Udono},
  \citenamefont {Sugimoto}, \citenamefont {Kaneko},\ and\ \citenamefont
  {Ohta}}]{Udono2022PRB}%
  \BibitemOpen
  \bibfield  {author} {\bibinfo {author} {\bibfnamefont {M.}~\bibnamefont
  {Udono}}, \bibinfo {author} {\bibfnamefont {K.}~\bibnamefont {Sugimoto}},
  \bibinfo {author} {\bibfnamefont {T.}~\bibnamefont {Kaneko}}, \ and\ \bibinfo
  {author} {\bibfnamefont {Y.}~\bibnamefont {Ohta}},\ }\href {\doibase
  10.1103/PhysRevB.105.L241108} {\bibfield  {journal} {\bibinfo  {journal}
  {Phys. Rev. B}\ }\textbf {\bibinfo {volume} {105}},\ \bibinfo {pages}
  {L241108} (\bibinfo {year} {2022})}\BibitemShut {NoStop}%
\bibitem [{\citenamefont {Gr\r{a}n{\"a}s}\ \emph {et~al.}(2020)\citenamefont
  {Gr\r{a}n{\"a}s}, \citenamefont {Vaskivskyi}, \citenamefont {Thunstr{\"o}m},
  \citenamefont {Ghimire}, \citenamefont {Knut}, \citenamefont
  {S{\"o}derstr{\"o}m}, \citenamefont {Kjellsson}, \citenamefont {Turenne},
  \citenamefont {Engel}, \citenamefont {Beye}, \citenamefont {Lu},
  \citenamefont {Reid}, \citenamefont {Schlotter}, \citenamefont {Coslovich},
  \citenamefont {Hoffmann}, \citenamefont {Kolesov}, \citenamefont
  {Sch{\"u}{\ss}ler-Langeheine}, \citenamefont {Styervoyedov}, \citenamefont
  {Tancogne-Dejean}, \citenamefont {Sentef}, \citenamefont {Reis},
  \citenamefont {Rubio}, \citenamefont {Parkin}, \citenamefont {Karis},
  \citenamefont {Nordgren}, \citenamefont {Rubensson}, \citenamefont
  {Eriksson},\ and\ \citenamefont {D{\"u}rr}}]{vaskivskyi2020}%
  \BibitemOpen
  \bibfield  {author} {\bibinfo {author} {\bibfnamefont {O.}~\bibnamefont
  {Gr\r{a}n{\"a}s}}, \bibinfo {author} {\bibfnamefont {I.}~\bibnamefont
  {Vaskivskyi}}, \bibinfo {author} {\bibfnamefont {P.}~\bibnamefont
  {Thunstr{\"o}m}}, \bibinfo {author} {\bibfnamefont {S.}~\bibnamefont
  {Ghimire}}, \bibinfo {author} {\bibfnamefont {R.}~\bibnamefont {Knut}},
  \bibinfo {author} {\bibfnamefont {J.}~\bibnamefont {S{\"o}derstr{\"o}m}},
  \bibinfo {author} {\bibfnamefont {L.}~\bibnamefont {Kjellsson}}, \bibinfo
  {author} {\bibfnamefont {D.}~\bibnamefont {Turenne}}, \bibinfo {author}
  {\bibfnamefont {R.~Y.}\ \bibnamefont {Engel}}, \bibinfo {author}
  {\bibfnamefont {M.}~\bibnamefont {Beye}}, \bibinfo {author} {\bibfnamefont
  {J.}~\bibnamefont {Lu}}, \bibinfo {author} {\bibfnamefont {A.~H.}\
  \bibnamefont {Reid}}, \bibinfo {author} {\bibfnamefont {W.}~\bibnamefont
  {Schlotter}}, \bibinfo {author} {\bibfnamefont {G.}~\bibnamefont
  {Coslovich}}, \bibinfo {author} {\bibfnamefont {M.}~\bibnamefont {Hoffmann}},
  \bibinfo {author} {\bibfnamefont {G.}~\bibnamefont {Kolesov}}, \bibinfo
  {author} {\bibfnamefont {C.}~\bibnamefont {Sch{\"u}{\ss}ler-Langeheine}},
  \bibinfo {author} {\bibfnamefont {A.}~\bibnamefont {Styervoyedov}}, \bibinfo
  {author} {\bibfnamefont {N.}~\bibnamefont {Tancogne-Dejean}}, \bibinfo
  {author} {\bibfnamefont {M.~A.}\ \bibnamefont {Sentef}}, \bibinfo {author}
  {\bibfnamefont {D.~A.}\ \bibnamefont {Reis}}, \bibinfo {author}
  {\bibfnamefont {A.}~\bibnamefont {Rubio}}, \bibinfo {author} {\bibfnamefont
  {S.~S.~P.}\ \bibnamefont {Parkin}}, \bibinfo {author} {\bibfnamefont
  {O.}~\bibnamefont {Karis}}, \bibinfo {author} {\bibfnamefont
  {J.}~\bibnamefont {Nordgren}}, \bibinfo {author} {\bibfnamefont {J.~E.}\
  \bibnamefont {Rubensson}}, \bibinfo {author} {\bibfnamefont {O.}~\bibnamefont
  {Eriksson}}, \ and\ \bibinfo {author} {\bibfnamefont {H.~A.}\ \bibnamefont
  {D{\"u}rr}},\ }\href {https://arxiv.org/abs/2008.11115} {\bibfield  {journal}
  {\bibinfo  {journal} {arXiv:2008.11115}\ } (\bibinfo {year}
  {2020})}\BibitemShut {NoStop}%
\bibitem [{\citenamefont {Bionta}\ \emph {et~al.}(2021)\citenamefont {Bionta},
  \citenamefont {Haddad}, \citenamefont {Leblanc}, \citenamefont {Gruson},
  \citenamefont {Lassonde}, \citenamefont {Ibrahim}, \citenamefont {Chaillou},
  \citenamefont {\'Emond}, \citenamefont {Otto}, \citenamefont
  {Jim\'enez-Gal\'an}, \citenamefont {Silva}, \citenamefont {Ivanov},
  \citenamefont {Siwick}, \citenamefont {Chaker},\ and\ \citenamefont
  {L\'egar\'e}}]{Bionta2021PRR}%
  \BibitemOpen
  \bibfield  {author} {\bibinfo {author} {\bibfnamefont {M.~R.}\ \bibnamefont
  {Bionta}}, \bibinfo {author} {\bibfnamefont {E.}~\bibnamefont {Haddad}},
  \bibinfo {author} {\bibfnamefont {A.}~\bibnamefont {Leblanc}}, \bibinfo
  {author} {\bibfnamefont {V.}~\bibnamefont {Gruson}}, \bibinfo {author}
  {\bibfnamefont {P.}~\bibnamefont {Lassonde}}, \bibinfo {author}
  {\bibfnamefont {H.}~\bibnamefont {Ibrahim}}, \bibinfo {author} {\bibfnamefont
  {J.}~\bibnamefont {Chaillou}}, \bibinfo {author} {\bibfnamefont
  {N.}~\bibnamefont {\'Emond}}, \bibinfo {author} {\bibfnamefont {M.~R.}\
  \bibnamefont {Otto}}, \bibinfo {author} {\bibfnamefont {A.}~\bibnamefont
  {Jim\'enez-Gal\'an}}, \bibinfo {author} {\bibfnamefont {R.~E.~F.}\
  \bibnamefont {Silva}}, \bibinfo {author} {\bibfnamefont {M.}~\bibnamefont
  {Ivanov}}, \bibinfo {author} {\bibfnamefont {B.~J.}\ \bibnamefont {Siwick}},
  \bibinfo {author} {\bibfnamefont {M.}~\bibnamefont {Chaker}}, \ and\ \bibinfo
  {author} {\bibfnamefont {F.~m.~c.}\ \bibnamefont {L\'egar\'e}},\ }\href
  {\doibase 10.1103/PhysRevResearch.3.023250} {\bibfield  {journal} {\bibinfo
  {journal} {Phys. Rev. Research}\ }\textbf {\bibinfo {volume} {3}},\ \bibinfo
  {pages} {023250} (\bibinfo {year} {2021})}\BibitemShut {NoStop}%
\bibitem [{\citenamefont {Imada}\ \emph {et~al.}(1998)\citenamefont {Imada},
  \citenamefont {Fujimori},\ and\ \citenamefont {Tokura}}]{Tokura_RMP}%
  \BibitemOpen
  \bibfield  {author} {\bibinfo {author} {\bibfnamefont {M.}~\bibnamefont
  {Imada}}, \bibinfo {author} {\bibfnamefont {A.}~\bibnamefont {Fujimori}}, \
  and\ \bibinfo {author} {\bibfnamefont {Y.}~\bibnamefont {Tokura}},\ }\href
  {\doibase 10.1103/RevModPhys.70.1039} {\bibfield  {journal} {\bibinfo
  {journal} {Rev. Mod. Phys.}\ }\textbf {\bibinfo {volume} {70}},\ \bibinfo
  {pages} {1039} (\bibinfo {year} {1998})}\BibitemShut {NoStop}%
\bibitem [{\citenamefont {Dagotto}(1994)}]{Dagotto1994RMP}%
  \BibitemOpen
  \bibfield  {author} {\bibinfo {author} {\bibfnamefont {E.}~\bibnamefont
  {Dagotto}},\ }\href {\doibase 10.1103/RevModPhys.66.763} {\bibfield
  {journal} {\bibinfo  {journal} {Rev. Mod. Phys.}\ }\textbf {\bibinfo {volume}
  {66}},\ \bibinfo {pages} {763} (\bibinfo {year} {1994})}\BibitemShut
  {NoStop}%
\bibitem [{Note1()}]{Note1}%
  \BibitemOpen
  \bibinfo {note} {Supplemental Material [url], which includes Ref. \cite
  {Werner2018PRB}.}\BibitemShut {Stop}%
\bibitem [{\citenamefont {Georges}\ \emph {et~al.}(1996)\citenamefont
  {Georges}, \citenamefont {Kotliar}, \citenamefont {Krauth},\ and\
  \citenamefont {Rozenberg}}]{Georges1996}%
  \BibitemOpen
  \bibfield  {author} {\bibinfo {author} {\bibfnamefont {A.}~\bibnamefont
  {Georges}}, \bibinfo {author} {\bibfnamefont {G.}~\bibnamefont {Kotliar}},
  \bibinfo {author} {\bibfnamefont {W.}~\bibnamefont {Krauth}}, \ and\ \bibinfo
  {author} {\bibfnamefont {M.~J.}\ \bibnamefont {Rozenberg}},\ }\href {\doibase
  10.1103/RevModPhys.68.13} {\bibfield  {journal} {\bibinfo  {journal} {Rev.
  Mod. Phys.}\ }\textbf {\bibinfo {volume} {68}},\ \bibinfo {pages} {13}
  (\bibinfo {year} {1996})}\BibitemShut {NoStop}%
\bibitem [{\citenamefont {Eckstein}\ and\ \citenamefont
  {Werner}(2010)}]{Eckstein2010b}%
  \BibitemOpen
  \bibfield  {author} {\bibinfo {author} {\bibfnamefont {M.}~\bibnamefont
  {Eckstein}}\ and\ \bibinfo {author} {\bibfnamefont {P.}~\bibnamefont
  {Werner}},\ }\href {\doibase 10.1103/PhysRevB.82.115115} {\bibfield
  {journal} {\bibinfo  {journal} {Phys. Rev. B}\ }\textbf {\bibinfo {volume}
  {82}},\ \bibinfo {pages} {115115} (\bibinfo {year} {2010})}\BibitemShut
  {NoStop}%
\bibitem [{\citenamefont {Aoki}\ \emph {et~al.}(2014)\citenamefont {Aoki},
  \citenamefont {Tsuji}, \citenamefont {Eckstein}, \citenamefont {Kollar},
  \citenamefont {Oka},\ and\ \citenamefont {Werner}}]{Aoki2013}%
  \BibitemOpen
  \bibfield  {author} {\bibinfo {author} {\bibfnamefont {H.}~\bibnamefont
  {Aoki}}, \bibinfo {author} {\bibfnamefont {N.}~\bibnamefont {Tsuji}},
  \bibinfo {author} {\bibfnamefont {M.}~\bibnamefont {Eckstein}}, \bibinfo
  {author} {\bibfnamefont {M.}~\bibnamefont {Kollar}}, \bibinfo {author}
  {\bibfnamefont {T.}~\bibnamefont {Oka}}, \ and\ \bibinfo {author}
  {\bibfnamefont {P.}~\bibnamefont {Werner}},\ }\href {\doibase
  10.1103/RevModPhys.86.779} {\bibfield  {journal} {\bibinfo  {journal} {Rev.
  Mod. Phys.}\ }\textbf {\bibinfo {volume} {86}},\ \bibinfo {pages} {779}
  (\bibinfo {year} {2014})}\BibitemShut {NoStop}%
\bibitem [{\citenamefont {Sandholzer}\ \emph {et~al.}(2019)\citenamefont
  {Sandholzer}, \citenamefont {Murakami}, \citenamefont {G\"org}, \citenamefont
  {Minguzzi}, \citenamefont {Messer}, \citenamefont {Desbuquois}, \citenamefont
  {Eckstein}, \citenamefont {Werner},\ and\ \citenamefont
  {Esslinger}}]{Kilian2019PRL}%
  \BibitemOpen
  \bibfield  {author} {\bibinfo {author} {\bibfnamefont {K.}~\bibnamefont
  {Sandholzer}}, \bibinfo {author} {\bibfnamefont {Y.}~\bibnamefont
  {Murakami}}, \bibinfo {author} {\bibfnamefont {F.}~\bibnamefont {G\"org}},
  \bibinfo {author} {\bibfnamefont {J.}~\bibnamefont {Minguzzi}}, \bibinfo
  {author} {\bibfnamefont {M.}~\bibnamefont {Messer}}, \bibinfo {author}
  {\bibfnamefont {R.}~\bibnamefont {Desbuquois}}, \bibinfo {author}
  {\bibfnamefont {M.}~\bibnamefont {Eckstein}}, \bibinfo {author}
  {\bibfnamefont {P.}~\bibnamefont {Werner}}, \ and\ \bibinfo {author}
  {\bibfnamefont {T.}~\bibnamefont {Esslinger}},\ }\href {\doibase
  10.1103/PhysRevLett.123.193602} {\bibfield  {journal} {\bibinfo  {journal}
  {Phys. Rev. Lett.}\ }\textbf {\bibinfo {volume} {123}},\ \bibinfo {pages}
  {193602} (\bibinfo {year} {2019})}\BibitemShut {NoStop}%
\bibitem [{\citenamefont {Sch\"{u}ler}\ \emph {et~al.}(2020)\citenamefont
  {Sch\"{u}ler}, \citenamefont {Gole\v{z}}, \citenamefont {Murakami},
  \citenamefont {Bittner}, \citenamefont {Herrmann}, \citenamefont {Strand},
  \citenamefont {Werner},\ and\ \citenamefont {Eckstein}}]{Nessi2020}%
  \BibitemOpen
  \bibfield  {author} {\bibinfo {author} {\bibfnamefont {M.}~\bibnamefont
  {Sch\"{u}ler}}, \bibinfo {author} {\bibfnamefont {D.}~\bibnamefont
  {Gole\v{z}}}, \bibinfo {author} {\bibfnamefont {Y.}~\bibnamefont {Murakami}},
  \bibinfo {author} {\bibfnamefont {N.}~\bibnamefont {Bittner}}, \bibinfo
  {author} {\bibfnamefont {A.}~\bibnamefont {Herrmann}}, \bibinfo {author}
  {\bibfnamefont {H.~U.}\ \bibnamefont {Strand}}, \bibinfo {author}
  {\bibfnamefont {P.}~\bibnamefont {Werner}}, \ and\ \bibinfo {author}
  {\bibfnamefont {M.}~\bibnamefont {Eckstein}},\ }\href {\doibase
  https://doi.org/10.1016/j.cpc.2020.107484} {\bibfield  {journal} {\bibinfo
  {journal} {Computer Physics Communications}\ }\textbf {\bibinfo {volume}
  {257}},\ \bibinfo {pages} {107484} (\bibinfo {year} {2020})}\BibitemShut
  {NoStop}%
\bibitem [{\citenamefont {Werner}\ \emph {et~al.}(2017)\citenamefont {Werner},
  \citenamefont {Strand}, \citenamefont {Hoshino},\ and\ \citenamefont
  {Eckstein}}]{Werner2017}%
  \BibitemOpen
  \bibfield  {author} {\bibinfo {author} {\bibfnamefont {P.}~\bibnamefont
  {Werner}}, \bibinfo {author} {\bibfnamefont {H.~U.~R.}\ \bibnamefont
  {Strand}}, \bibinfo {author} {\bibfnamefont {S.}~\bibnamefont {Hoshino}}, \
  and\ \bibinfo {author} {\bibfnamefont {M.}~\bibnamefont {Eckstein}},\ }\href
  {\doibase 10.1103/PhysRevB.95.195405} {\bibfield  {journal} {\bibinfo
  {journal} {Phys. Rev. B}\ }\textbf {\bibinfo {volume} {95}},\ \bibinfo
  {pages} {195405} (\bibinfo {year} {2017})}\BibitemShut {NoStop}%
\bibitem [{\citenamefont {Martinez}\ and\ \citenamefont
  {Horsch}(1991)}]{Martinez1991PRB}%
  \BibitemOpen
  \bibfield  {author} {\bibinfo {author} {\bibfnamefont {G.}~\bibnamefont
  {Martinez}}\ and\ \bibinfo {author} {\bibfnamefont {P.}~\bibnamefont
  {Horsch}},\ }\href {\doibase 10.1103/PhysRevB.44.317} {\bibfield  {journal}
  {\bibinfo  {journal} {Phys. Rev. B}\ }\textbf {\bibinfo {volume} {44}},\
  \bibinfo {pages} {317} (\bibinfo {year} {1991})}\BibitemShut {NoStop}%
\bibitem [{\citenamefont {Sangiovanni}\ \emph {et~al.}(2006)\citenamefont
  {Sangiovanni}, \citenamefont {Toschi}, \citenamefont {Koch}, \citenamefont
  {Held}, \citenamefont {Capone}, \citenamefont {Castellani}, \citenamefont
  {Gunnarsson}, \citenamefont {Mo}, \citenamefont {Allen}, \citenamefont {Kim},
  \citenamefont {Sekiyama}, \citenamefont {Yamasaki}, \citenamefont {Suga},\
  and\ \citenamefont {Metcalf}}]{Sangiovanni2006}%
  \BibitemOpen
  \bibfield  {author} {\bibinfo {author} {\bibfnamefont {G.}~\bibnamefont
  {Sangiovanni}}, \bibinfo {author} {\bibfnamefont {A.}~\bibnamefont {Toschi}},
  \bibinfo {author} {\bibfnamefont {E.}~\bibnamefont {Koch}}, \bibinfo {author}
  {\bibfnamefont {K.}~\bibnamefont {Held}}, \bibinfo {author} {\bibfnamefont
  {M.}~\bibnamefont {Capone}}, \bibinfo {author} {\bibfnamefont
  {C.}~\bibnamefont {Castellani}}, \bibinfo {author} {\bibfnamefont
  {O.}~\bibnamefont {Gunnarsson}}, \bibinfo {author} {\bibfnamefont {S.-K.}\
  \bibnamefont {Mo}}, \bibinfo {author} {\bibfnamefont {J.~W.}\ \bibnamefont
  {Allen}}, \bibinfo {author} {\bibfnamefont {H.-D.}\ \bibnamefont {Kim}},
  \bibinfo {author} {\bibfnamefont {A.}~\bibnamefont {Sekiyama}}, \bibinfo
  {author} {\bibfnamefont {A.}~\bibnamefont {Yamasaki}}, \bibinfo {author}
  {\bibfnamefont {S.}~\bibnamefont {Suga}}, \ and\ \bibinfo {author}
  {\bibfnamefont {P.}~\bibnamefont {Metcalf}},\ }\href {\doibase
  10.1103/PhysRevB.73.205121} {\bibfield  {journal} {\bibinfo  {journal} {Phys.
  Rev. B}\ }\textbf {\bibinfo {volume} {73}},\ \bibinfo {pages} {205121}
  (\bibinfo {year} {2006})}\BibitemShut {NoStop}%
\bibitem [{\citenamefont {Lenar\ifmmode \check{c}\else
  \v{c}\fi{}i\ifmmode~\check{c}\else \v{c}\fi{}}\ and\ \citenamefont
  {Prelov\ifmmode~\check{s}\else \v{s}\fi{}ek}(2013)}]{Zala2013PRL}%
  \BibitemOpen
  \bibfield  {author} {\bibinfo {author} {\bibfnamefont {Z.}~\bibnamefont
  {Lenar\ifmmode \check{c}\else \v{c}\fi{}i\ifmmode~\check{c}\else
  \v{c}\fi{}}}\ and\ \bibinfo {author} {\bibfnamefont {P.}~\bibnamefont
  {Prelov\ifmmode~\check{s}\else \v{s}\fi{}ek}},\ }\href {\doibase
  10.1103/PhysRevLett.111.016401} {\bibfield  {journal} {\bibinfo  {journal}
  {Phys. Rev. Lett.}\ }\textbf {\bibinfo {volume} {111}},\ \bibinfo {pages}
  {016401} (\bibinfo {year} {2013})}\BibitemShut {NoStop}%
\bibitem [{\citenamefont {Gole\ifmmode~\check{z}\else \v{z}\fi{}}\ \emph
  {et~al.}(2014)\citenamefont {Gole\ifmmode~\check{z}\else \v{z}\fi{}},
  \citenamefont {Bon\ifmmode~\check{c}\else \v{c}\fi{}a}, \citenamefont
  {Mierzejewski},\ and\ \citenamefont {Vidmar}}]{Denis2014PRB}%
  \BibitemOpen
  \bibfield  {author} {\bibinfo {author} {\bibfnamefont {D.}~\bibnamefont
  {Gole\ifmmode~\check{z}\else \v{z}\fi{}}}, \bibinfo {author} {\bibfnamefont
  {J.}~\bibnamefont {Bon\ifmmode~\check{c}\else \v{c}\fi{}a}}, \bibinfo
  {author} {\bibfnamefont {M.}~\bibnamefont {Mierzejewski}}, \ and\ \bibinfo
  {author} {\bibfnamefont {L.}~\bibnamefont {Vidmar}},\ }\href {\doibase
  10.1103/PhysRevB.89.165118} {\bibfield  {journal} {\bibinfo  {journal} {Phys.
  Rev. B}\ }\textbf {\bibinfo {volume} {89}},\ \bibinfo {pages} {165118}
  (\bibinfo {year} {2014})}\BibitemShut {NoStop}%
\bibitem [{\citenamefont {Eckstein}\ and\ \citenamefont
  {Werner}(2016)}]{Eckstein2016}%
  \BibitemOpen
  \bibfield  {author} {\bibinfo {author} {\bibfnamefont {M.}~\bibnamefont
  {Eckstein}}\ and\ \bibinfo {author} {\bibfnamefont {P.}~\bibnamefont
  {Werner}},\ }\href@noop {} {\bibfield  {journal} {\bibinfo  {journal}
  {Scientific Reports}\ }\textbf {\bibinfo {volume} {6}},\ \bibinfo {pages}
  {21235} (\bibinfo {year} {2016})}\BibitemShut {NoStop}%
\bibitem [{\citenamefont {Vidal}(2003)}]{Vidal2003PRL}%
  \BibitemOpen
  \bibfield  {author} {\bibinfo {author} {\bibfnamefont {G.}~\bibnamefont
  {Vidal}},\ }\href {\doibase 10.1103/PhysRevLett.91.147902} {\bibfield
  {journal} {\bibinfo  {journal} {Phys. Rev. Lett.}\ }\textbf {\bibinfo
  {volume} {91}},\ \bibinfo {pages} {147902} (\bibinfo {year}
  {2003})}\BibitemShut {NoStop}%
\bibitem [{\citenamefont {Kilen}\ \emph {et~al.}(2020)\citenamefont {Kilen},
  \citenamefont {Kolesik}, \citenamefont {Hader}, \citenamefont {Moloney},
  \citenamefont {Huttner}, \citenamefont {Hagen},\ and\ \citenamefont
  {Koch}}]{Kilen2020PRL}%
  \BibitemOpen
  \bibfield  {author} {\bibinfo {author} {\bibfnamefont {I.}~\bibnamefont
  {Kilen}}, \bibinfo {author} {\bibfnamefont {M.}~\bibnamefont {Kolesik}},
  \bibinfo {author} {\bibfnamefont {J.}~\bibnamefont {Hader}}, \bibinfo
  {author} {\bibfnamefont {J.~V.}\ \bibnamefont {Moloney}}, \bibinfo {author}
  {\bibfnamefont {U.}~\bibnamefont {Huttner}}, \bibinfo {author} {\bibfnamefont
  {M.~K.}\ \bibnamefont {Hagen}}, \ and\ \bibinfo {author} {\bibfnamefont
  {S.~W.}\ \bibnamefont {Koch}},\ }\href {\doibase
  10.1103/PhysRevLett.125.083901} {\bibfield  {journal} {\bibinfo  {journal}
  {Phys. Rev. Lett.}\ }\textbf {\bibinfo {volume} {125}},\ \bibinfo {pages}
  {083901} (\bibinfo {year} {2020})}\BibitemShut {NoStop}%
\bibitem [{\citenamefont {Becker}\ \emph {et~al.}(1988)\citenamefont {Becker},
  \citenamefont {Fragnito}, \citenamefont {Cruz}, \citenamefont {Fork},
  \citenamefont {Cunningham}, \citenamefont {Henry},\ and\ \citenamefont
  {Shank}}]{Becker1988PRL}%
  \BibitemOpen
  \bibfield  {author} {\bibinfo {author} {\bibfnamefont {P.~C.}\ \bibnamefont
  {Becker}}, \bibinfo {author} {\bibfnamefont {H.~L.}\ \bibnamefont
  {Fragnito}}, \bibinfo {author} {\bibfnamefont {C.~H.~B.}\ \bibnamefont
  {Cruz}}, \bibinfo {author} {\bibfnamefont {R.~L.}\ \bibnamefont {Fork}},
  \bibinfo {author} {\bibfnamefont {J.~E.}\ \bibnamefont {Cunningham}},
  \bibinfo {author} {\bibfnamefont {J.~E.}\ \bibnamefont {Henry}}, \ and\
  \bibinfo {author} {\bibfnamefont {C.~V.}\ \bibnamefont {Shank}},\ }\href
  {\doibase 10.1103/PhysRevLett.61.1647} {\bibfield  {journal} {\bibinfo
  {journal} {Phys. Rev. Lett.}\ }\textbf {\bibinfo {volume} {61}},\ \bibinfo
  {pages} {1647} (\bibinfo {year} {1988})}\BibitemShut {NoStop}%
\bibitem [{\citenamefont {Nagai}\ \emph {et~al.}(2021)\citenamefont {Nagai},
  \citenamefont {Uchida}, \citenamefont {Kusaba}, \citenamefont {Endo},
  \citenamefont {Miyata},\ and\ \citenamefont {Tanaka}}]{Nagai2022}%
  \BibitemOpen
  \bibfield  {author} {\bibinfo {author} {\bibfnamefont {K.}~\bibnamefont
  {Nagai}}, \bibinfo {author} {\bibfnamefont {K.}~\bibnamefont {Uchida}},
  \bibinfo {author} {\bibfnamefont {S.}~\bibnamefont {Kusaba}}, \bibinfo
  {author} {\bibfnamefont {T.}~\bibnamefont {Endo}}, \bibinfo {author}
  {\bibfnamefont {Y.}~\bibnamefont {Miyata}}, \ and\ \bibinfo {author}
  {\bibfnamefont {K.}~\bibnamefont {Tanaka}},\ }\href@noop {} {\enquote
  {\bibinfo {title} {Effect of incoherent electron-hole pairs on high harmonic
  generation in atomically thin semiconductors},}\ } (\bibinfo {year} {2021}),\
  \Eprint {http://arxiv.org/abs/2112.12951} {arXiv:2112.12951 [physics.optics]}
  \BibitemShut {NoStop}%
\bibitem [{\citenamefont {Strand}\ \emph {et~al.}(2017)\citenamefont {Strand},
  \citenamefont {Gole\ifmmode~\check{z}\else \v{z}\fi{}}, \citenamefont
  {Eckstein},\ and\ \citenamefont {Werner}}]{Hugo2017PRB}%
  \BibitemOpen
  \bibfield  {author} {\bibinfo {author} {\bibfnamefont {H.~U.~R.}\
  \bibnamefont {Strand}}, \bibinfo {author} {\bibfnamefont {D.}~\bibnamefont
  {Gole\ifmmode~\check{z}\else \v{z}\fi{}}}, \bibinfo {author} {\bibfnamefont
  {M.}~\bibnamefont {Eckstein}}, \ and\ \bibinfo {author} {\bibfnamefont
  {P.}~\bibnamefont {Werner}},\ }\href {\doibase 10.1103/PhysRevB.96.165104}
  {\bibfield  {journal} {\bibinfo  {journal} {Phys. Rev. B}\ }\textbf {\bibinfo
  {volume} {96}},\ \bibinfo {pages} {165104} (\bibinfo {year}
  {2017})}\BibitemShut {NoStop}%
\bibitem [{\citenamefont {Werner}\ \emph {et~al.}(2018)\citenamefont {Werner},
  \citenamefont {Strand}, \citenamefont {Hoshino}, \citenamefont {Murakami},\
  and\ \citenamefont {Eckstein}}]{Werner2018PRB}%
  \BibitemOpen
  \bibfield  {author} {\bibinfo {author} {\bibfnamefont {P.}~\bibnamefont
  {Werner}}, \bibinfo {author} {\bibfnamefont {H.~U.~R.}\ \bibnamefont
  {Strand}}, \bibinfo {author} {\bibfnamefont {S.}~\bibnamefont {Hoshino}},
  \bibinfo {author} {\bibfnamefont {Y.}~\bibnamefont {Murakami}}, \ and\
  \bibinfo {author} {\bibfnamefont {M.}~\bibnamefont {Eckstein}},\ }\href
  {\doibase 10.1103/PhysRevB.97.165119} {\bibfield  {journal} {\bibinfo
  {journal} {Phys. Rev. B}\ }\textbf {\bibinfo {volume} {97}},\ \bibinfo
  {pages} {165119} (\bibinfo {year} {2018})}\BibitemShut {NoStop}%
\end{thebibliography}%

\end{document}